\documentclass[fleqn,usenatbib]{mnras}

\usepackage{newtxtext,newtxmath}

\usepackage[T1]{fontenc}
\usepackage{ae,aecompl}

\usepackage{float}
\usepackage{array}
\usepackage{relsize}
\usepackage{graphicx}	
\usepackage{amsmath}	
\usepackage[normalem]{ulem} 

\title[MARVELS Chemo-kinematics]{Chemo-kinematics of the Milky Way from the SDSS-III MARVELS Survey}

\author[N. Grieves et al.]{Nolan Grieves,$^{1}$\thanks{E-mail: ngrieves@ufl.edu}
Jian Ge,$^{1}$
Neil Thomas,$^{2}$
Kevin Willis,$^{1}$
Bo Ma,$^{1}$
Diego Lorenzo-Oliveira,$^{3,4}$
\newauthor
A. B. A. Queiroz,$^{5,4}$
Luan Ghezzi,$^{6}$
Cristina Chiappini,$^{7,4}$
Friedrich Anders,$^{7,4}$
\newauthor
Let\'{\i}cia Dutra-Ferreira,$^{5,4}$
Gustavo F. Porto de Mello,$^{8,4}$
Bas\'{\i}lio X. Santiago,$^{5,4}$
\newauthor
Luiz N. da Costa,$^{6,4}$
Ricardo L. C. Ogando,$^{6,4}$
E. F. del Peloso,$^{4}$
Jonathan C. Tan,$^{9,1}$
\newauthor
Donald P. Schneider,$^{10,11}$
Joshua Pepper,$^{12}$
Keivan G. Stassun,$^{13}$
Bo Zhao,$^{1}$
\newauthor
Dmitry Bizyaev,$^{14,15}$
and Kaike Pan$^{14}$
\\
$^{1}$Department of Astronomy, University of Florida, Gainesville, FL 32611, USA\\
$^{2}$Department of Astronautical Engineering, United States Air Force Academy, CO 80840, USA\\
$^{3}$Universidade de S\~ao Paulo, Departamento de Astronomia IAG/USP, Rua do Mat\~ao 1226, Cidade Universit\'aria, S\~ao Paulo, SP 05508-900, Brazil\\
$^{4}$Laborat\'orio Interinstitucional de e-Astronomia-LIneA, Rua Gereral Jos\'e Cristino 77, S\~ao Crist\'ov\~ao, Rio de Janeiro, RJ 20921-400, Brazil\\
$^{5}$Instituto de F\'\i sica, Universidade Federal do Rio Grande do  Sul, Caixa Postal 15051,Porto Alegre, RS - 91501-970, Brazil\\
$^{6}$Observat\'orio Nacional, Rua General Jos\'e Cristino 77, S\~ao Crist\'ov\~ao, Rio de Janeiro, RJ 20921-400, Brazil\\
$^{7}$Leibniz-Institut f\"ur Astrophysik Potsdam, An der Sternwarte 16, 14482 Potsdam, Germany\\
$^{8}$Observat\'orio do Valongo, Universidade Federal do Rio de Janeiro,  Ladeira do Pedro Ant\^onio 43, Rio de Janeiro, RJ 20080-090, Brazil\\
$^{9}$Department of Astronomy, University of Virginia, Charlottesville, VA 22904\\
$^{10}$Department of Astronomy and Astrophysics, The Pennsylvania State University, University Park, PA 16802\\
$^{11}$Center for Exoplanets and Habitable Worlds, The Pennsylvania State University, University Park, PA 16802\\
$^{12}$Department of Physics, Lehigh University, 16 Memorial Drive East, Bethlehem, PA, 18015, USA\\
$^{13}$Vanderbilt University, Physics \& Astronomy Department, 6301 Stevenson Center Ln., Nashville, TN 37235\\
$^{14}$Apache Point Observatory and New Mexico State University, P.O. Box 59, Sunspot, NM, 88349-0059, USA\\
$^{15}$Sternberg Astronomical Institute, Moscow State University, Moscow, Russia\\
}

\date{Accepted 2018 August 31. Received 2018 August 7; in original form 2018 April 4}

\pubyear{2018}

\begin{document}
\label{firstpage}
\pagerange{\pageref{firstpage}--\pageref{lastpage}}
\maketitle

\begin{abstract}
Combining stellar atmospheric parameters, such as effective temperature, surface gravity, and metallicity, with barycentric radial velocity data provides insight into the chemo-dynamics of the Milky Way and our local Galactic environment. We analyze 3075 stars with spectroscopic data from the Sloan Digital Sky Survey III (SDSS-III) MARVELS radial velocity survey and present atmospheric parameters for 2343 dwarf stars using the spectral indices method, a modified version of the equivalent width method. We present barycentric radial velocities for   a sample of  2610 stars with  a median uncertainty  of 0.3 km s$^{-1}$. We determine stellar ages using two independent methods and calculate ages for  2335  stars with a maximum-likelihood isochronal age-dating method and for  2194  stars with a Bayesian age-dating method. Using previously published parallax data we compute Galactic orbits and space velocities for  2504  stars to explore stellar populations based on kinematic and age parameters. This study combines good ages and exquisite velocities to explore local chemo-kinematics of the Milky Way, which complements many of the recent studies of giant stars with the APOGEE survey, and we find our results to be in agreement with current chemo-dynamical models of the Milky Way. Particularly, we find from our metallicity distributions and velocity-age relations of a kinematically-defined thin disk that the metal rich end has stars of all ages, even after we clean the sample of highly eccentric stars, suggesting that radial migration plays a key role in the metallicity scatter of the thin disk. All stellar parameters and kinematic data derived in this work are catalogued and published online in machine-readable form.
\end{abstract}



\section{Introduction} \label{sec:intro}

Studying the positions, kinematics, and chemical compositions of Galactic stars allows insight into the formation and evolution of the Milky Way \citep[e.g.,][]{Majewski1993,Freeman2002,Nordstrom2004,Rix2013}. Specifically, obtaining precise stellar atmospheric parameters and absolute (barycentric) radial velocities for stars in the local solar neighborhood is critical in understanding our Galactic environment. Solar-type stars are ideal for investigating the chemical evolution of the solar neighborhood and the overall Galaxy as their atmospheric compositions remain relatively unchanged during their long lifetimes, allowing investigation into a substantial fraction of the Milky Way's history. Combining this data with stellar radial velocities generates information on the chemo-dynamics of stars and ongoing processes in the Galaxy. In addition, obtaining kinematic and atmospheric information of the host stars of extra-solar planets is crucial to understanding the varying conditions in which planets can form and survive.

Recently, large surveys using multi-fiber spectrographs have helped to illuminate the history of the Milky Way and characterize large populations of stars. Specifically the Sloan Digital Sky Survey \citep[SDSS;][]{York2000} and its legacy surveys have produced several large-scale spectroscopic studies designed to precisely characterize large populations of stars and the Milky Way's structure and evolution. The Sloan Extension for Galactic Understanding and Exploration \citep[SEGUE;][]{Yanny2009} and its continuation SEGUE-2 investigated the Milky Way's structure by observing over 358,000 stars covering 2500 deg$^{2}$ of sky with a spectral resolution of R $\equiv$ $\lambda$/$\Delta\lambda$ $\approx$ 1800. In order to gain insight into the Galaxy's dynamical structure and chemical history, the Apache Point Observatory Galactic Evolution Experiment \citep[APOGEE;][]{Majewski2017} observed over 100,000 evolved late-type stars spanning the Galactic disk, bulge, and halo with a spectral resolution of R $\sim$ 22,000 in the infrared (1.51-1.70 $\mu$m). 

Here we study stellar kinematics and characteristics using spectra from the Sloan Digital Sky Survey III \citep[SDSS-III;][]{Eisenstein2011} Multi-object APO Radial Velocity Exoplanet Large-area Survey \citep[MARVELS;][]{Ge2008} taken with the SDSS 2.5-m telescope at the Apache Point Observatory
\citep{Gunn2006}. MARVELS used a fibre-fed dispersed fixed delay interferometer (DFDI) combined with a medium resolution \citep[R $\sim$ 11,000;][]{Ge2009} spectrograph to observe $\sim$5,500 stars with a goal of characterizing short-to-intermediate period giant planets in a large and homogenous sample of stars. The MARVELS survey complements APOGEE in that it focused on observing FGK dwarf stars in the optical (5000 - 5700 \AA) rather than red giants in the infrared. \citet{Grieves2017} compares the latest MARVELS radial velocity set from the University of Florida Two Dimensional \citep[UF2D;][]{Thomas2015} data processing pipeline to previous MARVELS pipeline results, while \citet{Alam2015} presents an overview of previous MARVELS data reductions. 

We present a new radial velocity data set from the MARVELS survey using an independent spectral wavelength solution pipeline \citep{Thomas2015}. The wavelength solutions from this new MARVELS pipeline allow determination of absolute radial velocities. These measurements can produce accurate Galactic space velocities when parallax and proper motion measurements are available, especially in view of the  recent  second \textit{Gaia} data release \citep[\textit{Gaia} DR2;][]{Gaia2018}. We present space velocities when these data are available. We also derive stellar atmospheric parameters (log $\textit{g}$, T$_{\text{eff}}$, and [Fe/H]) as well as mass and radius values for the dwarf stars in our sample using spectral indices (specific spectral regions combining multiple absorption lines into broad and blended features). \citet{Ghezzi2014} used the spectral indices method to obtain accurate atmospheric parameters for 30 stars using MARVELS spectra. We extend this work using the spectral indices method to determine atmospheric parameters of all MARVELS dwarf stars with robust spectra in the latest University of Florida One Dimensional (UF1D) pipeline \citep{Thomas2016}. 

In $\S$ \ref{sec:specind} we describe the spectral indices method and its application to the MARVELS spectra. In $\S$ \ref{sec:atmpar} we present our atmospheric parameters for MARVELS dwarf stars, compare these results to previous surveys, and provide our dwarf and giant star classifications. We describe our method to obtain absolute radial velocities, and compare these values to previous surveys in $\S$ \ref{sec:absrvs}. In $\S$ \ref{sec:galvel} we determine Galactic space velocities and Galactic orbital parameters for our absolute radial velocity stars that have external parallax and proper motion values. In $\S$ \ref{sec:ages} we determine ages for a sample of our stars. In $\S$ \ref{sec:distance} we present the distances for our sample.  In $\S$ \ref{sec:results} we discuss our results and investigate the Galactic chemo-kinematics of these stars and distributions of their metallicities, ages, and other characteristics. In $\S$ \ref{sec:conc} we summarize our conclusions.

\section{The Spectral Indices Method} \label{sec:specind}

As is the case for many recent large-scale spectroscopic surveys such as SEGUE, APOGEE, the RAdial Velocity Experiment  \citep[RAVE;][]{Steinmetz2006}, or the LAMOST Experiment for  Galactic Understanding and Exploration \citep[LEGUE;][]{Zhao2012}, MARVELS operates at a moderate spectral resolution to obtain a larger sample than would be possible with higher resolution instruments. However, accurate stellar characterization and atmospheric parameters are difficult to obtain with moderate resolution spectra because spectral features are subject to a high degree of blending. This severe blending of atomic lines and spectral features render it unfeasible to perform classical spectroscopic methods, e.g., \citet{Sousa2014}, that depend on measurements of the equivalent widths (EWs) of individual lines. Therefore, many surveys with low to moderate resolution have employed the spectral synthesis technique to obtain atmospheric parameters such as SEGUE \citep{Lee2008, Smolinski2011}, APOGEE \citep{GarciaPerez2016}, RAVE \citep{Kunder2017}, and LAMOST \citep{Wu2011,Wu2014}. However, as detailed in \citet{Ghezzi2014} the spectral synthesis method has a number of drawbacks, including a dependency on the completeness and accuracy of atomic line databases, the need to accurately determine broadening parameters (instrument profile, macro turbulence, and rotational velocities), and parameters are often more correlated than results obtained from classical model atmosphere analysis.

\citet{Ghezzi2014} developed the spectral indices method as an alternative approach to the spectral synthesis technique to obtain accurate atmospheric parameters for low to moderate resolution spectra. Spectral indices are specific spectral regions that have multiple absorption lines formed by similar chemical species blended into broad features. \citet{Ghezzi2014} specifically selected indices that are dominated by either neutral iron-peak species or ionized species, which have properties that allow the determination of T$_{\text{eff}}$, [Fe/H], and log \textit{g}. \citet{Ghezzi2014} selected 96 potential indices (80 dominated by neutral iron-peak species and 16 dominated by ionized species) through detailed inspection of FEROS Ganymede spectra \citep{Ribas2010} over the wavelength range 5100-5590 \AA\ at both the original resolution (R $\sim$ 48,000) and sampling and spectra downgraded to mimic the MARVELS resolution (R $\sim$ 11,000) and sampling. 

After initial spectral indices were identified, \citet{Ghezzi2014} calibrated and validated the spectral indices method while simultaneously creating a pipeline for MARVELS spectra, which involved four steps: continuum normalization, EW measurement, calibration construction, and atmospheric parameter determination. The normalization process was automated for MARVELS spectra and uses reduced, defringed, 1D Doppler-corrected spectra as input and fits a number of 1D Legendre polynomials to the continuum points of each spectra. 

Normalized spectra are input for the next step, which measures the EWs for specified indices by direct integration of their profiles. \citet{Ghezzi2014} created a set of calibrations that allow characterization of atmospheric parameters based solely on spectral indices through a multivariate analysis of both measured EWs of the spectral indices and precise atmospheric parameters (T$_{\text{eff}}$, [Fe/H], and log \textit{g}) derived from detailed and homogeneous high-resolution spectra for a set of calibration stars; four spectral indices were removed during this calibration determination due to poor fitting of these features. 

The final code in the pipeline determines atmospheric parameters and their associated uncertainties using the measured EWs and the previously determined calibrations. Atmospheric parameters are measured based on the minimization of the reduced chi-square between measured EWs and theoretical EWs that were calculated for each point of a 3D grid of atmospheric parameters in the following intervals: 4700 K $\le$ T$_{\text{eff}}$ $\le$ 6600 K, with 10 K steps; $-$0.90 $\le$ [Fe/H] $\le$ 0.50, with 0.02 dex steps; and 3.50 $\le$ log \textit{g} $\le$4.70, with 0.05 dex steps. For details on the spectral indices pipeline and method see section 4 of \citet{Ghezzi2014}.

Finally \citet{Ghezzi2014} tested the method specifically for MARVELS with a validation sample of 30 MARVELS stars that had high resolution spectra obtained from other instruments and subsequent high resolution analysis of atmospheric parameters. Each MARVELS star has two sets of spectra due to the interferometer which creates two ``beams". Each set of spectra are analyzed separately and two sets of parameters for each star are combined using a simple average and uncertainties are obtained through an error propogation. During the MARVELS validation process \citet{Ghezzi2014} found that only 64 of the 92 indices produced accurate atmospheric parameters. The final average offsets and and 1$\sigma$ Gaussian dispersions (standard deviations) obtained by \citet{Ghezzi2014} using these 64 indices for atmospheric parameters of 30 stars obtained from the spectral indices method with MARVELS spectra compared to high-resolution analysis are $-$28 $\pm$ 81 K for $\Delta$T$_{\text{eff}}$, 0.02 $\pm$ 0.05 for $\Delta$[Fe/H], and $-$0.07 $\pm$ 0.15 for $\Delta$log \textit{g}. 

\section{Atmospheric Parameters} \label{sec:atmpar}

\begin{figure}
      	\centering
	\includegraphics[width=0.78\linewidth]{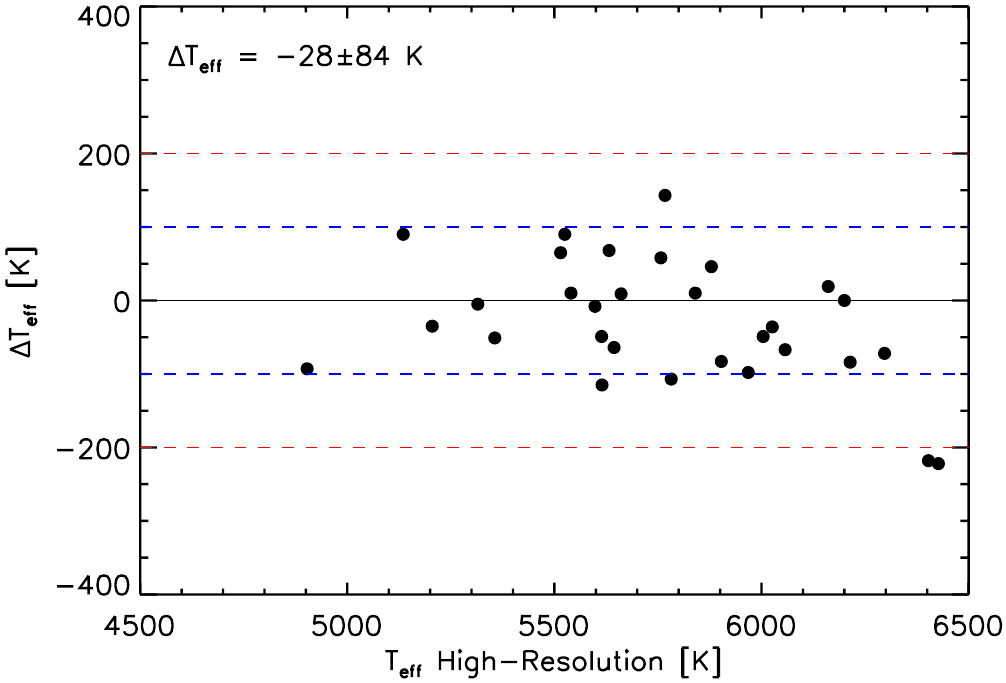}
	\includegraphics[width=0.76\linewidth]{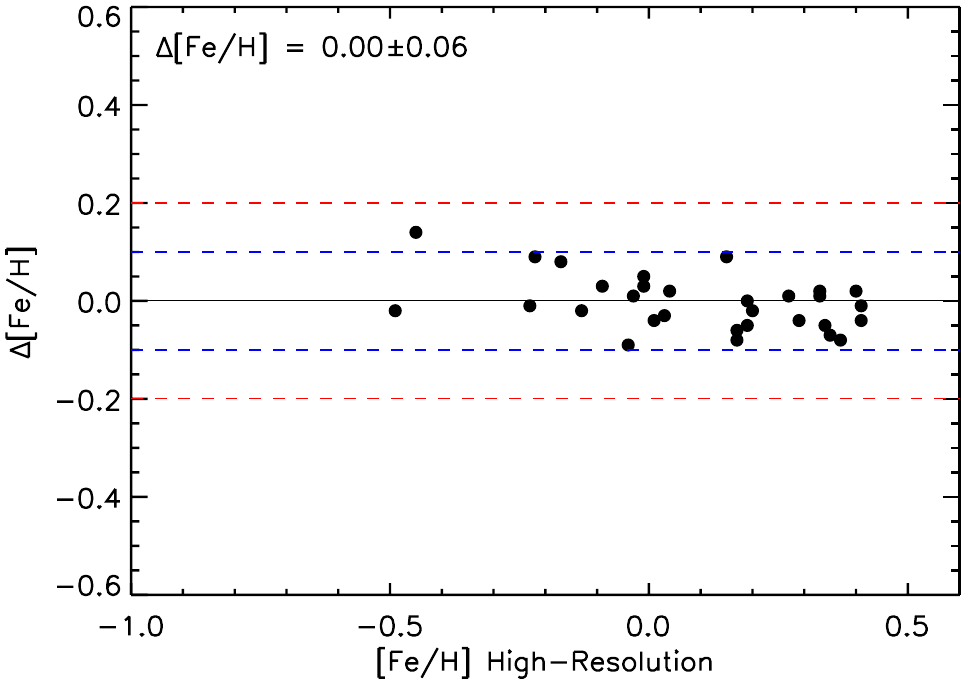}
	\includegraphics[width=0.78\linewidth]{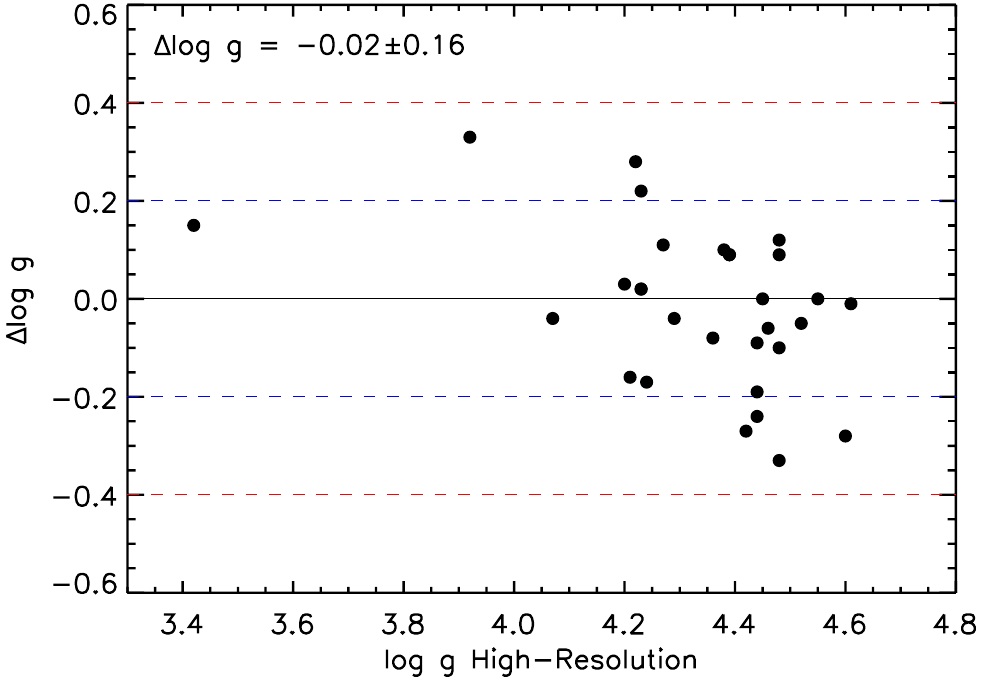}
\caption{Comparison between the 30 validation stars with high-resolution analysis results found in tables 1 and 2 in \citet{Ghezzi2014} and our current results with the spectral indices method and new MARVELS spectra. Mean offsets ($\Delta$ $\equiv$ high-resolution $-$ MARVELS) and their standard deviations are shown in the upper left of each plot, which give our approximate precision for these atmospheric parameters.}
\label{fig:refstars}
\end{figure}

This work uses input spectra that were pre-processed by the newest UF1D pipeline \citep{Thomas2015, Thomas2016}, while \citet{Ghezzi2014} used input spectra that were pre-processed with the older CCF+DFDI MARVELS pipeline released in the SDSS DR11 \citep{Alam2015}. This new process was previously described in \citet{Grieves2017} who obtained stellar parameters for 10 brown dwarf host stars. Here we obtain similar results as \citet{Grieves2017} when analyzing \citet{Ghezzi2014}'s 30 MARVELS validation stars with the newest UF1D input spectra; the results are displayed in figure \ref{fig:refstars}: $-$28 $\pm$ 84 K for $\Delta$T$_{\text{eff}}$, 0.00 $\pm$ 0.06 for $\Delta$[Fe/H], and $-$0.02 $\pm$ 0.16 for $\Delta$log \textit{g}.  We note a possible systematic trend in log \textit{g} when compared to high-resolution results in figure \ref{fig:refstars}, which may affect parameters derived using this data such as stellar ages and distances. We discuss possible affects further in section \ref{sec:ages}.  

We determine the mass (M$_{\star}$) and radius (R$_{\star}$) of each star from T$_{\text{eff}}$, [Fe/H], and log \textit{g} using the empirical polynomial relations of \citet{Torres2010a}, which were derived from a sample of eclipsing binaries with precisely measured masses and radii. We estimate the uncertainties in M$_{\star}$ and R$_{\star}$  by propagating the uncertainties in T$_{\text{eff}}$, [Fe/H], and log \textit{g} using the covariance matrices of the \citet{Torres2010a} relations (kindly provided by G. Torres). Approximate spectral classifications were determined from a star's T$_{\text{eff}}$ and its associated spectral type in table 5 of \citet{Pecaut2013}. 

\subsection{MARVELS Target Selection} 

A detailed knowledge of survey target selection and biases is required to investigate Galactic chemo-kinematics with survey data. We do not account for the selection function in this work, but here we give an overview of the MARVELS target selection, which mainly consists of FGK dwarf stars ideal for radial velocity planet surveys with $\sim$24\% giant stars observed as well. 

The MARVELS target selection process is described by \citet{Paegert2015}. Each MARVELS field consists of a circular field of view of seven square degrees with 60 stars selected for observation. The MARVELS survey observed 92 fields for a total of 1565 observations between 2008 October and 2012 July. Optical fibers for the instrument were changed in 2011 January and thus observations for MARVELS are divided into two different phases before and after this time, the ``initial" (Years 1-2) and ``final" (Years 3-4) phases, respectively. Due to ineffective giant star removal with the initial target selection process, the two phases consist of different target selection methods. MARVELS observed 44 fields in the initial phase and 48 fields in the final phase, with three fields overlapping both phases. 

Of the 92 plates in the overall MARVELS survey, only 56 plates were robustly observed 10 or more times, consisting of 3360 stars (60 stars per plate). We did not run 278 of these 3360 stars through the spectral indices pipeline due to various observational problems with these stars causing poor or missing observations for the majority or all observations. These issues include dead fibers, misplugged fibers, spectra suggesting the star is a spectroscopic binary, or unreasonable photon errors. This culling leaves 3082 stars (including 7 duplicates) that were run through the spectral indices pipeline. For the duplicated stars we use the average of the values from both plates to create one set of atmospheric parameters, creating a sample of 3075 unique stars.

For the 56 fields used in this study, 43  are from the initial phase and 13 are from the final phase with three initial phase fields observed in years 3-4 as well. The target selection methods for both phases were designed to observe FGK dwarfs with 7.6 $\le$ $V$ $\le$ 13.0, 3500 $<$ T$_{\text{eff}}$ $\le$ 6250 K, and log \textit{g} $>$ 3.5, and six (10\%) giant stars were selected for each field.

Both MARVELS phases used the GSC 2.3 and 2MASS catalogues to select the 1000 brightest V magnitude stars for each target field that were optimal for the survey. This included only stars in the MARVELS magnitude limits (7.6 $\le$ $V$ $\le$ 13.0), stars that were not clearly too hot ($J$ $-$ $K_{S}$ $\ge$ 0.29), stars that were projected to be in the field for at least 2 years, selecting brighter stars when stars were close together, and allowing $>$5 arc seconds of separation from V $<$ 9 stars \citep{Paegert2015}. The final 100 stars for each target field (60 plugged and 40 in reserve in case of collision with guide stars) were then selected by removing all but the six brightest giant stars, excluding hot stars (T$_{\text{eff}}$ $>$ 6250), and limiting F stars (3500 $\le$ T$_{\text{eff}}$ $\le$ 6250 K) to 40\% of all targets in the field \citep{Paegert2015}. Close binaries and known variable stars were also removed. Many MARVELS fields contain radial velocity reference stars (with known planets or RV stable stars), which were exempt from the target selection algorithms. 

The initial and final phase selection methods differ in the selection of the final 100 stars for each target field. The initial phase used a spectroscopic snapshot taken by the SDSS double spectrograph, mainly used for SEGUE, to derive T$_{\text{eff}}$, log \textit{g}, and [Fe/H] using a modified version of the SEGUE Stellar Parameter Pipeline (SSPP). These stellar parameters were used to perform the final cuts of removing giant and hot stars and limiting F stars; however, the SSPP pipeline misidentified cool giants as dwarfs causing a giant contamination rate of $\sim$21\% \citep{Paegert2015}. The final phase replaced the SSPP parameters method with a giant cut based on reduced proper motion, detailed in section \ref{sec:rpmj}, and estimates of the effective temperatures using the infrared flux method \citep[IRFM,][]{Casagrande2010}. \citet{Paegert2015} estimated the giant contamination rate for the final phase to only be 4\%, and including the 10\% of stars designated to be giants \citet{Paegert2015} estimated that 31\% of the stars in the initial phase are giants, while 14\% are giants in the final phase. MARVELS does not exclude subgiants (3.5 $\le$ log \textit{g} $\le$ 4.1) and they are included in the ``dwarf" sample.

The initial and final phases of the MARVELS survey also differed in target field selection. Target field selection for the initial phase was designed to find fields with radial velocity reference stars, fields without reference stars were chosen to provide large target densities of stars with 7.5 $<$ $V$ $<$ 13, and 11 fields were chosen such that they were centered on one of the 21 KEPLER photometry fields. However, the final phase was required to share target fields with APOGEE, which placed the fields outside of the galactic plane and required excluding stars if they were too close to APOGEE targets. This requirement caused the final phase stars to be dimmer on average, with a shift in the peak $V$ magnitude distribution from around 11.25 mag for the initial phase to 11.55 mag for the final phase \citep{Paegert2015}. Figure 7 of \citet{Paegert2015} shows the distribution of the MARVELS target fields on the sky in galactic coordinates. The 3075 stars considered in this study consist of 2367 stars in the initial phase (175 of which were also observed in the final phase with plates HD4203, HD46375, and HIP14810) and 708 stars in the final phase. 

\subsection{MARVELS Main-sequence stellar sample} \label{sec:sample}

\begin{figure}
	\centering
	\includegraphics[width=0.72\linewidth]{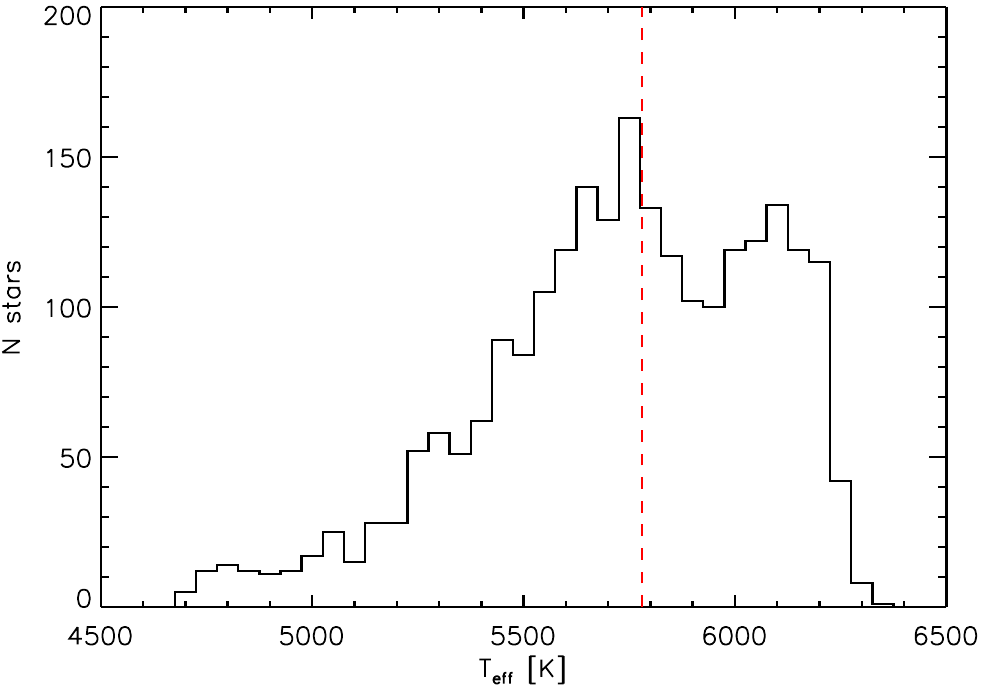}
	\includegraphics[width=0.72\linewidth]{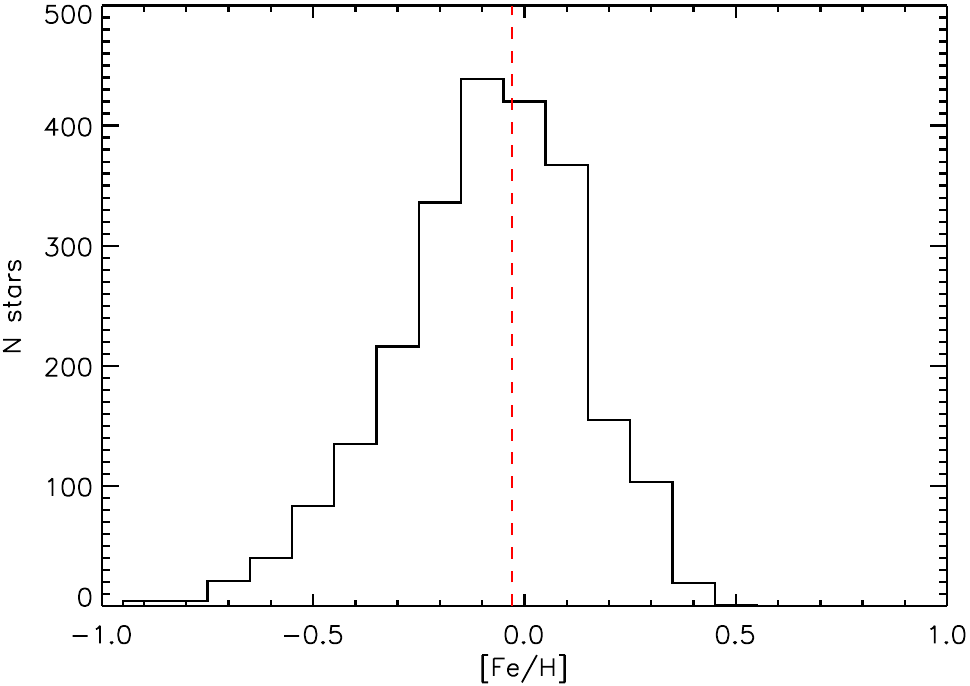}
	\includegraphics[width=0.72\linewidth]{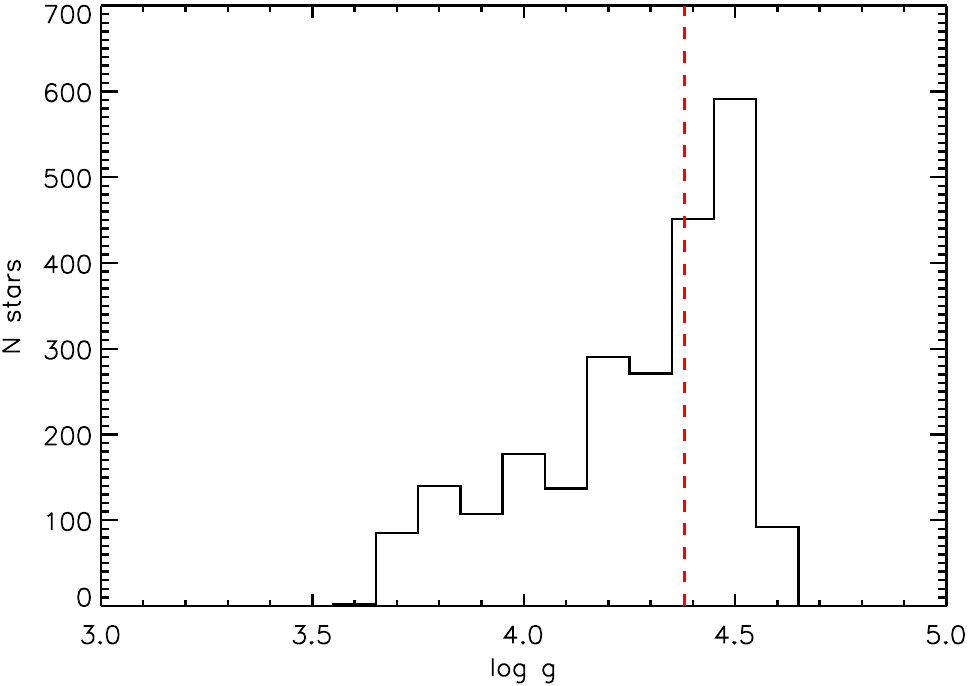}
\caption{Distributions of atmospheric parameters for the 2343 MARVELS main-sequence sample obtained from the spectral indices method. The red dashed vertical lines display the median values of 5780 K for T$_{\text{eff}}$, 4.38 for log $\textit{g}$, and $-$0.03 for [Fe/H].}
\label{fig:mardist}
\end{figure}

As stated in section 2.1 of \citet{Ghezzi2014}, the spectral indices pipeline was optimized for dwarf stars. Giant stars have considerably different spectra from those of dwarfs and subgiants, and thus a proper analysis would require a different distinct set of spectral indices. Therefore, the parameters obtained for giant stars with this pipeline should not be considered reliable. The spectral indices pipeline was also optimized for a certain range of temperatures and metallicities. Specifically, the pipeline automatically flags any star with parameters that lie outside  the range of 3.5 to 4.7 for log $\textit{g}$, 4700 to 6000 K for T$_{\text{eff}}$, or $-$0.9 to 0.5 for [Fe/H]. In our sample of 3075 stars 535 were flagged including 510 outside the log $\textit{g}$ range and 53 outside the T$_{\text{eff}}$ range, leaving a total of 2540 stars. To further avoid unreliable stellar parameters we only present measurements for dwarf stars according to the definition from \citet{Ciardi2011}, where a star is considered to be a dwarf if the surface gravity is greater than the value specified in the following algorithm:
\[
    \text{log } g \ge 
\begin{cases}
    3.5 & \text{if }   \text{T$_{\text{eff}}$} \ge 6000 \text{ K} \\
    4.0 & \text{if }   \text{T$_{\text{eff}}$} \le 4250 \text{ K} \\
    5.2 - 2.8 \times 10^{-4}\text{ T$_{\text{eff}}$}  & \text{if } 4250 <  \text{T$_{\text{eff}}$} < 6000 \text{ K}. \\
\end{cases}
\]

Of the 2540 stars with no flags, this algorithm designates 2343 stars as dwarfs. We set these 2343 stars as our final MARVELS main-sequence or dwarf stellar sample. Figure \ref{fig:mardist} displays the distributions of T$_{\text{eff}}$, log $\textit{g}$, and [Fe/H] for this sample, which has median values of 5780 K for T$_{\text{eff}}$, 4.38 for log $\textit{g}$, and $-$0.03 for [Fe/H]. 

\subsection{Comparison to RPM$_{J}$ cut designations} \label{sec:rpmj}

\begin{figure}
      	\centering
	\includegraphics[width=0.9\linewidth]{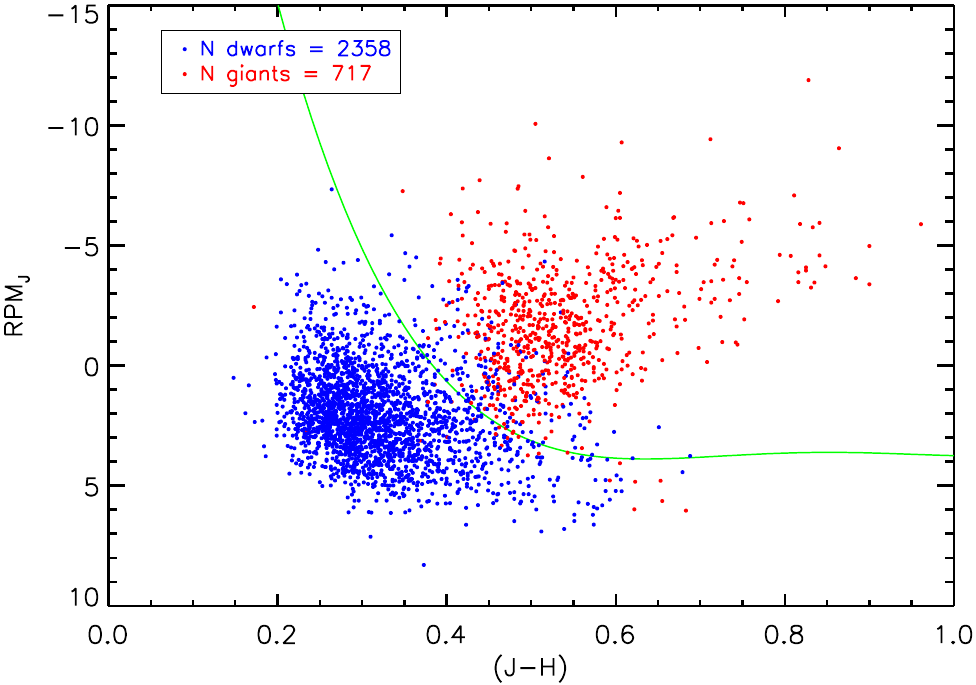}
\caption{$J$-band reduced proper motion (RPM$_{J}$) vs $J-H$ color for the 3075 stars in the MARVELS survey that have `robust' spectra and were submitted to the spectral indices pipeline. The green line indicates the RPM$_{J}$ cut to determine giant or dwarf/subgiant designation. Stars above the green line are designated as RPM$_{J}$ giants and those below as dwarf/subgiants. In this sample we classify 2230 stars as RPM$_{J}$ dwarf/subgiants and 845 as RPM$_{J}$ giants. Blue circles represent stars designated as dwarfs using the  \citet{Ciardi2011} criteria and red circles represent giant designation with this same criteria.}
\label{fig:rpmjcut}
\end{figure}

Previous MARVELS studies \citep[e.g., ][]{Paegert2015,Grieves2017} have used a $J$-band reduced proper motion (RPM$_{J}$) constraint to assign giant or dwarf/subgiant designations for stars in the MARVELS sample. RPM$_{J}$ values are computed as follows:
\begin{equation}
\mu = \sqrt{(\cos d *  \mu_{r})^{2} + \mu_{d}^{2}}
\end{equation}
\begin{equation}
\text{RPM}_{J} = J + 5 \log(\mu),
\end{equation}
where $J$ is the star's 2MASS Survey \citep{Skrutskie2006} $J$-band magnitude and $\mu_{r}$, $\mu_{d}$, and $d$ are Guide Star Catolog 2.3 \citep[GSC 2.3;][]{Lasker2008} proper motions in right ascension and declination (in arc seconds per year) and declination, respectively. An empirical RPM$_{J}$ cut described in \citet{Collier2007} is applied:
\begin{equation}
\begin{split}
y = -58 + 313.42(J - H) - 583.6(J - H)^{2} \\
 + 473.18(J - H^{3} - 141.25(J - H)^{4},
\end{split}
\end{equation} 
where $H$ is the star's 2MASS Survey $H$-band magnitude. Stars with RPM$_{J}$ $\leq$ $y$ are regarded as RPM$_{J}$-dwarfs and stars with RPM$_{J}$ $>$ $y$ as RPM$_{J}$-giants. \citet{Paegert2015} found this method to have a giant contamination rate of $\sim$4\% and that subgiants are mixed with the ``dwarf" sample at a level of 20\%-40\%. 

We compare the RPM$_{J}$ cut method to our designation based on \citet{Ciardi2011}'s definition using both T$_{\text{eff}}$ and log $\textit{g}$ values. In our initial sample of 3075 stars, 2230 were classified as RPM$_{J}$ dwarf/subgiants and 845 as RPM$_{J}$ giants. Using the new \citet{Ciardi2011} definition we designate 2358 stars as dwarfs and 717 as giants; including 692 giants in the 845 RPM$_{J}$ giant sample and 2205 dwarfs in the 2231 RPM$_{J}$ dwarf/subgiant sample. The RPM$_{J}$ cut method is more likely to designate stars as giants, as 153 of the RPM$_{J}$ giant stars were designated as dwarfs and only 25 RPM$_{J}$ dwarf/subgiant stars were designated as giants, which gives the RPM$_{J}$ cut method a $\sim$1\% giant contamination rate. Figure \ref{fig:rpmjcut} shows the RPM$_{J}$ cut for our sample with \citet{Ciardi2011} designations in blue and red.

\subsection{Comparison to other Surveys} \label{sec:sample2}

\begin{figure}
      	\centering
	\includegraphics[width=0.82\linewidth]{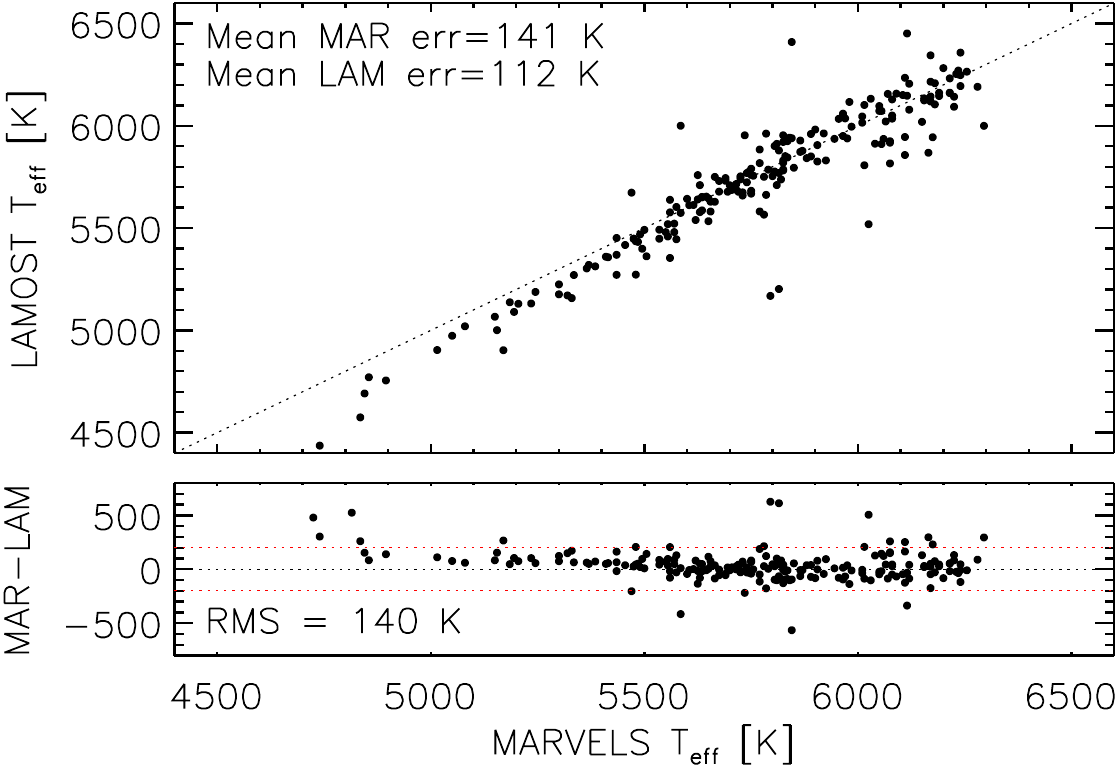}
	\includegraphics[width=0.82\linewidth]{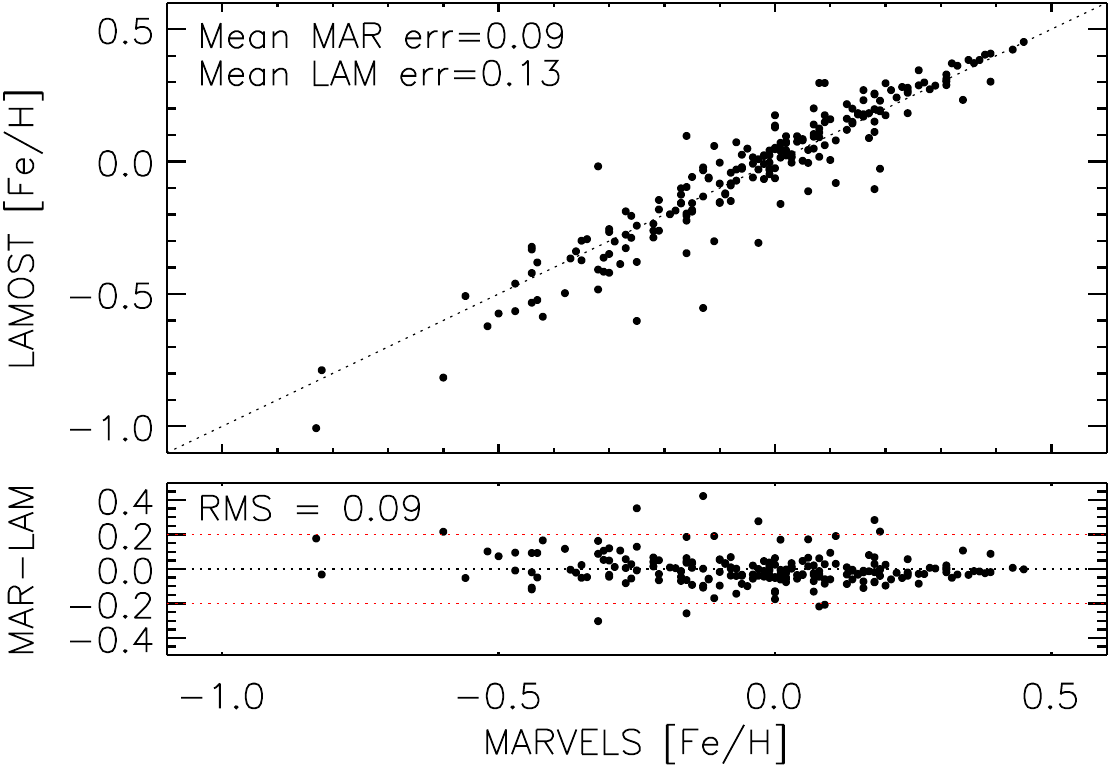}
	\includegraphics[width=0.84\linewidth]{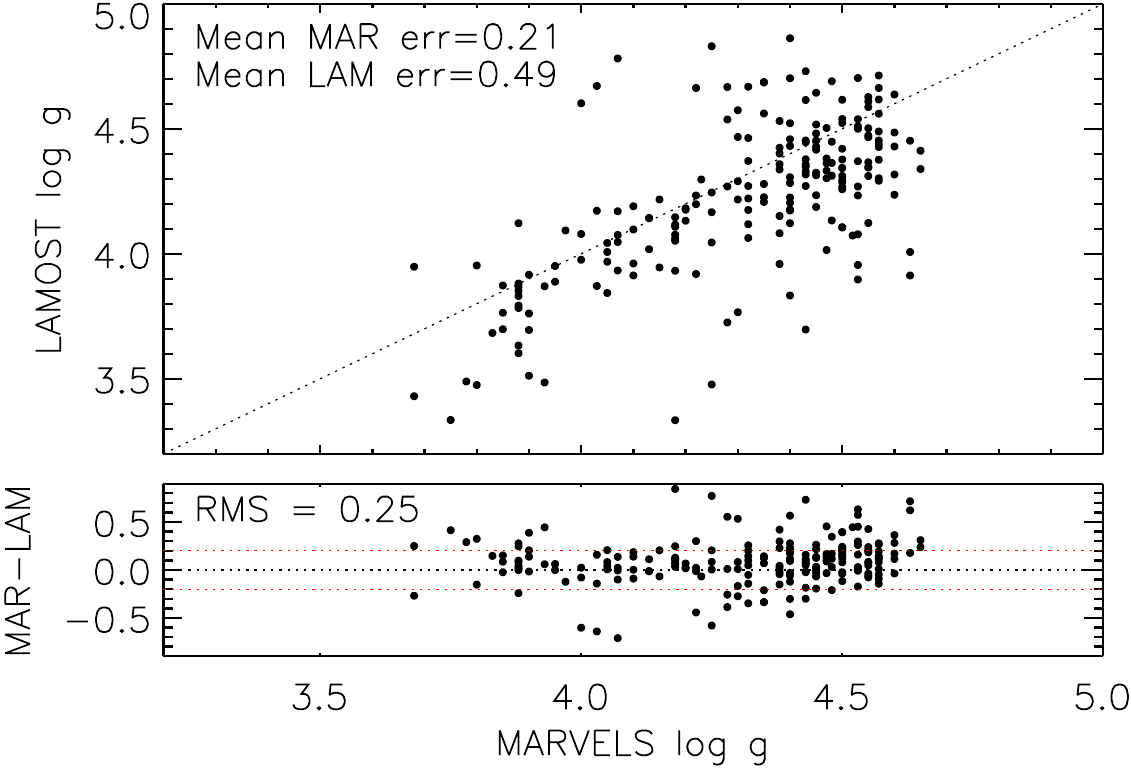}
\caption{Comparison between atmospheric parameters obtained from LAMOST and the MARVELS spectral indices method for the 206 stars in both samples. Mean errors on the plots show the mean of the errors for each survey for this specific sample of stars. Red dashed lines in the residuals show $\pm$200 K for T$_{\text{eff}}$ and $\pm$0.2 dex for [Fe/H] and log \textit{g}. Offsets and standard deviations are $\Delta$T$_{\text{eff}}$ = 35 $\pm$ 136 K, $\Delta$[Fe/H] = 0.00 $\pm$ 0.09, $\Delta$log \textit{g} = 0.09 $\pm$ 0.24.}
\label{fig:clamost}
\end{figure}

\begin{figure}
      	\centering
	\includegraphics[width=0.82\linewidth]{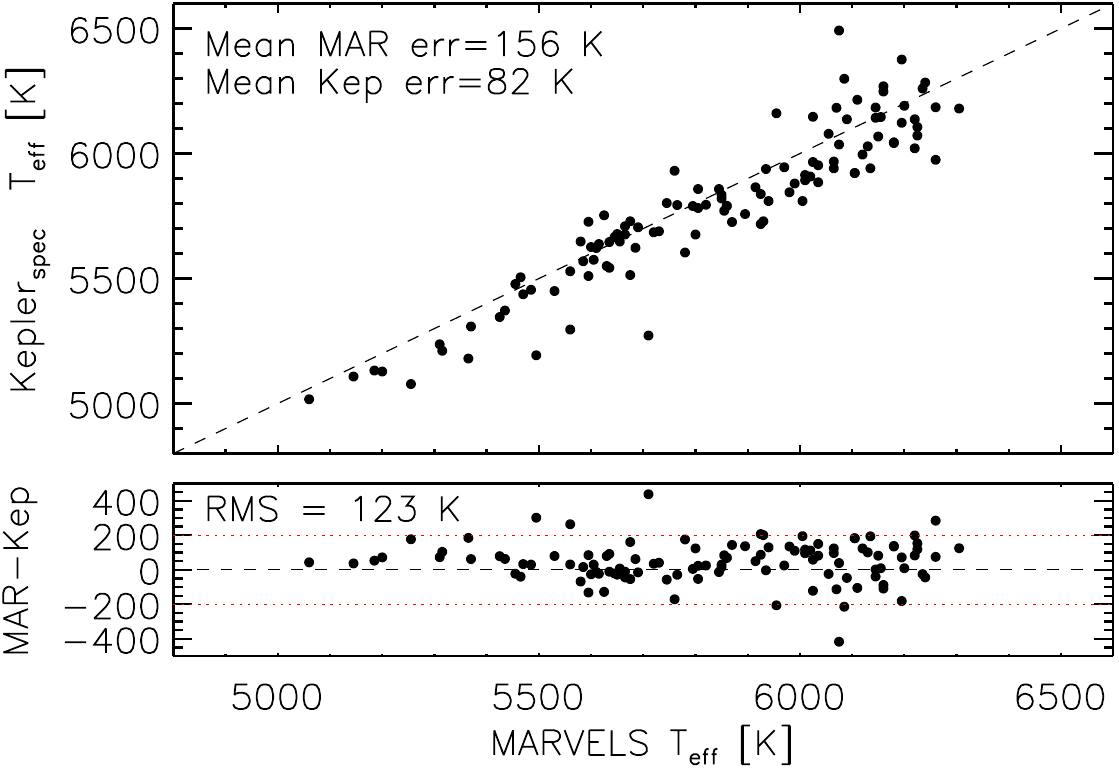}
	\includegraphics[width=0.82\linewidth]{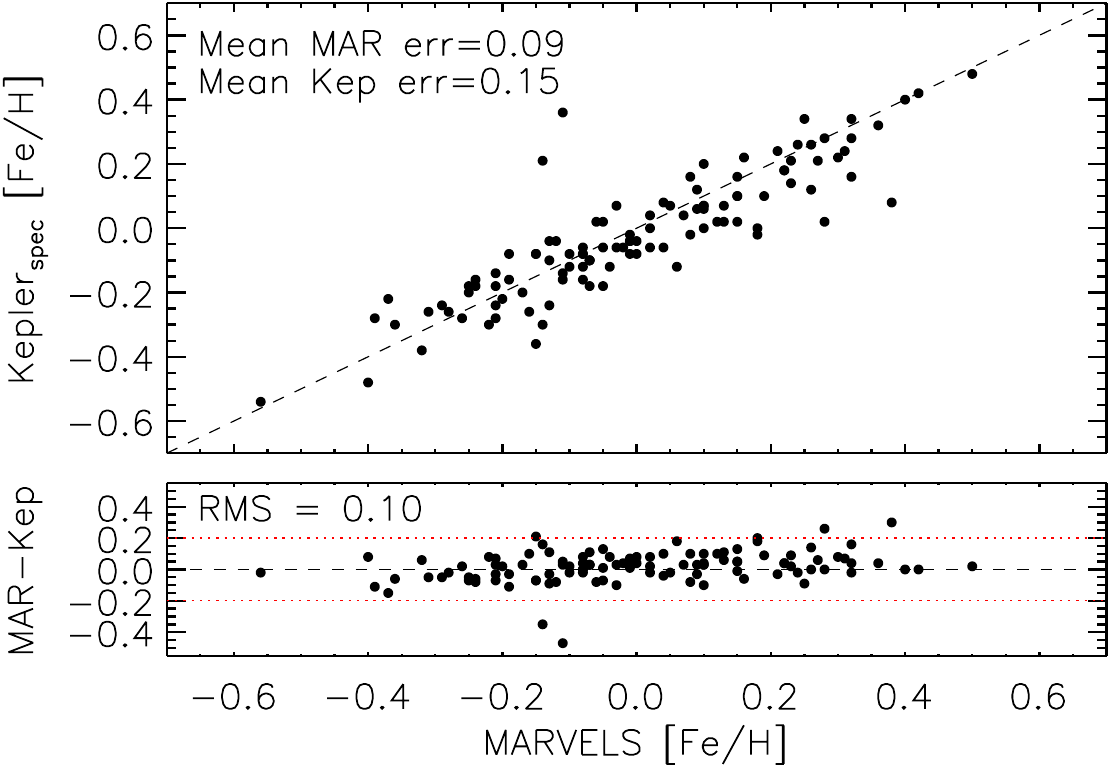}
	\includegraphics[width=0.84\linewidth]{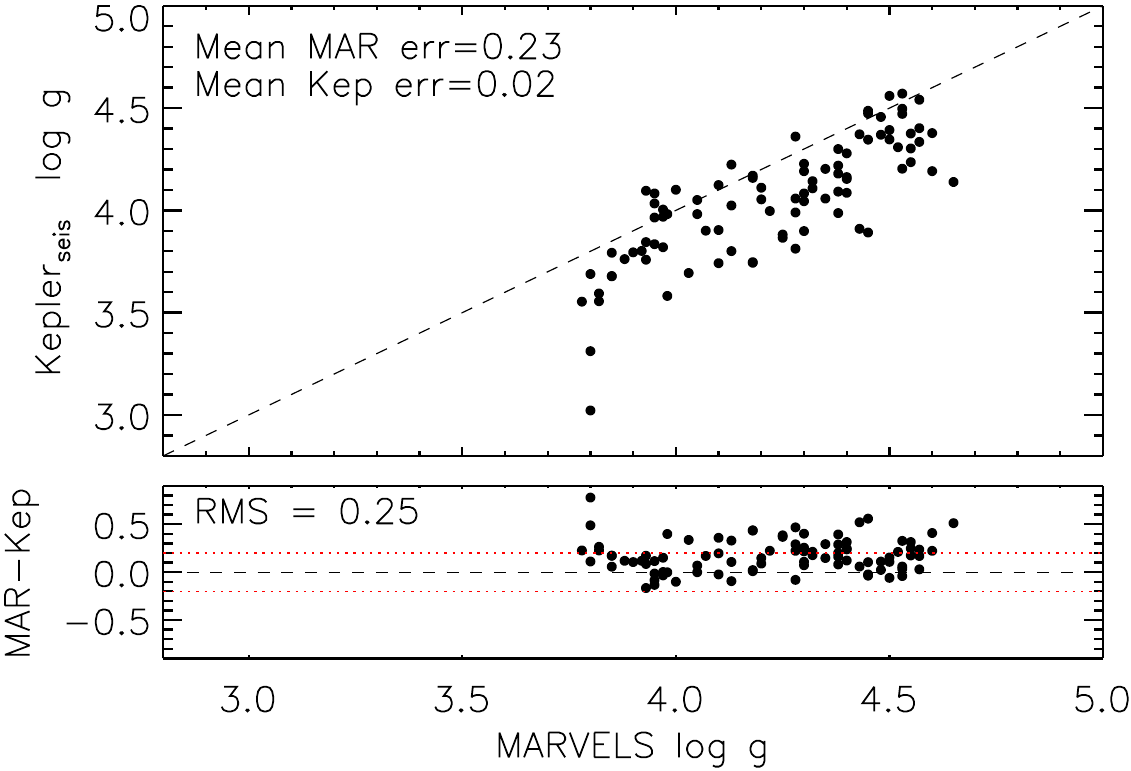}
\caption{Comparison between atmospheric parameters obtained from \textit{Kepler} and the MARVELS spectral indices method. \textit{Top and Middle}: Comparison of 112 \textit{Kepler} stars with spectroscopically derived T$_{\text{eff}}$ and [Fe/H] parameters. \textit{Bottom}: Comparison of 91 \textit{Kepler} stars with asteroseismic log \textit{g} parameters. Red dashed lines in the residuals show $\pm$200 K for T$_{\text{eff}}$ and $\pm$0.2 dex for [Fe/H] and log \textit{g}. Offsets and standard deviations are $\Delta$T$_{\text{eff}}$ = $-$47 $\pm$ 115 K, $\Delta$[Fe/H] = $-$0.02 $\pm$ 0.10, and $\Delta$log \textit{g} = $-$0.18 $\pm$ 0.17. We discuss the significant systematic offset with log \textit{g} in section \ref{sec:ast_logg}}
\label{fig:ckepler}
\end{figure}

Of our sample of 2343 dwarf stars with atmospheric parameters, 206 are in the LAMOST DR2 data set, where the stellar parameters are estimated by the LAMOST pipeline \citep{Wu2011,Wu2014}. Figure \ref{fig:clamost} compares results of these two surveys. Our results agree to the LAMOST results within the errors of both surveys. The offsets (MARVELS $-$ LAMOST) for the atmospheric parameters are $\Delta$T$_{\text{eff}}$ = 35 $\pm$ 136 K, $\Delta$[Fe/H] = 0.00 $\pm$ 0.09, and $\Delta$log \textit{g} = 0.09 $\pm$ 0.24.

We also compare our measurements to the latest \textit{Kepler} stellar properties \citep[DR25; ][]{Mathur2017}. For the T$_{\text{eff}}$ and [Fe/H] parameters we compare our results to stars that have been spectroscopically analyzed and given the highest priority by \citet{Mathur2017}, with errors of $\sim$0.15 dex for both T$_{\text{eff}}$ and [Fe/H]. A total of 112 stars are in common; the T$_{\text{eff}}$ and [Fe/H] parameters in the \textit{Kepler} data release are displayed in figure \ref{fig:ckepler}. The differences of these parameters are within expected errors of both surveys with offsets (MARVELS $-$ \textit{Kepler}) of $\Delta$T$_{\text{eff}}$ = $-$47 $\pm$ 115 K and $\Delta$[Fe/H] = $-$0.02 $\pm$ 0.10.

When comparing surface gravities to \textit{Kepler} parameters we use stars that were derived from asteroseismology, which are the highest priority values for \citet{Mathur2017} with errors of $\sim$0.03 dex for log \textit{g}. There are 91 overlapping stars in our sample with the \textit{Kepler} asteroseismology suface gravities. Our results have a systematic offset of  $\Delta$log \textit{g} = -0.18 $\pm$ 0.17; however, this level of offset is often found between surface gravity results derived from spectroscopic means and asteroiseismology \citep[e.g.,][]{Meszaros2013,Mortier2014}. We address this offset and present a correction in section \ref{sec:ast_logg}. 

\subsubsection{Asteroseismology Surface Gravity Correction} \label{sec:ast_logg}

\begin{figure}
      	\centering
	\includegraphics[width=0.72\linewidth]{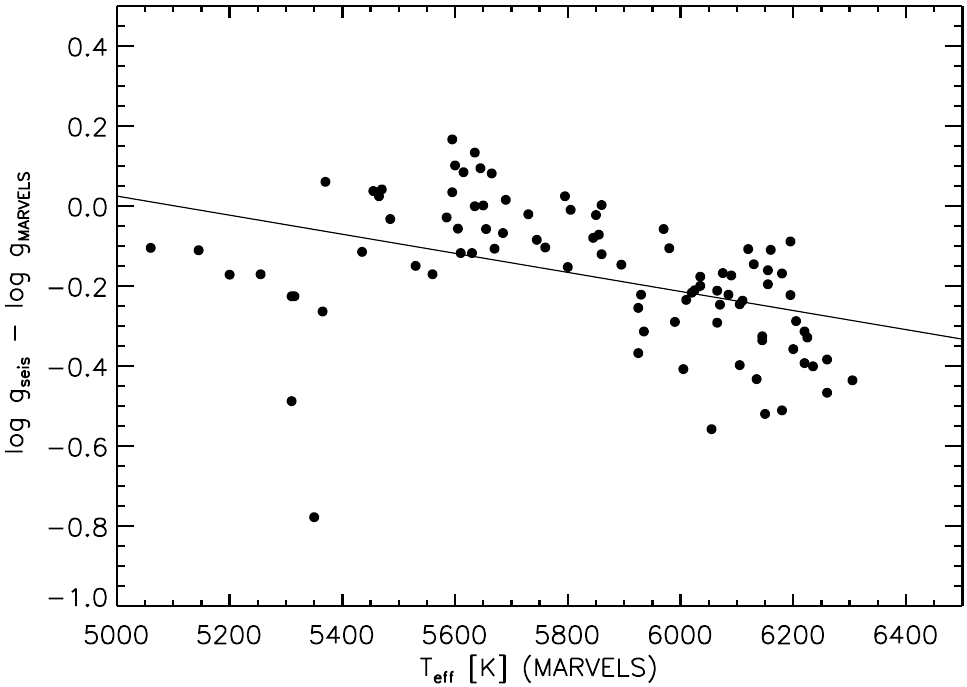}
\caption{Surface gravity difference (asteroseismic $\text{--}$ spectroscopic) versus the MARVELS effective temperature. The solid black curve indicates the linear fit in equation \ref{eq:logg_cor} for these 91 stars. This fit may be applied as a calibration for the log \textit{g} values presented in this work.}
\label{fig:logg_cor}
\end{figure}

Previous studies \citep[e.g., ][]{Torres2012,Meszaros2013,Huber2013,Mortier2013,Mortier2014,Heiter2015,Valentini2017} have demonstrated that surface gravity measurements derived from spectroscopic methods do not typically match higher-quality measurements obtained from other non-spectroscopic methods, such as asteroseismology or stellar models. \citet{Mortier2014} present a correction for spectroscopically-derived surface gravities using an effective temperature dependence. Here we present a similar correction for our results using the 91 stars in our sample overlapping with \textit{Kepler} asteroseismology data. Figure \ref{fig:logg_cor} displays our comparison of surface gravities with effective temperature, which is fit by the following linear relation:
\begin{equation}
\text{log} \ g_{\text{seis}} - \text{log} \ g_{\text{MARV}} = -2.38 \pm 0.54 \ \times \ 10^{-4}  \cdot \text{T}_{\text{eff}} \ 
+ \ 1.22  \pm 0.31,
\label{eq:logg_cor}
\end{equation}
which is similar to the correction found by \citet{Mortier2014}. After applying this correction the offset is $\Delta$log \textit{g} = 0.00 $\pm$ 0.16, which is within our expected surface gravity uncertainties. We do not apply this correction to the surface gravities derived in this work and present our results as a purely spectroscopic analysis. However, the correction may be applied if desired using equation \ref{eq:logg_cor} and the MARVELS log \textit{g} and T$_{\text{eff}}$ values presented in this work.

\subsection{Solar Twins} \label{sec:solartwin}

 Solar twins are special targets for investigating how our Sun and similar stars formed and evolved, and often solar twin samples allow for more precise determination of chemical abundances and fundamental parameters \citep[e.g.,][]{Ramirez2014,Nissen2015}. As previously demonstrated these stars can create a high quality data set ideal for testing planet formation, stellar composition, and galaxy formation theory. Our MARVELS stars includes a sample of `solar twins' defined as those that have T$_{\text{eff}}$, log \textit{g} and [Fe/H] inside the intervals $\pm$100 K, $\pm$0.1 [cgs], and 0.1 dex, respectively, around the solar values (5777 K, 4.44 [cgs], 0 dex).  This consists of 61 stars, which may be ideal for high precision follow-up observations, although our sample is somewhat fainter than previous high-precision studies, with a mean V magnitude of 10.8.

\section{Absolute Radial Velocities} \label{sec:absrvs}

\subsection{Wavelength Solution}

We derive absolute radial velocities (RVs) using the 1D wavelength calibrated MARVELS spectra. We adopt a separate wavelength calibration technique than that used to derive the UF1D relative RVs. The spectral processing and RV determination for the UF1D relative RVs is detailed in \citet{Thomas2016}; wavelength solutions for each beam are based on single observations of ThAr spectra and manual calibrations using Image Reduction and Analysis Facility \citep[IRAF;][]{Tody1993} software, which does not account for changes over time. Here we use a completely autonomous wavelength calibration technique to derive wavelength solutions for individual observations. An overview of this technique is provided below, with full details described in \citet{Thomas2015}. 

The moderate resolution ($R$ $\sim$ 11,000) of the MARVELS instrument causes almost all visible spectral features to be blended. This blending precludes the determination of a wavelength solution by matching individual features to a known line list as might be possible at higher resolution. Individual stellar observation wavelength solutions are based on matching spectral features in MARVELS spectra to a high-resolution template spectrum convolved down to MARVELS resolution. The master template spectrum is based on a solar atlas obtained from the National Solar Observatory data archive convolved to the approximate resolution of MARVELS. We expect some errors in using a solar template to calibrate the range of stellar types and metallicities in this survey; however, this issue is manageable as all survey stars are FGK types and there is no source for actual high-resolution wavelength calibrated spectra for each of the $\sim$3,000 stars in the survey. 

Individual templates for each beam are created by cropping the master template to a beam's approximate wavelength coverage determined from the ThAr calibrations. Significant spectral features that can be matched to the template are identified in each MARVELS spectra. Each feature's centroid is determined resulting in $\sim$200-300 unique features with pixel locations for each stellar spectrum; MARVELS spectra have $\sim$0.15 $\AA$/pixel. Several processing steps are used to roughly align the template spectrum to each observation and match features based on pattern recognition. Mismatched features are identified and discarded, and poor performing features not present in several observations or yielding varying results are rejected. These steps allow each feature and its corresponding pixel location to be assigned a precise rest wavelength from its matching template feature. Wavelengths are also corrected for Earth's known barycentric velocity. These results are then interpolated to integer pixel locations across the CCD to yield the wavelength solution for each individual spectrum. 

Tungsten Iodine (TIO) spectra are wavelength calibrated via the same technique using a high resolution ($R$ = 200,000) TIO spectrum of the MARVELS iodine cell obtained at the Coud{\'e} fed spectrograph at Kitt Peak as a master template. Comparing the TIO calibration wavelength solution and the observed stellar solution yields the Doppler shift in wavelength, allowing absolute RV determination. 

\subsection{Radial Velocities}

RVs can be directly determined from the TIO and stellar wavelength solution via equation \ref{eq:rv}:
\begin{equation}
\label{eq:rv}
\text{RV} = \frac{c \ (\lambda_{\text{TIO}} - \lambda_{\text{STAR}}) }{\lambda_{\text{TIO}}}
\end{equation}
where $c$ is the speed of light and $\lambda$ is wavelength. The mean value of equation \ref{eq:rv} across all pixels is a single observation's RV. A final absolute RV for each star is obtained by averaging the RVs of all observations from both beams. We obtain error estimates by calculating the RMS of the mean-subtracted RVs for these observations. Previous studies \citep[e.g.,][]{Holtzman2015} often consider stars exhibiting RV scatter on the level of $\sim$500 m s$^{-1}$ to be RV variable. We remove likely RV variable stars from our sample as these RVs are likely unreliable. Considering our relatively large absolute RV error values for stable stars ($\sim$300 m s$^{-1}$), we set an RV variability cut based on both absolute RV errors and RV RMS values from our latest and more accurate (for relative velocities) UF2D pipeline, detailed in \citet{Thomas2015} and \citet{Grieves2017}. Stars exhibiting RV RMS values greater than 500 m s$^{-1}$ in both the UF2D RVs and absolute RVs are not presented. We plan to publish binary stars in an upcoming MARVELS binary paper. This 500 m s$^{-1}$ cut removes 465 of the 3075 stars in our sample, leaving a final sample of 2610 stars with absolute RVs from MARVELS. 

\subsection{Zero Point Offset}

\begin{table}
    \caption{Comparison of MARVELS Absolute RVs to \citet{Chubak2012}}
     \label{tab:refrv}
    \begin{center}
   \begin{tabular}{lccc}
    \hline
   \hline
	 Star &  RV$_{\text{MARVELS}}$ 	& RV$_{\text{Chubak}}$   & $\Delta$RV  \\	
	 	& (km s$^{-1}$)   &  (km s$^{-1}$)    & (km s$^{-1}$)  \\	
	 	\hline		
HIP 32769 & -52.838 $\pm$ 0.412 & -52.417 $\pm$ 0.103 & -0.421 \\ 
HAT-P-3 & -23.201 $\pm$ 0.332 & -23.372 $\pm$ 0.155 & 0.171 \\ 
HD 17156 & -3.213 $\pm$ 0.498 & -3.207 $\pm$ 0.110 & -0.006 \\ 
HD 219828 & -24.086 $\pm$ 0.341 & -24.104 $\pm$ 0.075 & 0.018 \\ 
HD 4203 & -14.263 $\pm$ 0.485 & -14.092 $\pm$ 0.131 & -0.171 \\ 
HD 43691 & -29.172 $\pm$ 0.542 & -28.916 $\pm$ 0.045 & -0.256 \\ 
HD 46375 & -0.927 $\pm$ 0.452 & -0.906 $\pm$ 0.095 & -0.021 \\ 
HIP 32892 & 23.777 $\pm$ 0.418 & 23.587 $\pm$ 0.073 & 0.190 \\ 
HD 49674 & 12.013 $\pm$ 0.395 & 12.034 $\pm$ 0.148 & -0.021 \\ 
HD 68988 & -69.502 $\pm$ 0.479 & -69.383 $\pm$ 0.153 & -0.119 \\ 
HD 80355 & -6.572 $\pm$ 0.342 & -6.714 $\pm$ 0.065 & 0.142 \\ 
HD 80606 & 3.611 $\pm$ 0.438 & 3.948 $\pm$ 0.241 & -0.337 \\ 
HD 88133 & -3.855 $\pm$ 0.187 & -3.454 $\pm$ 0.119 & -0.401 \\ 
HD 9407 & -33.235 $\pm$ 0.275 & -33.313 $\pm$ 0.124 & 0.078 \\ 
HIP 14810 & -5.121 $\pm$ 0.681 & -4.971 $\pm$ 0.300 & -0.150 \\ 
HD 173701 & -45.568 $\pm$ 0.296 & -45.630 $\pm$ 0.093 & 0.062 \\ 
WASP-1 & -13.284 $\pm$ 0.451 & -13.430 $\pm$ 0.089 & 0.146 \\ 
HD 1605 & 10.069 $\pm$ 0.473 & 9.775 $\pm$ 0.104 & 0.294 \\
	\hline
       \end{tabular}   \\
       \end{center}
	\footnotesize{Comparison of 18 MARVELS stars in the \citet{Chubak2012} sample. This sample has a mean $\Delta$RV (RV$_{\text{MARVELS}}$ - RV$_{\text{Chubak}}$) of 0.044 km s$^{-1}$ and a standard deviation of 0.210 km s$^{-1}$. These values are displayed in figure \ref{fig:comp_rvs}.}
\end {table}    

As detailed in previous studies, e.g., \citet{Nidever2002}, barycentric RVs have several technical challenges that are not associated with obtaining relative RVs needed to identify planets and substellar companions. Barycentric RVs experience a gravitational redshift when leaving the stellar photosphere introducing spurious velocities. Barycentric RVs must also account for a transverse Doppler effect (time dilation), subphotospheric convection (granulation), macro turbulence, stellar rotation, pressure shifts, oscillations, and activity cycles, with granulation (convective blueshift) having the greatest effect on RV measurements \citep{Nidever2002}. As in previous studies, e.g., \citet{Nidever2002}, \citet{Chubak2012}, we correct for gravitational redshift and convective blueshift to first order by using the known RV of the Sun to set the zero point for the stellar RV measurements.

\begin{figure}
      	\centering
	\includegraphics[width=0.75\linewidth]{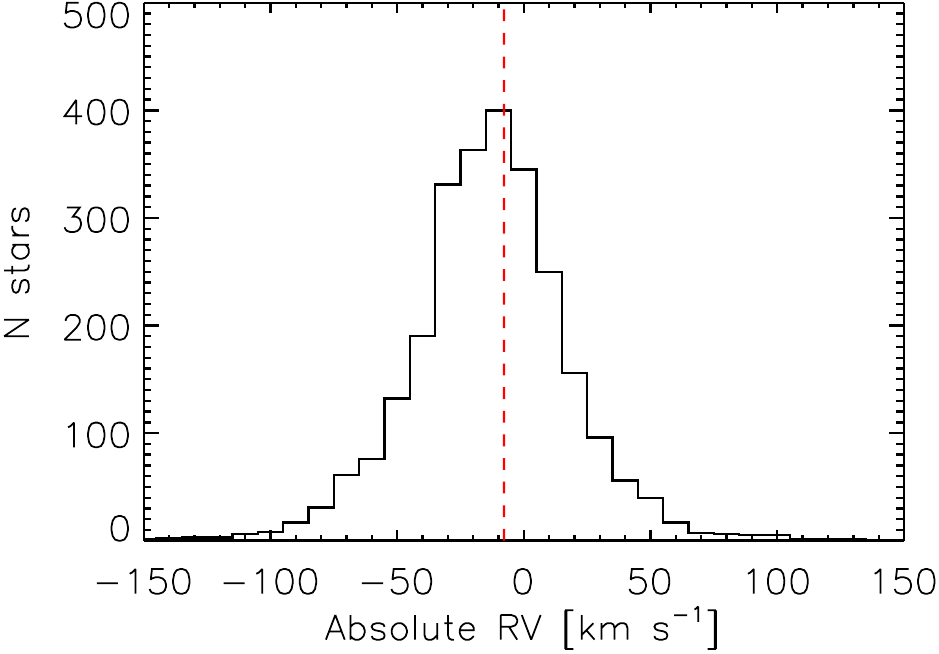}
	\includegraphics[width=0.75\linewidth]{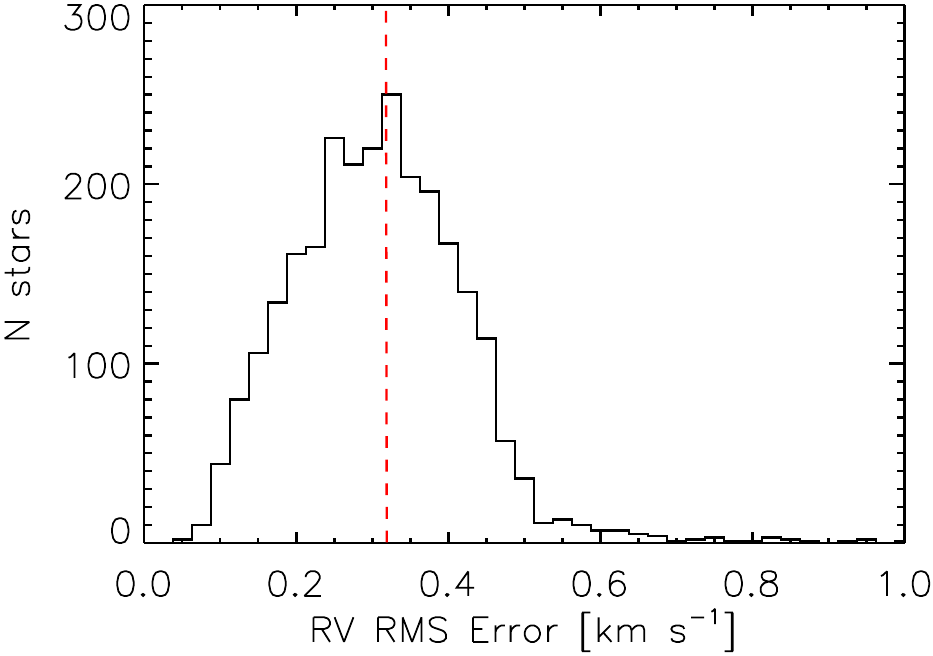}
\caption{\textit{Top}: Distribution of RVs for the 2610 MARVELS stars in our absolute RV sample. The red vertical dashed line shows the median RV for our sample, $-$7.862 km s$^{-1}$. \textit{Bottom}: Distribution of the RMS errors of the absolute RVs. The red vertical dashed line shows the median RV error for our sample, 0.319 km s$^{-1}$.}
\label{fig:abs_rv_dist}
\end{figure}

MARVELS obtained solar spectra by observing the sky during midday using the same fiber path from the calibration unit to the instrument as the TIO, we designate these observations as `SKY' spectra. MARVELS obtained 175 SKY spectra throughout the survey, which have a mean photon limit of 12.3 m s$^{-1}$ for all beams \citep{Thomas2016}. The mean difference between solar RVs and the barycentric velocity are calculated given the time and location of each observation and we take a mean of all offsets for all beams. The overall mean offset is 573 m s$^{-1}$, which we set as our zero point and subtract this value from all stellar RVs. Figure \ref{fig:abs_rv_dist} shows the final sample of MARVELS RVs along with their RMS errors. 

\subsection{Comparison to Previous Surveys}

\begin{figure}
      	\centering
	\includegraphics[width=0.85\linewidth]{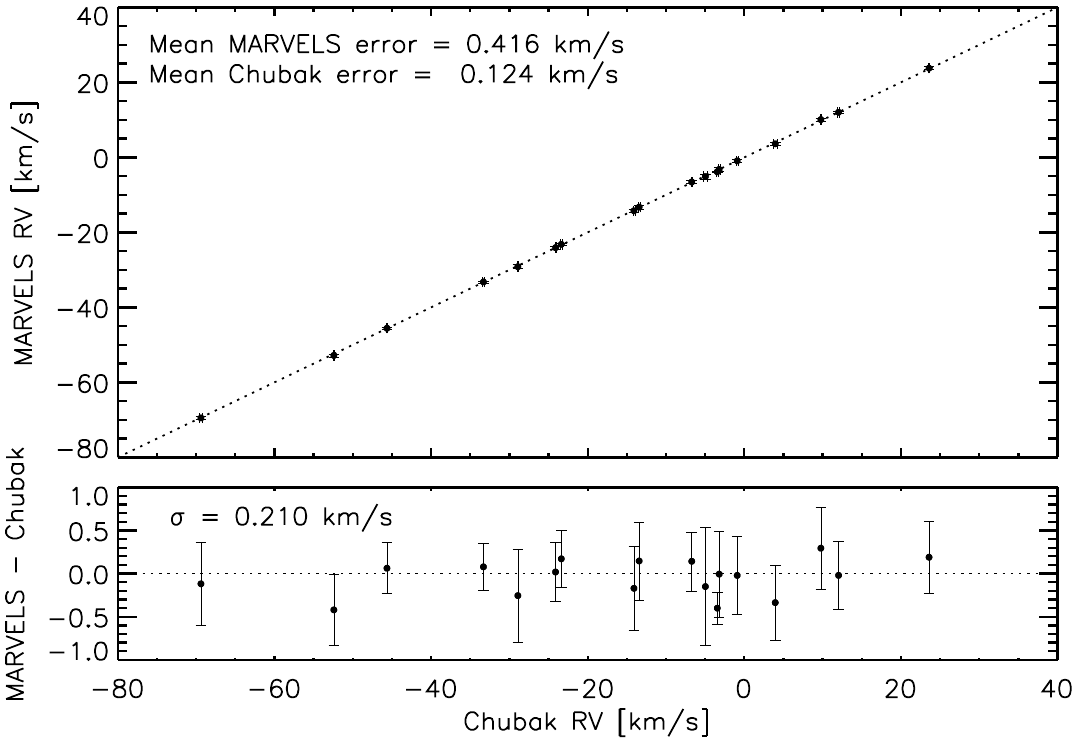}
	\includegraphics[width=0.85\linewidth]{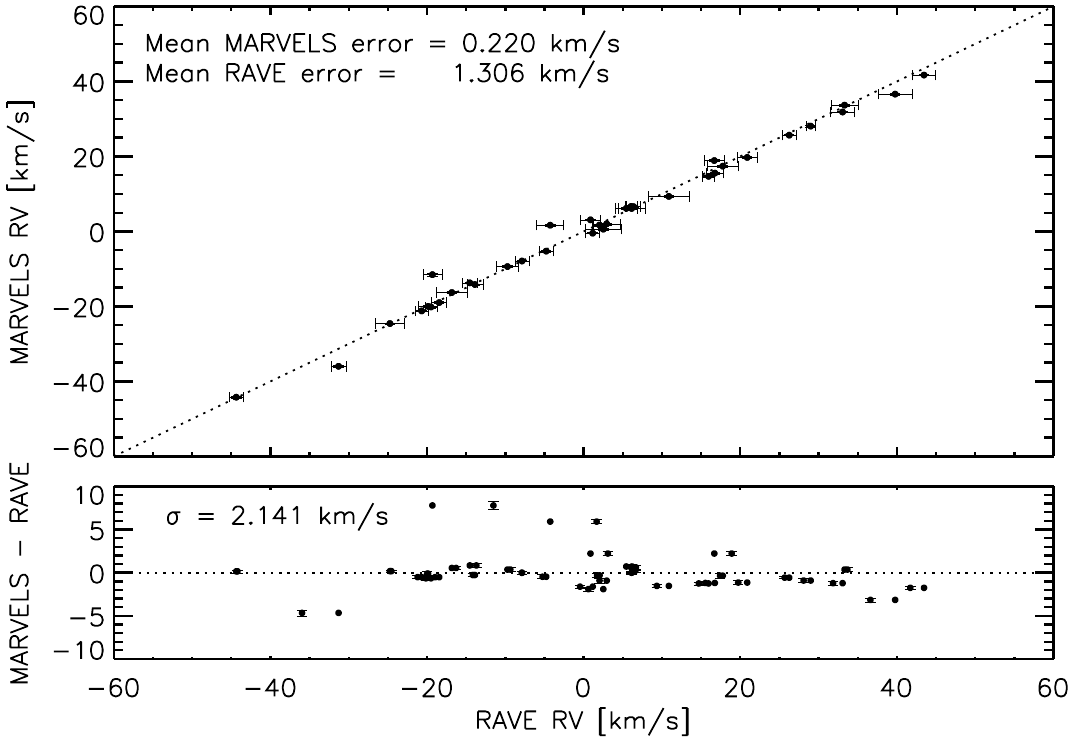}
	\includegraphics[width=0.85\linewidth]{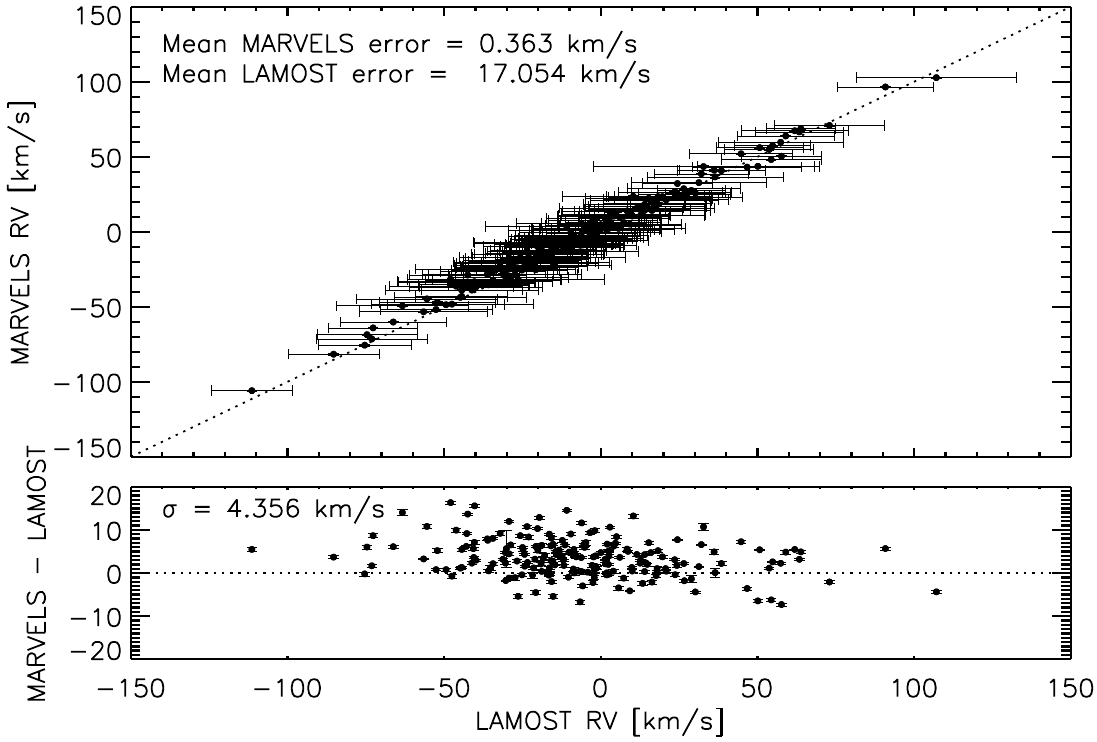}
\caption{Absolute RVs for MARVELS compared to other surveys. \textit{Top}: 18 stars in common between the MARVELS absolute RV sample and the catalogue by \citet{Chubak2012}, which represents a reliable sample of stars with average errors of only 0.124 km s$^{-1}$. These stars have a mean offset of 0.044 $\pm$ 0.210 km s$^{-1}$. \textit{Middle}: 36 stars in our sample in common with the RAVE DR5 sample, which have a mean offset of 0.081 $\pm$ 2.141 km s$^{-1}$. \textit{Bottom}: 195 stars in our RV sample that overlap with the LAMOST DR2 sample yielding a mean offset of 3.5 $\pm$ 4.4 km s$^{-1}$.}
\label{fig:comp_rvs}
\end{figure}

\citet{Chubak2012} constructed a catalogue of absolute velocities to share the zero point of \citet{Nidever2002}. Table \ref{tab:refrv} presents 18 stars in our absolute RV MARVELS sample that overlap in the catalogue created by \citet{Chubak2012} using observations from the Keck and Lick observatories (only considering stars with more than one observation and errors presented in \citet{Chubak2012}. These 18 stars have a mean offset of 0.044 $\pm$ 0.210 km s$^{-1}$, which is in agreement with the mean MARVELS error of these stars (0.416 km s$^{-1}$) and the mean error \citet{Chubak2012} assigned for these stars (0.124 km s$^{-1}$). The RVs of these 18 stars are displayed in figure \ref{fig:comp_rvs}. We also compare our RV results for stars overlapping in the RAVE DR5 catalogue and the LAMOST DR2 catalogue. A total of 36 stars in our sample overlap with the RAVE DR5 sample; the mean offset is 0.081 $\pm$ 2.141 km s$^{-1}$. The mean error of the 36 MARVELS RVs for these stars is 0.220 km s$^{-1}$ while the RAVE RVs have a mean error of 1.306 km s$^{-1}$. There are 195 stars in our RV sample that overlap with the LAMOST DR2 sample, which yield a mean offset of 3.5 $\pm$ 4.4 km s$^{-1}$; the MARVELS errors for this sample (mean error = 0.363 km s$^{-1}$) are significantly lower than the errors LAMOST assigns to their RV values for these stars (mean error = 17.054 km s$^{-1}$). The RVs of these overlapping RAVE and LAMOST stars are presented in figure \ref{fig:comp_rvs}.

\section{Kinematics} \label{sec:galvel}

\subsection{Galactic Space Velocities} \label{sec:galspacevel}

To understand the kinematic nature of a star its Galactic space velocity components U (radially inwards towards the Galactic Center), V (along the direction of Galactic rotation), and W (vertically upwards toward the North Galactic Pole) can be calculated given a star's proper motion, radial velocity, and parallax  e.g., \citet{Johnson1987}. Using parallax and proper motion values from  the second \textit{Gaia} data release \citep[\textit{Gaia} DR2; ][]{Gaia2018}  and the Hipparcos mission \citep{vanLeeuwen2007} we were able to obtain space velocity components for  2504  stars in our sample of 2610 stars with absolute radial velocities. The calculated space velocities (U$_{\text{LSR}}$, V$_{\text{LSR}}$, W$_{\text{LSR}}$) are related to the local standard of rest (LSR; the velocity of a fictional particle that moves around the plane of the Milky Way on the closed orbit in the plane that passes through the present location of the Sun) by adding the Sun's velocity components relative to the LSR (U$_{\odot}$, V$_{\odot}$, W$_{\odot}$) = (11.10, 12.24, 7.25) km s$^{-1}$ from \citet{Schonrich2010}. 

We present the Galactocentric Cartesian coordinates $x$, $y$, $z$ and Galactocentric radii $R_{Gal}$ for stars with distance values using the \texttt{Galactocentric} python code developed by Astropy \citep{Astropy2018} assuming $R_{0}$ = 8.2 kpc for the Sun's distance from the Galactic Center.  Distances are determined from the available parallax data. 

We also computed Galactocentric velocities in a cylindrical reference frame. In this case the velocities components are $V_{R}$ (radial), $V_{\phi}$ (rotational), and $V_{Z}$ (vertical) defined as positive with increasing $R$, $\phi$, and $Z$, with $Z$ pointing towards the North Galactic Pole and $V_{Z}$ = W$_{\text{LSR}}$. $V_{R}$ and $V_{\phi}$ are computed as follows:

\begin{equation}
V_{R} = [x \cdot \text{U}_{\text{LSR}} + y \cdot (\text{V}_{\text{LSR}} + v_{circ}^{\odot})]/R_{Gal}
\label{eq:vr_gal}
\end{equation}
\begin{equation}
V_{\phi} = -[x \cdot  (\text{V}_{\text{LSR}} + v_{circ}^{\odot}) - y \cdot \text{U}_{\text{LSR}}]/R_{Gal},
\label{eq:vphi_gal}
\end{equation}
where $x$ and $y$ are Galactocentric Cartesian coordinates, $R_{Gal}$ is the Galactocentric radius ($R_{Gal}$ = $\sqrt{x^2 + y^2}$), U$_{\text{LSR}}$ and V$_{\text{LSR}}$ are the associated space-velocity components in the Galactic cardinal directions relative to the LSR as described above, and $v_{circ}^{\odot}$ = 238 km s$^{-1}$ is the circular rotation velocity at the Sun's position, which is in line with recent estimates \cite[e.g.,][]{Bland-Hawthorn2016}. We also compute the guiding-center radius of a stellar orbit, which we computed using the approximation $R_{guide}$ = $\frac{V_{\phi} \cdot R_{Gal}}{v_{circ}^{\odot}}$ similar to \citet{Anders2017a}.

\subsection{Galactic Orbits} \label{sec:galorb}

We determine galactic orbits for each star using its full phase-space information: right ascension, declination, distance, proper motion, and radial velocity. We calculate orbits using the Python module \textit{galpy} \citep{Bovy2015} ref assuming a standard Milky Way type potential consisting of an NFW-type dark matter halo \citep{Navarro1997}, a Miyamoto-Nagai disc \citep{Miyamoto1975}, and a Hernquist stellar bulge \citep{Hernquist1990}, which achieves a flat rotation curve for the model Galaxy. We again assume $R_{0}$ = 8.2 kpc for the Sun's distance from the Galactic Center and $v_{circ}^{\odot}$ = 238 km s$^{-1}$ for the circular rotation velocity at the Sun's position. We adopt \citet{Schonrich2010}'s values for the solar motion with respect to the LSR (U$_{\odot}$, V$_{\odot}$, W$_{\odot}$) = (11.10, 12.24, 7.25) km s$^{-1}$. The stellar motions are integrated with scipy routine \texttt{odeint} over 10 Gyr in 10,000 steps. From the integrated Galactic orbits we are able to characterize the median orbital eccentricity $e$, the median Galactocentric radius $R_{med}$, and the maximum vertical amplitude $z_{max}$. Galactocentric coordinates and orbit parameters along with Galactic space velocities are presented in the online catalogue.

\section{Stellar Ages} \label{sec:ages}

We derive stellar ages for a subset of our stars using both the isochrone method \citep[e.g.,][]{Lachaume1999}, or the ``maximum-likelihood isochronal age-dating method", and the spectro-photometric distance code \texttt{StarHorse} \citep{Queiroz2018}, which is a ``Bayesian age-dating method". The maximum-likelihood isochronal age-dating method is inspired by the statistical approach used in \citet{Takeda2007}. In order to ensure smooth age probability distributions, we adopted a fine grid of evolutionary tracks from Yonsei-Yale Stellar Evolution Code \citep{Yi2003} with constant steps of 0.01 $M/M_{\odot}$ (0.40 $\leq$ $M/M_{\odot}$ $\leq$ 2.00), 0.05 dex (from $-$2.00 to $+$1.00 dex), 0.05 dex (from $+$0.00 to $+$0.40 dex) of mass, [Fe/H], and [$\alpha$/Fe], respectively. 

In the framework of the Bayesian probability theory, the set of available input parameters ({\bf X}) of each star is compared to its theoretical prediction given by the evolutionary tracks: $\mathbf{\Theta} \equiv (\mathrm{t}, M/M_{\odot},\mathrm{[Fe/H]},\mathrm{[\alpha/Fe]})$. We arbitrate solar $\mathrm{[\alpha/Fe]}$ for all sample stars with an except of metal-poor stars ([Fe/H] $\leq$ $-0.5$) in which $\mathrm{[\alpha/Fe]} = +0.3$ is assigned. The complete probability functions $P(\mathbf{\Theta}|\mathbf{X})$ along all possible evolutionary steps is defined by:  

\begin{equation}\label{eq:bay4}
P(\mathbf{\Theta}|\mathbf{X}) \propto P(\mathbf{X}|\mathbf{\Theta})P(\mathbf{\Theta}). 
\end{equation}

\begin{figure}
      	\centering
	\includegraphics[width=0.8\linewidth]{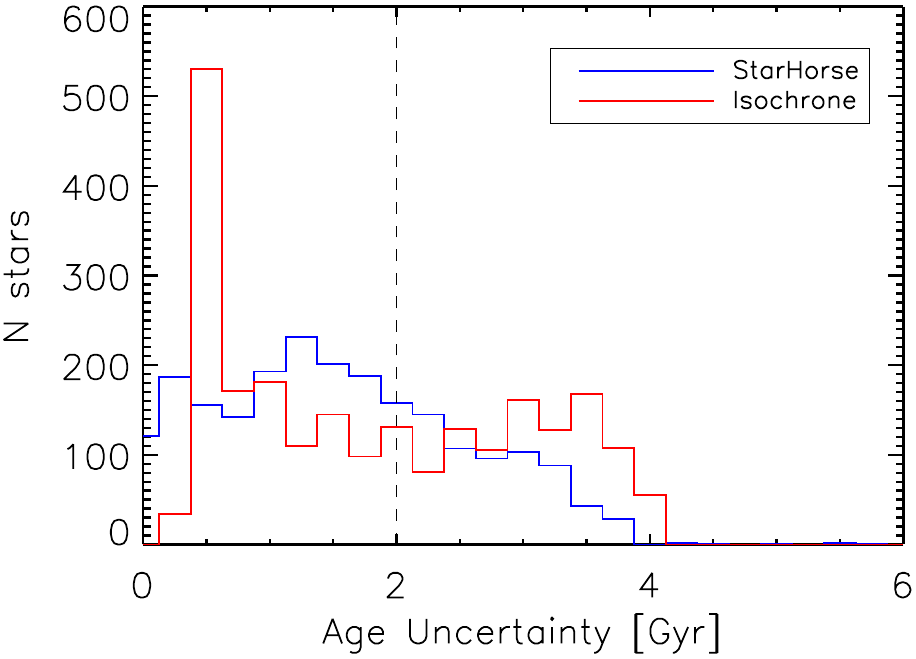}
\caption{Stellar age uncertainties (1$\sigma$ standard deviations) for both the \texttt{StarHorse} ages ( 2194  stars) and the maximum-likelihood isochronal age-dating method (Isochrone) ages ( 2335  stars).  The vertical dashed line displays the 2 Gyr age error cut for stars used in the analysis and results of this study.  }
\label{fig:age_dist}
\end{figure}

$P(\mathbf{X}|\mathbf{\Theta})$ is our likelihood function which is a combination of $N$ Gaussian probability distributions of each input atmospheric parameter ($v_j^{input}$) together with its respective uncertainty ($\sigma_j$). In the cases where the trigonometric parallaxes are available we build a set of $N =$ 4 input parameters given by: $\mathbf{X} \equiv (T_{\rm eff} \pm \sigma_{T_{\rm eff}}, \log~g \pm \sigma_{\log~g}, \mathrm{[Fe/H]} \pm \sigma_{\rm [Fe/H]}, \log(\mathrm{L/L_{\odot}}) \pm \sigma_{\log(\mathrm{L/L_{\odot}})})$. The stellar luminosities are calculated with equations 2 and 7 of \citet{Andrae2018} using published \textit{Gaia} DR2 G magnitudes, extinctions, and bolometric corrections. The luminosity errors are computed through 10$^4$ Monte Carlo simulations assuming Gaussian error distributions of parallax, photometry, and MARVELS $T_{\rm eff}$.  Alternatively, we restrict our approach to consider only $T_{\rm eff}$, $\log~g$, and $\mathrm{[Fe/H]}$ as input atmospheric parameters ($N =$ 3) for stars without distance information: 
\begin{equation}\label{eq:bay5}
\mathit{P}(\mathbf{X}|\mathbf{\Theta}) = \mathlarger{\prod}_{j=1}^N \frac{1}{\sqrt{2\pi}\,\sigma_j}\exp\left(-\frac{1}{2}\chi^2\right),
\end{equation}
where
\begin{equation}\label{eq:bay6}
\chi^2 = \mathlarger{\sum}_{j=1}^N \left(\frac{v_j^{input} - v_j^{theoretical}}{\sigma_j}\right)^2.
\end{equation}
$\mathit{P}(\mathbf{X}|\mathbf{\Theta})$ is the posterior probability function which is composed of a likelihood function and  the combination of collapsed prior probability functions of mass, metallicity, and age ($P(\mathbf{\Theta} \equiv P(t)\,P(M/M_\odot)\,P([Fe/H])$). The Salpeter-like IMF prior is $P$(mass) $\propto$ mass$^{-2.35}$  \citep{Takeda2007} and the metallicity prior is the same used by \citet{Casagrande2011} (also independent on the stellar age, see their Appendix A). Conservatively, we choose to adopt uniform prior distribution for the stellar ages from 0 to 14 Gyr. We stress that the mass and metallicity priors used in this approach are independent of any further age assumptions. So, we consider that our results are suitable for Galactic chemo-kinematic analysis presented in the following sections of this paper.  For each star, the posterior probability function is evaluated along every possible configuration in theoretical parameter space and integrated over all dimensions of $\mathbf{\Theta}$ but age:
\begin{equation}
\mathrm{\mathcal{P}(t) \propto \int\int P(\mathbf{\Theta}|\mathbf{X})\,d(M/M_\odot)\,d[Fe/H]}. 
\end{equation}
To save computational time, models older than 14 Gyr and outside $\pm$ 4 sigma domain defined by the observational uncertainties are ruled out from our calculations. Other relevant probability distributions of stellar properties can be also estimated in parallel to the age derivation such as mass, radius, radius and gravity at \textit{Zero/Terminal Age Main-Sequence}, and chance of the star being placed in the main-sequence or subgiant branch according to core H-exhaustion. We compute the mean age and the 5\%, 16\%, 50\%, 84\% and 95\% percentiles from the age probability distribution. We use the mean value of the probability distribution as our final age value for the maximum-likelihood isochronal age-dating method and determine stellar ages for  2335  stars. We show the age error distribution for the isochrone method in figure \ref{fig:age_dist}.

We derive a second set of independent stellar ages as well as interstellar extinctions and distances using a Bayesian age-dating method with the \texttt{StarHorse} code. \texttt{StarHorse} uses a bayesian approach that uses spectroscopic, photometric, and astrometric information as input to calculate the posterior probability distribution over a grid of stellar evolutionary models. The code has been extensively tested and validated for simulations and external samples, for a full description see \citet{Santiago2016, Queiroz2018}. This approach works similar to the maximum-likelihood isochronal age-dating method, assuming Gaussian errors and that the measured parameters are independent; the likelihood can also be written as in Equation \ref{eq:bay5}. The prior function used by \texttt{StarHorse}  for MARVELS stars includes only spatial priors,  as defined in \citet{Queiroz2018}, as well as a Chabrier initial mass function \citep{Chabrier2003},  and no metallicity prior on age.  For MARVELS stars we use T$_{\text{eff}}$, log $\textit{g}$, and [Fe/H] parameters derived in this work as well as previously published BVJHK magnitudes,  \textit{Gaia} DR2 G-band magnitudes and extinctions,  and parallax values as the set of measured parameters to obtain estimates of mass, age, distance, and V band extinction (A$_\text{V}$) using PARSEC 1.2S stellar models \citep{Bressan2012,Tang2014,Chen2015}. The PARSEC 1.2S stellar models employed in the calculation ranged in [M/H] (from -2.2 to 0.6 in steps of 0.02 dex) and in logarithmic age (from 7.5 to 10.13 in steps of 0.01 dex) delivering a broad range in the estimated ages. 

\begin{figure}
      	\centering
	\includegraphics[width=0.9\linewidth]{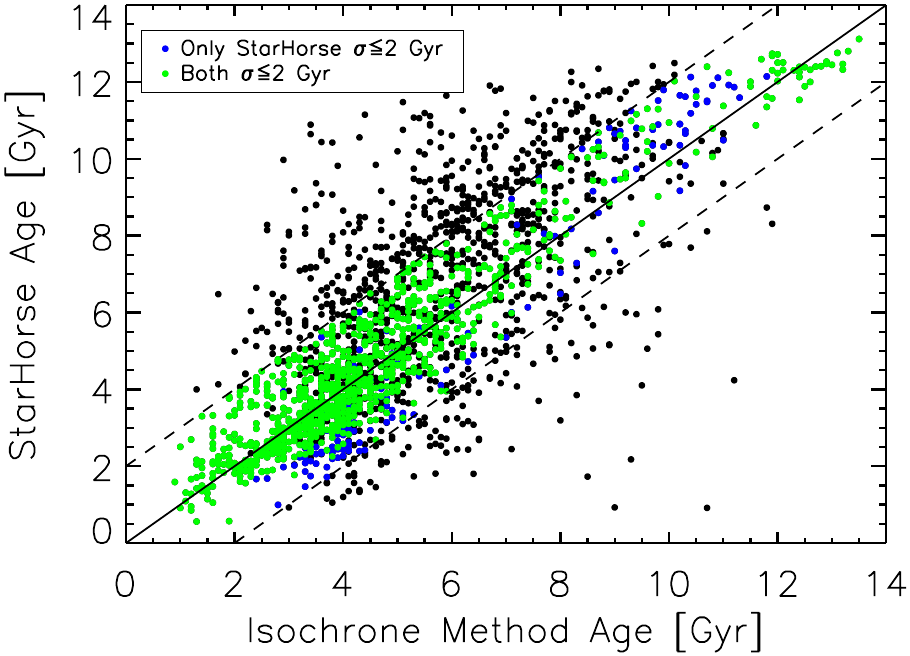}
\caption{Comparison of stellar ages determined from \texttt{StarHorse} and the isochrone method for  2186 stars; the overall comparison has a mean offset of 0.52 $\pm$ 1.72 Gyr. Colored symbols show stars used for our analysis and results, where both methods agree within 2 Gyr and at least one method has an error estimate below 2 Gyr. }
\label{fig:age_comp}
\end{figure}

We use the mean value of the outputted probability distribution from the \texttt{StarHorse} code as our final age values and determine ages using \texttt{StarHorse} for  2194  stars with atmospheric parameters, photometry, and parallax data. We also present the 5\%, 16\%, 50\%, 84\% and 95\% percentiles of the age probability distribution for the \texttt{StarHorse} ages  in the online catalogue . Figure \ref{fig:age_dist} shows the age error distribution for these stars.  Effects of systematic offsets between observed quantities (such as atmospheric parameters) and age, distance, and extinction results are discussed thoroughly in \citet{Santiago2016} and \citet{Queiroz2018}. \citet{Queiroz2018} found in general systematic errors affect estimated parameters typically less than $\pm$10 percent. Notably our data may have a systematic offset in log \textit{g} values, which could have a significant affect on the \texttt{StarHorse} distance estimates (presented in section \ref{sec:distance}), as log \textit{g} is the best parameter to discriminate between low-luminosity dwarfs and more luminous giants. Thus, an overestimate in log \textit{g} leads to an underestimate in the distances, and an underestimate in log \textit{g} leads to overestimating the distance. \citep{Santiago2016,Queiroz2018}.  

We compare the age results from these two methods in figure \ref{fig:age_comp} and find relatively good agreement between the two methods, which rely on different isochrones, with a mean offset (\texttt{StarHorse} - Isochrone) of  0.52 Gyr (or 6.0\%) and a dispersion 1.72 Gyr (or 32.4\%) for the 2186 stars with ages from both methods, where the percentages correspond to the difference in age estimates of each star divided by the mean age of the two methods. For our analysis we only consider ages for stars that have error estimates below 2 Gyr and the age from both methods are within 2 Gyr of each other. If both methods have age error estimate below 2 Gyr we take the mean of the two ages for the age of the star, which is the case for 981 stars. For another 144 stars, only the \texttt{StarHorse} age error estimate is below 2 Gyr, so only this method's age estimate is analyzed. This creates a sample of 1125 stars with ages for our analyses. We display this sample in figure \ref{fig:age_comp}.  The data published online contains stellar ages from both methods as well as maximum-likelihood isochronal method main-sequence or sub-giant branch predictions and \texttt{StarHorse} interstellar extinctions and distances for available stars.

\section{Distances} \label{sec:distance}

\begin{figure}
      	\centering
	\includegraphics[width=0.48\linewidth]{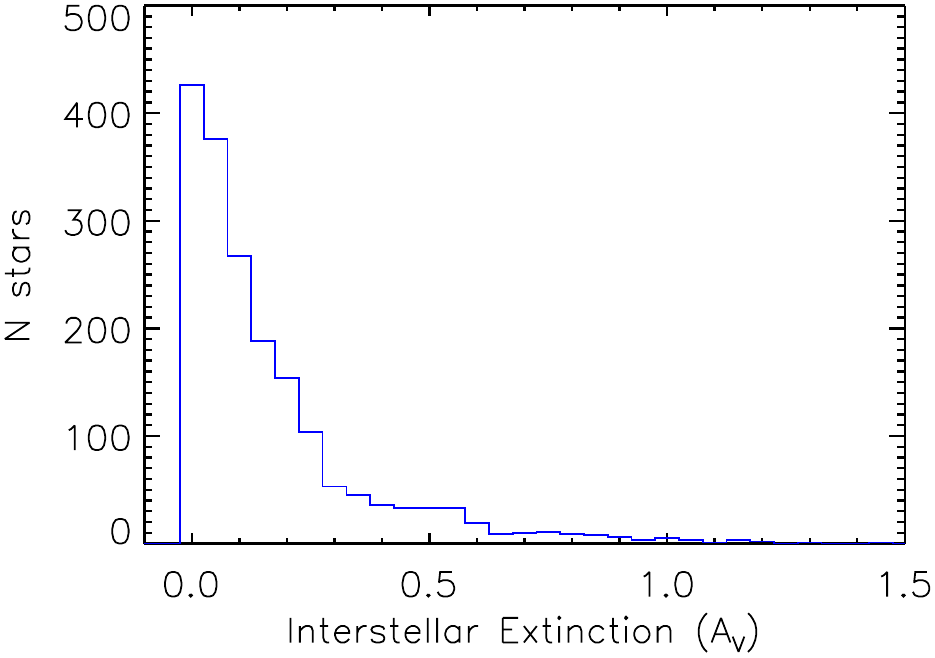}
         \includegraphics[width=0.48\linewidth]{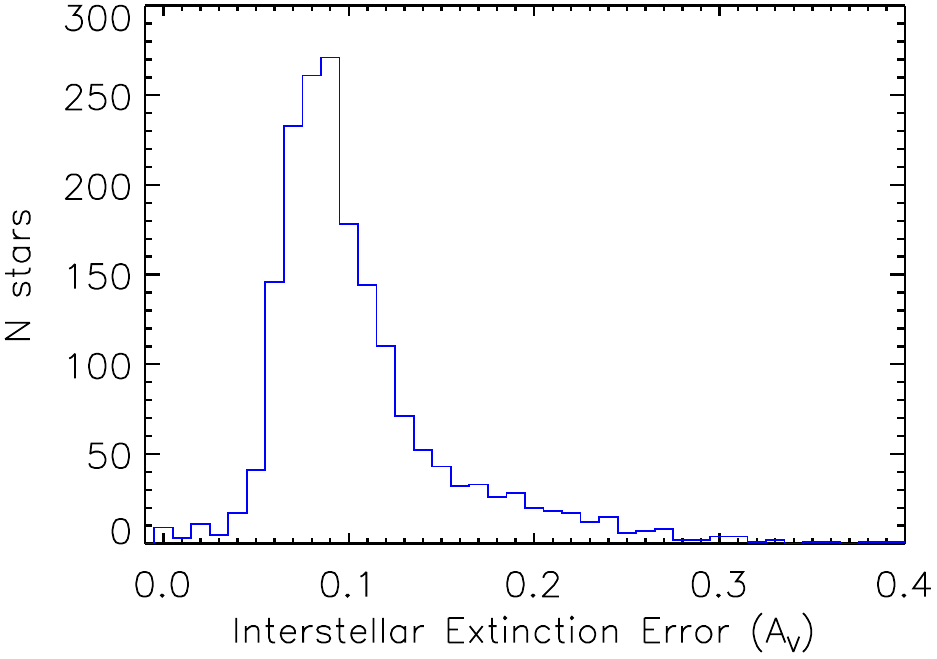}	
	\includegraphics[width=0.48\linewidth]{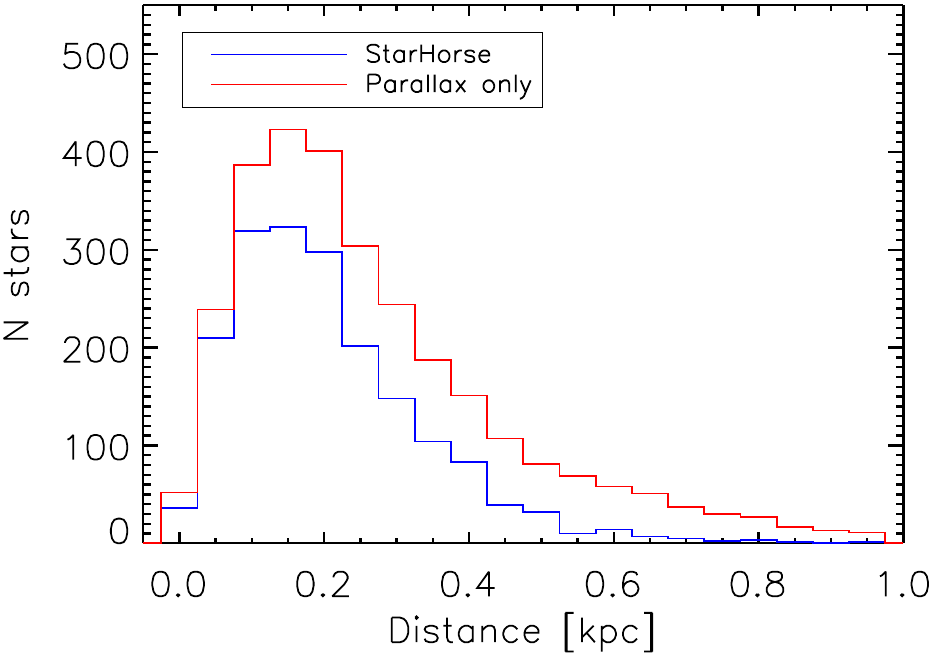}
	\includegraphics[width=0.48\linewidth]{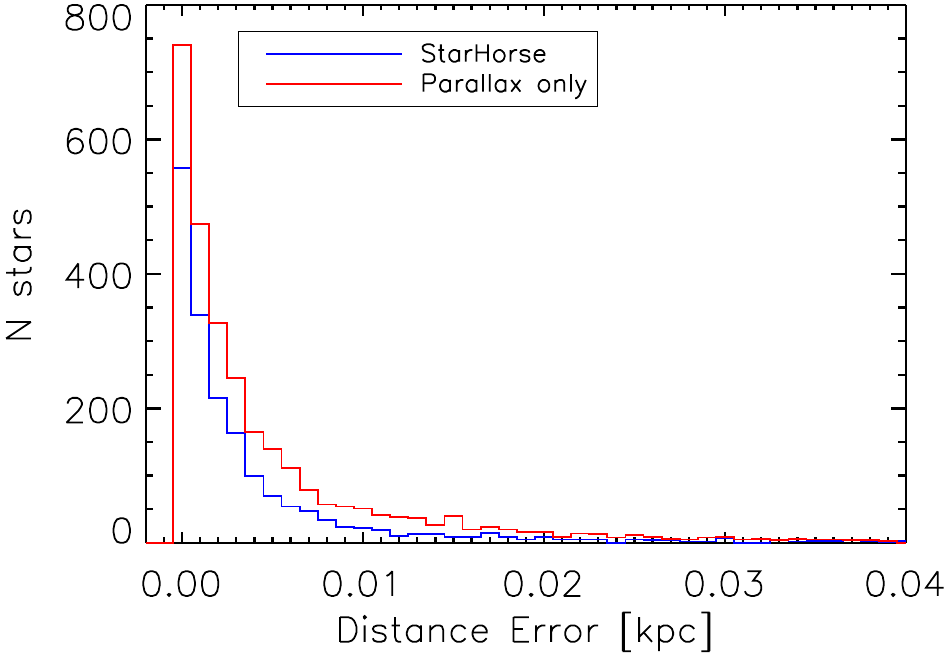}
\caption{\textit{Top}: V band interstellar extinction (A$_\text{V}$) distribution for  1841  stars derived using the \texttt{StarHorse} code and their associated error distribution. \textit{Bottom}: Distributions for distances  and their associated errors  obtained from the \texttt{StarHorse} code (blue,  1841 stars ) and distances using only parallax values (red,  2957 stars ).}
\label{fig:dist}
\end{figure}

Here we briefly present the distances determined for our sample of MARVELS stars.  We obtain distances for 2957 stars using parallax values from the \textit{Gaia} and Hipparcos missions. We also computed distances using the \texttt{StarHorse} code described in section \ref{sec:ages}, which account for interstellar extinction. However, as \texttt{StarHorse} is a statistical method, negative extinctions can arise when the photometry and the spectroscopic or astrometric parallax are slightly inconsistent. For these cases the interstellar extinction cannot be constrained well. Using the \texttt{StarHorse} code on 2194 stars to compute distances and extinctions, 353 were found to have negative extinction. We do not present the distances or extinctions for these stars in figure \ref{fig:dist}, giving a sample of 1841 stars with \texttt{StarHorse} distances and interstellar extinctions.  Figure \ref{fig:dist} displays interstellar extinction values obtained from the \texttt{StarHorse} code as well as distances from the \texttt{StarHorse} code and distances using only parallax values. Figure \ref{fig:dist} also displays the error distributions outputted from the \texttt{StarHorse} code and errors for parallax distances using propagation of error.  For our results and analysis we use the distances derived from parallax values. 

\section{Results} \label{sec:results}

In this section we analyze the results of the various parameters derived in this study, which are summarized in table \ref{tab:sumpar}. Previous studies revealed evidence that a sizable fraction of the geometrically-defined thick disk is chemically different from the thin disk \citep[e.g.,][]{Navarro2011,Fuhrmann2011,Adibekyan2013,Bensby2014}, and most of the thick disk population is kinematically hotter than the thin disk, suggesting that the thin and thick disk have a different physical origin. Analyses of kinematic data also suggest that there are substantial kinematical sub-structures in the Solar neighborhood associated with various stellar streams and moving groups \citep[e.g.,][]{Nordstrom2004,Bensby2014,Kushniruk2017}, but whether these structures are of Galactic or extragalactic origin remains uncertain. 

Kinematical sub-structures may be due to evaporated open clusters \citep[e.g.,][]{Eggen1996}, dynamical resonances within the Milky Way \citep[e.g.,][]{Dehnen1998}, or possibly from remnants of accreted satellite galaxies \citep[e.g.,][]{Navarro2004}. Specifically, one moving group or dynamical stream creation mechanism may be the resonant interaction between stars and the bar or spiral arms of the Milky Way. For example, the Hercules stream could be caused by the Sun being located just outside the bar's outer Lindblad resonance  \citep{Dehnen2000}. Nevertheless, kinematic groups retain information of various processes of the Milky Way's past and allow insight into the formation of the Galaxy. We investigate the metallicity distributions and other properties of the kinematically defined thin and thick disks as well as the two kinematical sub-structures identified in our study, the Hercules stream and the Arcturus moving group.

\begin{table}
\caption{Summary of data for MARVELS stars}
\label{tab:sumpar}
\begin{center}
\begin{tabular}{lc}
\hline
\hline
Parameter & N stars \\
\hline	
Stellar Sample	&	3075 \\	
Atmospheric Parameters	&	2343 \\
RVs				&	2610 \\
RVs \& Atmospheric Parameters	& 1971 \\
Galactic Space Velocities		&  2504  \\ 
Galactic orbital parameters	&	 2504  \\ 
 Distances \&  Galactic Cartesian coordinates &	 2957   \\ 
StarHorse Ages		&	 2194   \\ 
Isochrone Ages		&	 2335   \\ 
 Age analysis sample ($\sigma$$\le$2 Gyr)  &  1125    \\ 
\hline
\end{tabular}  
\end{center}
\end {table}    

\subsection{Defining Populations} \label{sec:kinpop}

\subsubsection{Kinematically-defined Populations}

To assign stars to a population in the Galaxy (e.g., thin disk, thick disk, or halo) we first adopt a purely kinematic approach by following the method outlined by \citet{Bensby2003, Bensby2005, Bensby2014} and assume the stellar populations in the thin disk, the thick disk, and the halo (as well as other stars that may not be kinematically associated with the thin or thick disk such as the Hercules stream or the Arcturus moving group) have Gaussian distributions,

\footnotesize
\begin{equation}
f\text{(U, V, W)} = k \cdot \text{exp} \bigg(-\frac{(U_{\text{LSR}} - U_{\text{asym}})^{2}}{2 \sigma_{\text{U}}^{2}}  
-\frac{(V_{\text{LSR}} - V_{\text{asym}})^{2}}{2 \sigma_{\text{V}}^{2}} 
-\frac{W_{\text{LSR}}^{2}}{2 \sigma_{\text{W}}^{2}} \bigg)
\label{eq:popdis}
\end{equation}
\normalsize
where the factor
\begin{equation}
k = \frac{1}{(2 \pi)^{3/2} \sigma_{\text{U}} \sigma_{\text{V}} \sigma_{\text{W}}}
\label{eq:popdis_k}
\end{equation}
normalizes the expression. $\sigma_{\text{U}}$, $\sigma_{\text{V}}$, and $\sigma_{\text{W}}$ are the characteristic velocity dispersions and V$_{\text{asym}}$ is the asymmetric drift. For our calculations we adopt the values from \citet{Bensby2014} found in table A.1 of their appendix for the velocity dispersions, rotational lags, and normalizations in the Solar neighborhood. To calculate the probability that a star belongs to a specific population such as the thin disk, thick disk, and stellar halo ($D$, $TD$, and $H$, respectively), the initial probability is multiplied by the observed fractions ($X$) of each population in the Solar neighborhood \citep{Bensby2014}. The final probability is then determined by dividing a given population's probability with another population's probability to obtain the relative probability between the two populations for that star. For instance, the relative probability between the thick disk (TD) and thin disk (D) for a star is determined by,

\begin{equation}
TD/D = \frac{X_{TD}}{X_{D}} \cdot \frac{f_{TD}}{f_{D}}.
\label{eq:tdd}
\end{equation}

\begin{figure}
      	\centering
	\includegraphics[width=0.85\linewidth]{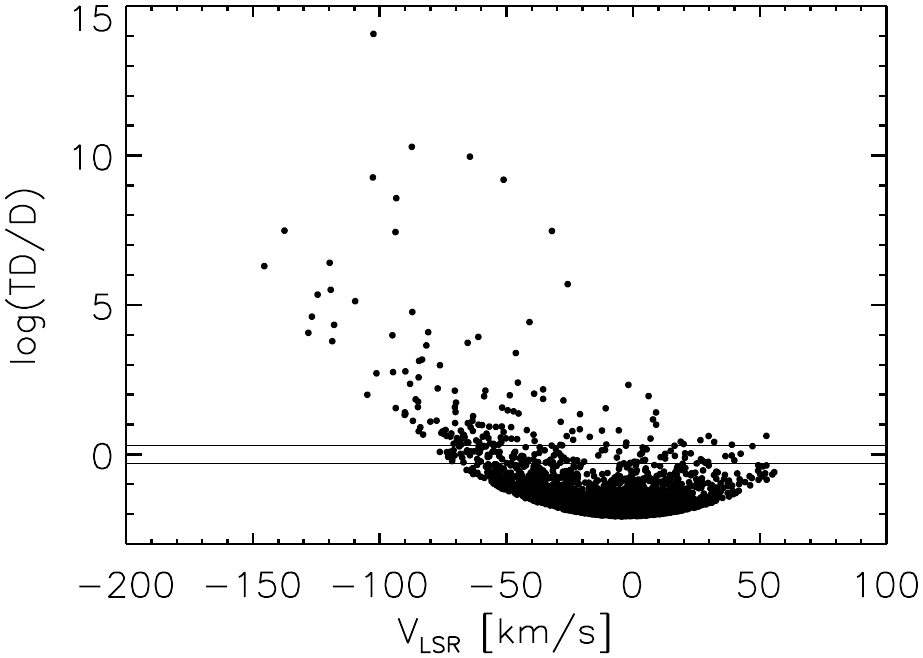}
\caption{Thick-to-thin disk probability ratios ($TD/D$) versus Galactic rotation velocity (V$_{\text{LSR}}$) for our sample of  2504  stars. The lower horizontal line displays the thin disk population cut ($TD/D < 0.5$,   2244  stars), where stars below this line are designated as thin disk stars. The upper horizontal line displays the thick disk population cut ($TD/D > 2$,  143  stars), where stars above this line are designated as thick disk stars. There are  117  stars between these two lines and are designated as thin/thick disk stars.}
\label{fig:v_thickthin}
\end{figure}

Following \citet{Bensby2014}, we assign stars to the thick disk if the thick disk probability is twice as large as the thin disk for that star ($TD/D > 2$). Stars are designated as populating the thin disk if $TD/D < 0.5$, and we label stars as `in between' the thin and thick disk if $0.5 < TD/D < 2$. Using this criteria there are  2244  thin disk stars,  143  thick disk stars, and  117  stars in between or thin/thick (trans) disk stars in our sample of  2504  stars.  Three  of these stars likely reside in the halo ($TD/H < 1$) and  17  stars may be part of the Hercules stream ($Her/TD > 2$ and $Her/D > 2$). A total of 19 stars in our sample could be associated with the Arcturus moving group with $-$115 $<$ V$_{\text{LSR}}$ $<$ $-$85 km s$^{-1}$. Figure \ref{fig:v_thickthin} displays the thick-to-thin disk probability ratios ($TD/D$) for our sample with the thin and thick disk cutoff values compared to their Galactic rotation velocity (V$_{\text{LSR}}$). 

\begin{figure}
      	\centering
	\includegraphics[width=0.99\linewidth]{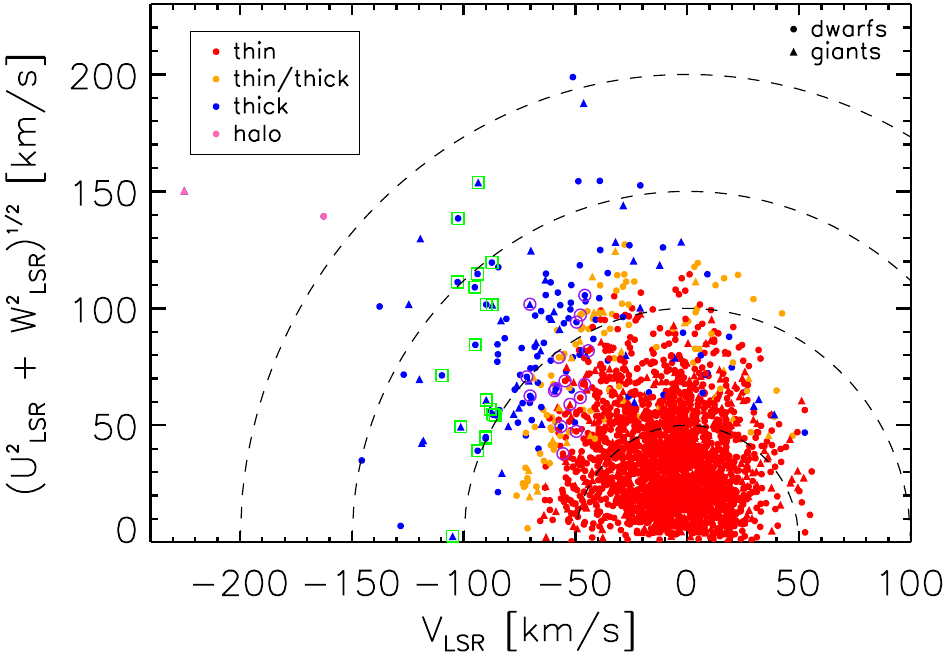}
\caption{Toomre diagram for our sample of  2504  MARVELS stars with Galactic space velocities. Dashed lines show constant values of total space velocity, $v_{tot}$ = (U$_{\text{LSR}}^{2}$ + V$_{\text{LSR}}^{2}$ + W$_{\text{LSR}}^{2}$)$^{1/2}$, in steps of 50 km s$^{-1}$. The stars' designation as members of the the thin, thin/thick, thick,  and halo  populations, based on probability ratios, are displayed as red, orange, blue,  and pink  symbols, respectively. Stars classified as dwarfs are represented with filled circles; those designated as giants are represented with filled triangles. Stars likely belonging to the Hercules stream based on kinematic probabilities ($Her/TD > 2$ and $Her/D > 2$) are surrounded by purple circles; stars that are candidates of the Arcturus moving group are surrounded by green squares.}
\label{fig:toomre}
\end{figure}

\begin{figure}
      	\centering
	\includegraphics[width=0.95\linewidth]{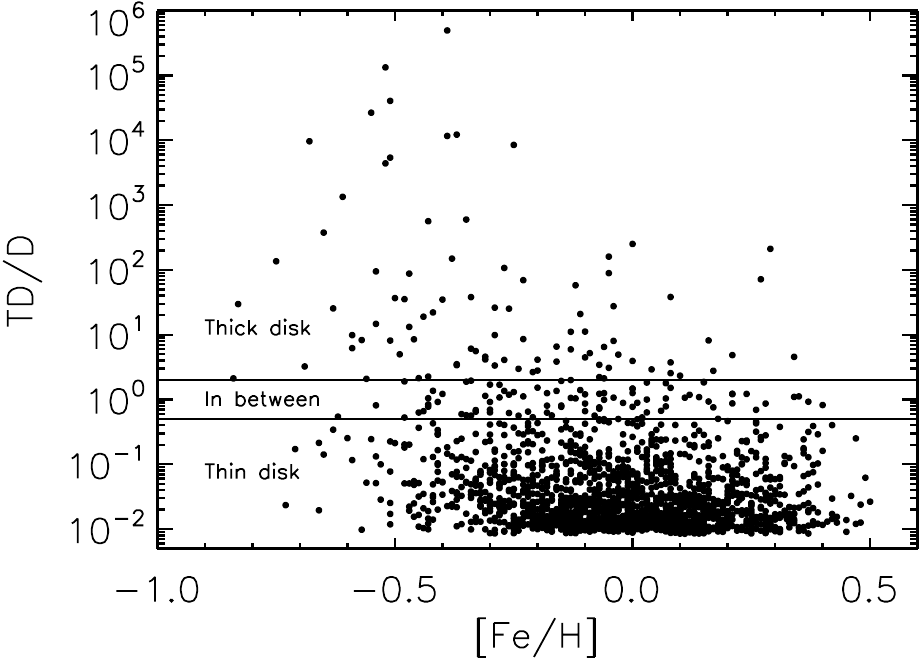}
\caption{Thick-to-thin disk probability ratios ($TD/D$) versus metallicity ([Fe/H]) for the  1878  stars with both Galactic space velocities and atmospheric parameters. As in figure \ref{fig:v_thickthin}, horizontal lines represent thin and thick disk populations cutoffs.}
\label{fig:feh_prob}
\end{figure}

\begin{figure}
      	\centering
	\includegraphics[width=0.95\linewidth]{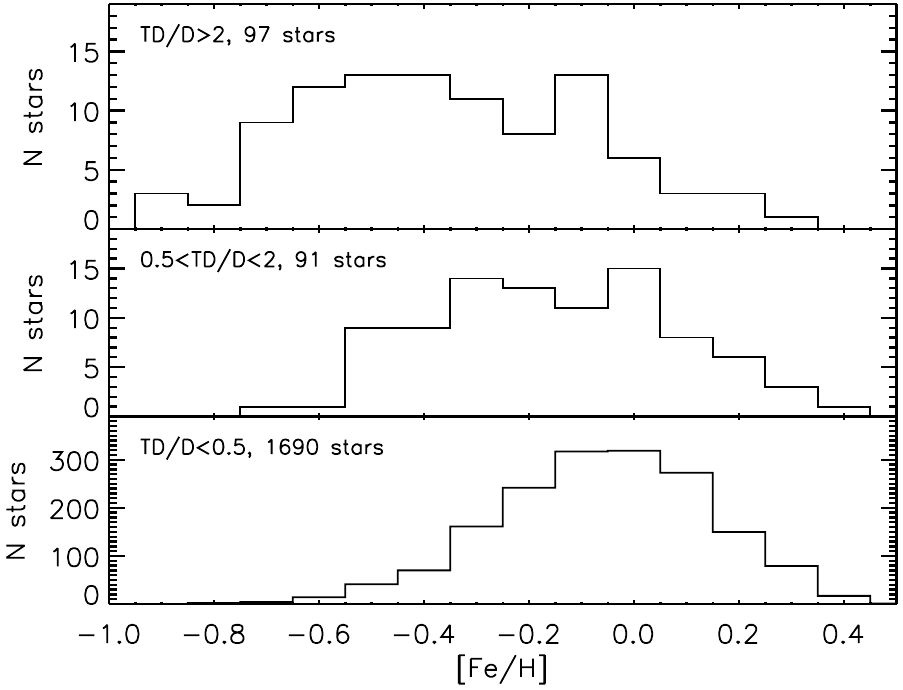}
\caption{Metallicity distributions for each of the thick (\textit{top}), thin/thick (\textit{middle}), and thin (\textit{bottom}) disk populations for our sample of   1878  stars with both Galactic space velocities and atmospheric parameters.}
\label{fig:feh_pops}
\end{figure}

A Toomre diagram, which represents the combined vertical and radial kinetic energies versus rotational energies, provides a different visual perspective to the kinematics of stars in the Galaxy and portrays a star's total velocity, $v_{tot}$ = (U$_{\text{LSR}}^{2}$ + V$_{\text{LSR}}^{2}$ + W$_{\text{LSR}}^{2}$)$^{1/2}$. Using total velocities allows rough approximations for population assignments where low-velocity stars ($v_{tot}$ $<$ 50 km s$^{-1}$) are typically thin disk stars, stars with 70 km s$^{-1}$ $<$ $v_{tot}$ $<$ 180 km s$^{-1}$ are likely thick disk stars, and stars with $v_{tot}$ $>$ 200 km s$^{-1}$ are likely halo stars \citep{Nissen2004, Bensby2014}. Figure \ref{fig:toomre} shows a Toomre diagram for our sample of  2504  stars with Galactic space velocities, with their population designation based on their relative population probabilities. Of the  2504  stars in our sample,  1701  dwarfs and  543  giants are in the thin disk,  92  dwarfs and  25  giants are in between the thin and thick disk cutoffs, and  98  dwarfs and  45  giants are in the thick disk.

Our sample contains  1878  dwarfs with Galactic space velocities and measured atmospheric parameters. Figure \ref{fig:feh_prob} shows the thick-to-thin disk probability ratios versus metallicity ([Fe/H]), and figure \ref{fig:feh_pops} displays the metallicity distributions for stars designated to the thick, thin/thick, or thin disk. The metallicity distributions in these three samples overlap, but a clear distinction in metallicity distribution is visible between the thick and thin disk stars. 

\subsubsection{Age-defined Populations}

\begin{figure}
      	\centering
	\includegraphics[width=0.9\linewidth]{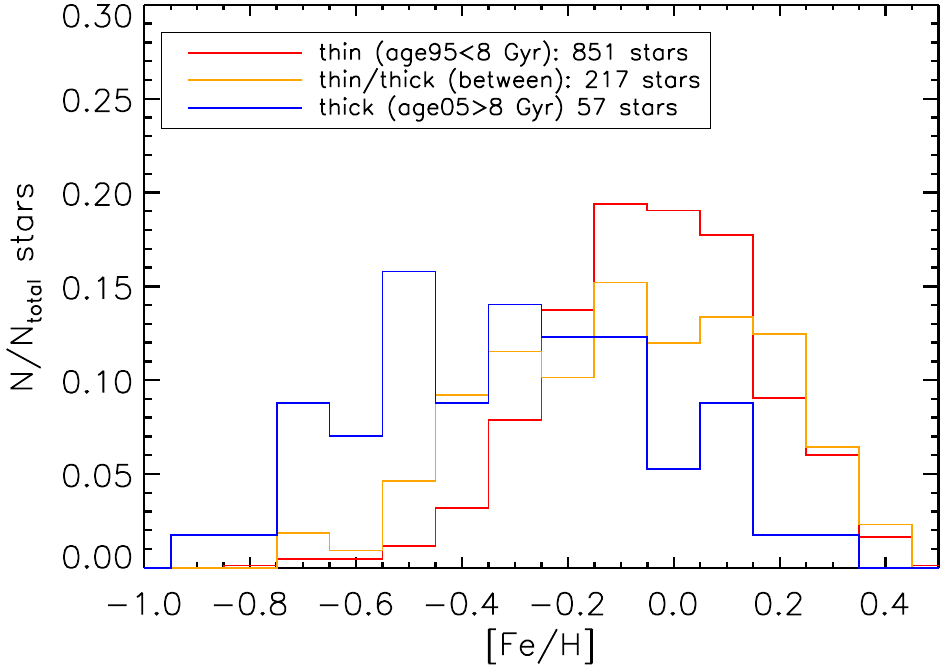}
	\centering
	\includegraphics[width=0.9\linewidth]{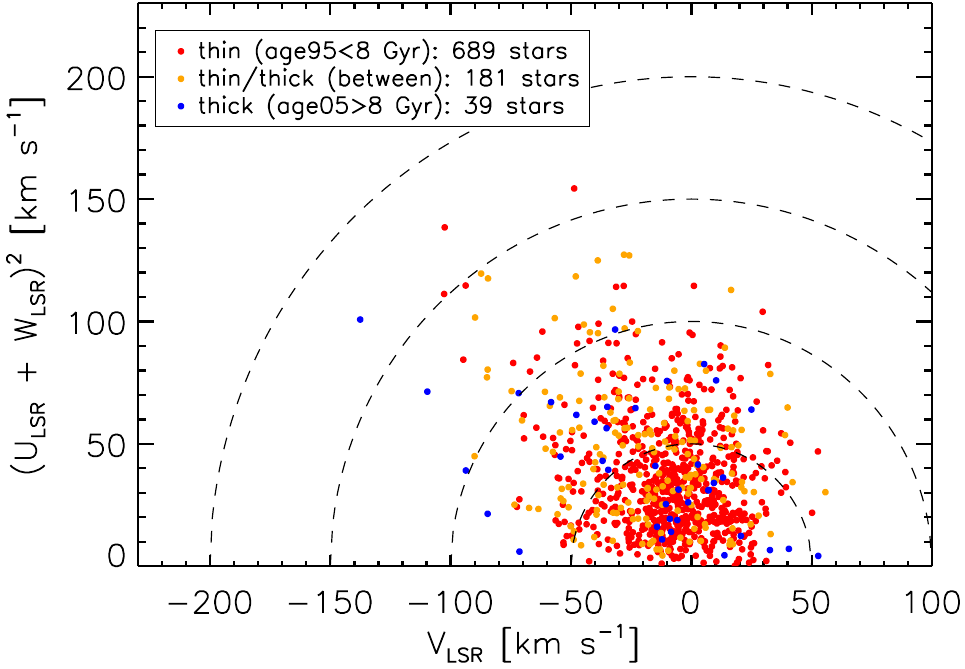}
\caption{\textit{Top}: Metallicity distributions for age-defined thin and thick disks based on probability distribution percentiles. \textit{Bottom}: Toomre diagram with age-defined thin and thick disks based on age probability distribution percentiles.}
\label{fig:agedefpop}
\end{figure}

We explore age-defined thin and thick disk populations using a 8 Gyr cut as prescribed by \citet{Haywood2013}. We assign age-defined populations using the age probability distribution percentiles  from 1125 stars with age errors less than 2 Gyr and where both methods agree within 2 Gyr . We define thin disk stars as having their 95th percentiles (age95) $<$ 8 Gyr, thick disk stars as having their 5th percentiles (age05) $>$ 8 Gyr, and thin/thick stars are those in between these two regimes. Figure \ref{fig:agedefpop} displays the metallicity distributions of these age-defined populations. Table \ref{tab:fehpop} displays the mean, standard deviation ($\sigma$), peak ($Pk$), skewness ($Sk$), median ($Md$), and the first $Q_{1}$ and third $Q_{3}$ quartiles (25th and 75th percentiles) of the [Fe/H] distributions for both the kinematically-defined and age-defined populations. Figure \ref{fig:agedefpop} also displays a Toomre diagram with stars color coded by age. Similar to \citet{Bensby2014} we find a large scatter across velocity space for both of these age distributions; in other words, both young and old stars can exhibit hot or cold kinematics.  However, some of this scatter may be introduced from uncertainties in our age estimates. For consistency we use the kinematically defined thin and thick disk for the rest of our analysis and results. 

\subsection{Kinematics \& Metallicities}

\begin{table}
\caption{[Fe/H] Distribution Parameters of Populations}
\label{tab:fehpop}
\begin{center}
Kinematically Defined Populations  \\
\hspace*{-1.0cm} \begin{tabular}{lcccccccc}
\hline
\hline
Kin. Pop. & N* & Mean & $\sigma$ & $Pk$ & $Sk$ & $Md$ & $Q_{1}$ & $Q_{3}$ \\
\hline		
thin 	&  1690 & $-$0.01 & 0.20 &  0.0 &  $-$0.25 & $-$0.01 & $-$0.14 &  0.12 \\
thin/thick &  91 &  $-$0.11 &  0.23 &  0.0 &  0.16 &  $-$0.01 &  $-$0.30 &  0.05 \\
thick &  97 &  $-$0.31 & 0.27 & $-$0.5 &  0.22 &  $-$0.34 & $-$0.51 &  $-$0.10 \\
Hercules &  14 &  $-$0.26 &  0.31 &  $-$0.1 &  $-$0.32 &  $-$0.10 &  $-$0.49 &  $-$0.05 \\
Arcturus & 13 &  $-$0.50 &  0.19 & $-$0.7 &   $-$0.47 &  $-$0.49 &  $-$0.64 &  $-$0.41 \\
\hline
\hline
\end{tabular} \\ 
Age Defined Populations  \\
\hspace*{-1.0cm} \begin{tabular}{lcccccccc}
\hline
\hline
Age. Pop. & N* & Mean & $\sigma$ & $Pk$ & $Sk$ & $Md$ & $Q_{1}$ & $Q_{3}$ \\
\hline		
thin &  851 &  0.01 &  0.20 &  $-$0.1 &  $-$0.28 &  0.01 &  $-$0.12 &  0.14 \\
thin/thick &  217 &  $-$0.04 &  0.25 &  $-$0.1 &  $-$0.20 &  $-$0.03 &  $-$0.25 &  0.16 \\
thick &  57 &  $-$0.26 &  0.27 &  $-$0.5 &  0.05 &  $-$0.26 &  $-$0.45 &  $-$0.05 \\
\hline
\hline
\end{tabular}  
\end{center}
\end{table}    

Table \ref{tab:fehpop} displays the median and quartile values of the kinematically-defined thin and thick disk, which exhibit a clear separation in metallicity between these kinematically-defined populations. The large metallicity dispersion in the thin disk can be explained by inward radial mixing of metal-poor and outward migration of metal-rich stars according to migration-based scenarios \citep{Sellwood2002,Roskar2008,Schonrich2009,Minchev2010,Brunetti2011} as well as ``blurring" \citep[i.e. wandering high eccentricity stars passing in the local volume, but with $R_{guide}$ outside the local snapshot of stars at the present time;][]{Schonrich2009,Minchev2013}. In section \ref{sec:blurclean} we examine the metallicity distribution for a blurring cleaned thin disk.

We also analyzed the metallicity distributions of the other two kinematically identified population groups, which are displayed in table \ref{tab:fehpop}. We find a large dispersion in the Arcturus stars; a metallicity homogeneity would suggest an origin from remnants of open clusters, but the lack of a clear chemical signature of the Arcturus stars indicates origins from dynamical perturbations within the Galaxy \citep[e.g.,][]{Williams2009,Ramya2012}. All 19 of our Arcturus moving group candidate stars are designated to the thick disk. \citet{Arifyanto2006} determined that the Arcturus group resided in the thick disk, and \citet{Bensby2014} found the general appearance ($\alpha$-abundance, age, and kinematics) to be similar to thick disk stars.  

\begin{figure}
      	\centering
	\includegraphics[width=0.85\linewidth]{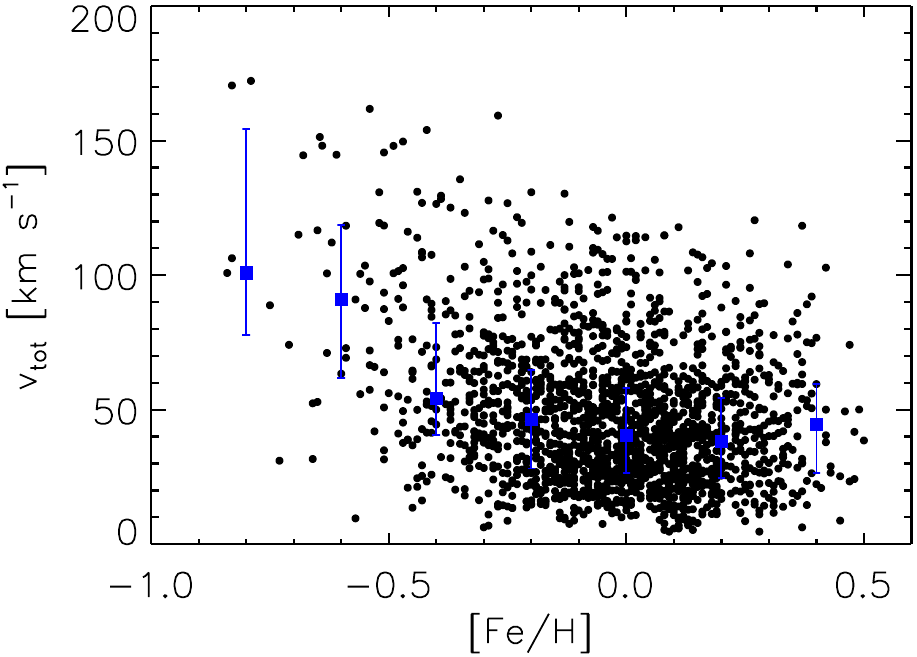}
	\includegraphics[width=0.85\linewidth]{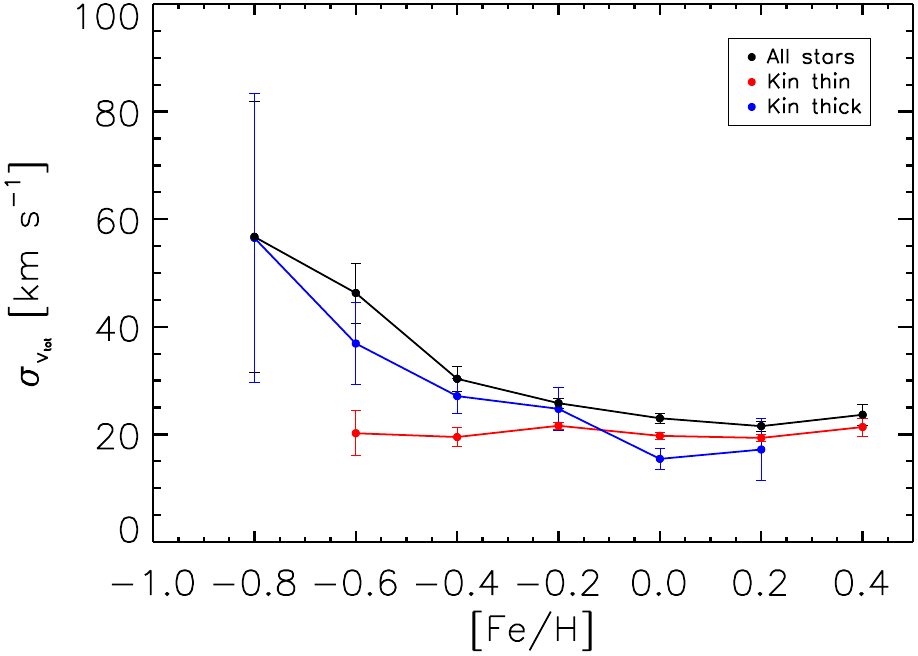}
\caption{\textit{Top}: Metallicity vs total space velocity, $v_{tot}$ = (U$_{\text{LSR}}^{2}$ + V$_{\text{LSR}}^{2}$ + W$_{\text{LSR}}^{2}$)$^{1/2}$, for our sample of 1878  stars with kinematic and atmospheric parameters. Blue squares show the median values with first and third quartiles as error bars for the data binned at 0.2 dex. \textit{Bottom}: Metallicity vs total space velocity dispersion binned at 0.2 dex. Black symbols represent all stars, while red and blue symbols represent stars in the kinematically-defined thin and thick disk, respectively. Errors were estimated using 1000 bootstrap realizations.}
\label{fig:vtotal_feh}
\end{figure}

The large spread in metallicity of the Hercules stream is consistent with  some  previous studies \citep[e.g.,][]{Raboud1998,Antoja2008} that reported large metallicity dispersions for this group. However, \citet{Arifyanto2006} and \citet{Kushniruk2017} reported that the Hercules stream primarily resided in the thin disk. \citet{Bensby2014} measured the abundance trend of the Hercules stars to be similar to the inner disk ($R_{Gal}$ = 4 to 7 kpc, $R_{Gal}$ $\equiv$ galactocentric radius), but the stars lay between the inner and outer ($R_{Gal}$ = 9 to 12 kpc) disk. The net velocity component of the Hercules stars is directed radially outwards from the Galactic center, suggesting origins at slightly smaller Galactocentric radii \citep{Bensby2014}, which is consistent with previous speculations that the Hercules stream stars originated in the inner parts of the Galaxy where they were kinematically heated by the central bar \citep[e.g.,][]{Dehnen2000}. Our Hercules stream stars have  seven  in the thin disk,  two  in the thin/thick disk, and  eight  in the thick disk. 

Using GALAH survey \citep{DeSilva2015} data of nearby stars, \citet{Quillen2018} found that the Hercules stream is most strongly seen in higher metallicity stars [Fe/H] > 0.2, and disagree with previous studies that found no significant metallicity preference for Hercules stream stars in the solar neighborhood. However, we use the (U,V,W) definitions for the population from \citet{Bensby2014}. As thoroughly discussed by \citet{Ramya2016}, this definition places the stream deeper among the thick disk stars, explaining the large contribution from the thick disk and lower metallicity in our proposed Hercules stream members. 

The top of figure \ref{fig:vtotal_feh} displays total space velocities, $v_{tot}$, vs metallicity to examine any possible trends. This figure reveals a possible small correlation between the two properties, with higher velocity stars likely having lower metallicity, as would be expected if these stars are kinematically different than the sun. The blue squares display the median values with first and third quartiles as error bars for the data binned at 0.2 dex. Overall the metallicity and total space velocity relationship appears relatively flat up to $\sim$50 km s$^{-1}$ with a negative trend in metallicity from 50 to 200 km s$^{-1}$. 

The bottom of figure \ref{fig:vtotal_feh} displays total space velocity dispersion vs metallicity binned at 0.2 dex. We found the dispersion for each bin by taking the standard deviation of the total space velocities and required at least four stars for each bin. Errors for dispersion were estimated by taking 1000 bootstrap realizations with replacement for each [Fe/H] bin and taking the overall standard deviation of all the standard deviations of each bootstrap as the error. We find that the total space velocity dispersion decreases with higher metallicity for thick disk stars, while the velocity dispersion for thin disk stars is relatively flat,  which is expected as our thin disk stars are selected to be kinematically cooler. 

\begin{figure}
      	\centering
	\includegraphics[width=0.85\linewidth]{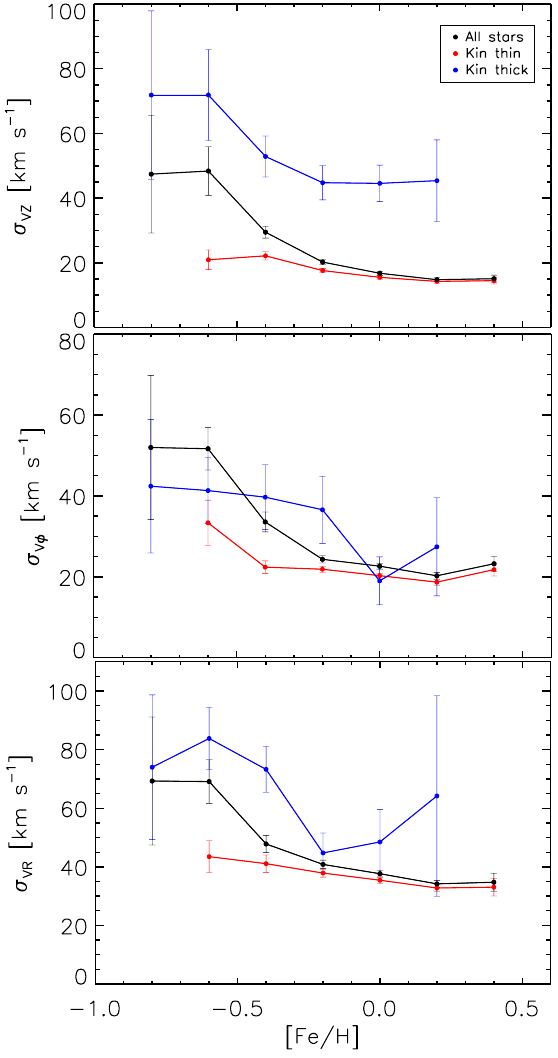}
\caption{Metallicity vs velocity dispersion binned at 0.2 dex for velocity dispersions of $V_{Z}$, $V_{\phi}$, and $V_{R}$ (from top to bottom). Black symbols represent all stars, while red and blue symbols represent stars in the kinematically-defined thin and thick disk, respectively. Errors were estimated using 1000 bootstrap realizations.}
\label{fig:vs_disp_feh}
\end{figure}

\begin{figure}
      	\centering
	\includegraphics[width=0.85\linewidth]{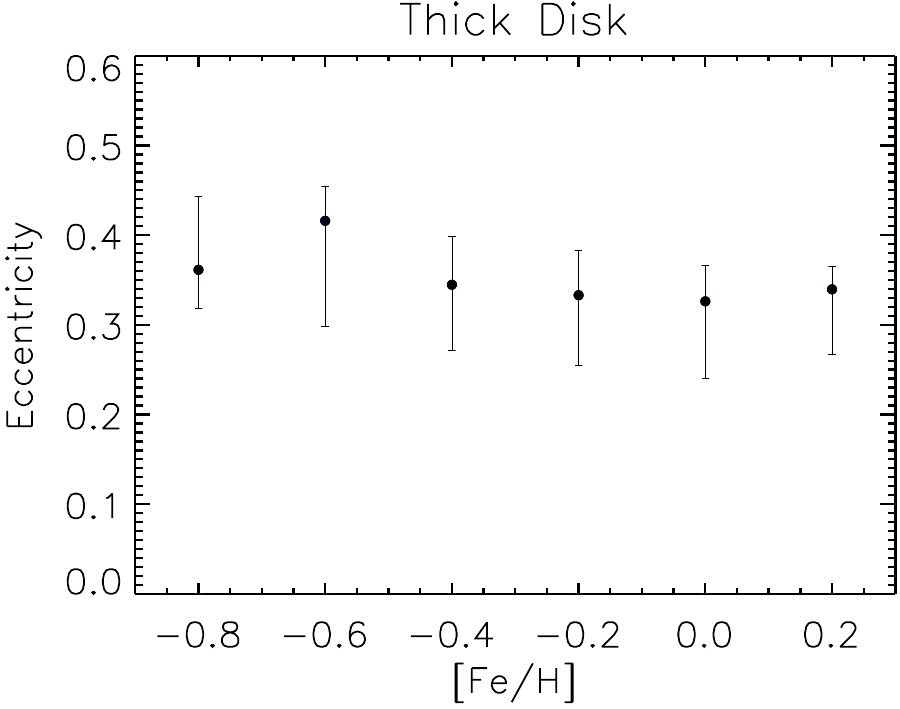}
\caption{Median eccentricity as a function of [Fe/H] for kinematically-defined thick disk stars binned at 0.2 dex. Error bars display the first and second quartiles of the eccentricity distribution in each [Fe/H] bin.}
\label{fig:ecc_feh_thick}
\end{figure}

Figure \ref{fig:vs_disp_feh} displays velocity dispersions for $V_{Z}$, $V_{\phi}$, and $V_{R}$ in 0.2 dex bins of [Fe/H]. Using the \textit{Gaia}-ESO Survey \citet{Recio-Blanco2014} examined velocity dispersions of FGK-type stars for their chemically separated thin and thick populations and found velocity dispersion are higher for the thick disk than the thin disk. \citet{Recio-Blanco2014} found no significant depence on either $\sigma_{VZ}$ or $\sigma_{V\phi}$ with metallicity, but that for the thick disk $\sigma_{V\phi}$ may increase with distance to the Galactic plane that is not visible for $\sigma_{VZ}$. We find that both the kinematically-defined thin disk and all stars exhibit a negative trend in $\sigma_{VZ}$, $\sigma_{V\phi}$, and $\sigma_{VR}$ with increasing metallicity. The kinematically-defined thick disk displays a negative trend with metallicity for $\sigma_{V\phi}$ as well. Differences between our results and those of \citet{Recio-Blanco2014} may arise from our sample being in a very local volume, whereas \citet{Recio-Blanco2014} is taking a larger volume where more stars from other radii are included. Thus, the dispersion and [Fe/H] relations are likely very dependent on sample selection.  We also note that differences in our results may arise from our kinematic population definitions, which are not the same as the chemical definitions used in other analyses such as \citet{Recio-Blanco2014}. By construction the thick disk stars shown here are kinematically hot and the thin disk stars kinematically cool, which is not necessarily the case for a thin and thick disk constructed chemically. 

Figure \ref{fig:ecc_feh_thick} displays median eccentricity as a function of [Fe/H] for kinematically-defined thick disk stars binned at 0.2 dex. \citet{Hayden2018} used 3000 stars selected from the fourth internal data release of the Gaia-ESO Survey and studied median eccentricity as function of [Fe/H] for their chemically-defined thick disk. As previously suggested by \citet{Recio-Blanco2014}, \citet{Hayden2018} found the eccentricity distribution to have a strong dependence on metallicity for the chemically-defined thick disk. For our kinematically-defined thick disk we do not find a trend of decreasing median eccentricity with [Fe/H], unlike \citet{Recio-Blanco2014} and \citet{Hayden2018}. However, this result is not surprising given our kinematic population selection instead of a chemical selection as done by the previously mentioned studies, which demonstrates a possible drawback of selecting populations kinematically. Differences may also arise due to different volume sizes of the two samples, as previously mentioned. 

\subsection{Ages \& Metallicities}

Stellar ages may likely be the best discriminator between the thin and thick disk \citep[e.g.,][]{Fuhrmann2011,Haywood2013,Bensby2014}; however, obtaining accurate stellar ages for a wide range of stars is notoriously difficult \citep[e.g.,][]{Soderblom2010}. Previous investigations find a clear age difference between the thin and thick disk and possible differences in the age-metallicity relations. \citet{Haywood2013} suggest that thin disk stars to tend to be less than 8 Gyr old, but metal-poor thin disk objects may have ages up to $\sim$10 Gyr. \citet{Haywood2013} find the thick disk formed over a period of 4-5 Gyr and the younger thick disk stars are 9-10 Gyr old. \citet{Haywood2013} find a correlation between age and metallicity for stars in the thick disk sequence and that the iron enrichment rate in the thick disk was about five times higher than the thin disk phase. However, \citet{Holmberg2007} find little or no variation in mean metallicity with age in the thin disk with a large and real scatter in [Fe/H] at all ages as well as no evidence for a significant age-metallicity relation in the thick disk. 

\citet{Kubryk2015} found that assuming the thick disk is composed of stars $>$9 Gyr leads to results consistent with most of the observed chemical and morphological properties of the thick disk. \citet{Kilic2017} used nearby ($<$ 40 pc) white dwarfs and derived ages of 6.8-7.0 Gyr for the thin disk and 8.7 $\pm$ 0.1 Gyr for the thick disk, and when using a deep proper motion catalog they derived ages of 7.4-8.2 Gyr for the thin disk and 9.5-9.9 Gyr for the thick disk. Notably, by combining results from the local and deep samples \citet{Kilic2017} find an age difference between the thin disk and thick disk to be 1.6 $\pm$ 0.4 Gyr.

\begin{table}
\caption{Age Distribution Parameters of Kinematic Populations}
\label{tab:agepop}
\begin{center}
\begin{tabular}{lccccccc}
\hline
\hline
Kin. Pop. & N* & Mean & $\sigma$  & $Md$ & $Q_{1}$ & $Q_{3}$ \\
& & [Gyr] & [Gyr] & [Gyr] & [Gyr] & [Gyr] \\
\hline		
thin &  842 &  4.6 &  2.4 &  4.0 &  3.1 &  5.5 \\
thin/thick &  31 &  6.7 &  2.8 &  5.9 &  4.2 &  9.4 \\
thick &  36 &  7.9 &  3.3 &  9.1 &  4.9 &  10.6 \\
Hercules &  6 &  8.9 &  3.8 &  10.5 &  4.6 &  12.4 \\
Arcturus &  9 &  7.9 &  3.5 &  9.2 &  4.9 &  10.8 \\
\hline
\end{tabular}  
\end{center}
\end{table}   

\begin{figure}
      	\centering
	\includegraphics[width=0.85\linewidth]{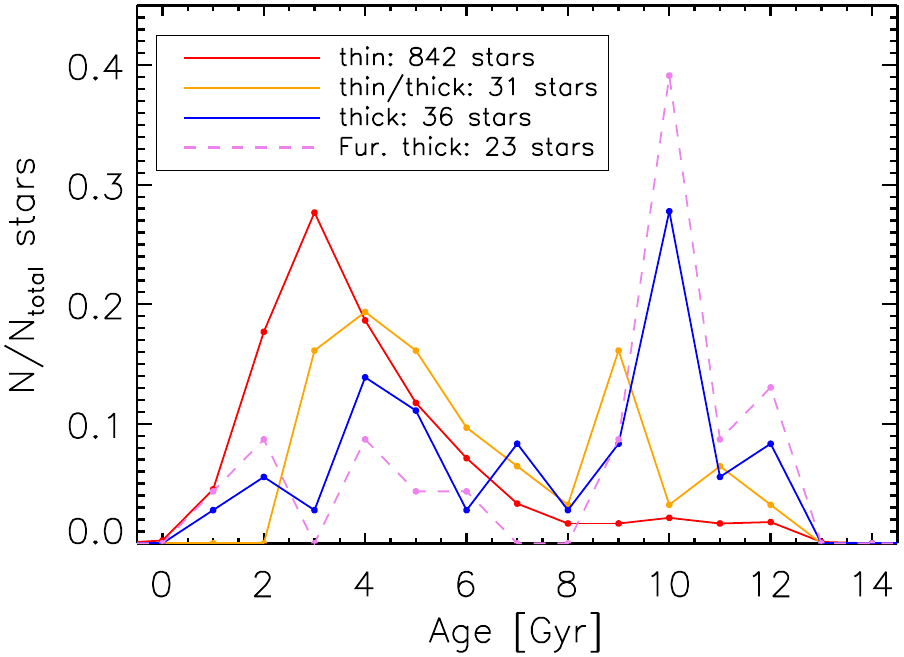}
\caption{Stellar age distributions for the kinematically-defined thin, thin/thick, and thick disk stars, as well as a Fuhrmann-like thick disk where the thick disk does not include stars with [Fe/H] $>$ -0.2 dex.}
\label{fig:pop_ages}
\end{figure}

As previously described, we create a final sample of ages  for analysis consisting of 1125 stars that have our most confident age estimates, where both methods agree to within 2 Gyr and error estimates are below 2 Gyr. We use these ages to examine the age properties of the kinematically-defined populations.   Figure \ref{fig:pop_ages} and table \ref{tab:agepop} display the kinematically-defined population age distributions. The thin and thick disk exhibit significant differences in age distributions,  as two clearly separated age peaks can be seen between the thin and thick disk populations in figure \ref{fig:pop_ages}.  We also show a ``Fuhrmann-like" thick disk \citep[e.g.,][]{Fuhrmann1998,Fuhrmann2004} where only low metallicity stars ([Fe/H] $>$ -0.2 dex) are included in the sample. We find that the Fuhrmann-like thick disk is generally older than the kinematically-defined thick disk that includes all metallicities.

Previous studies found that stars associated with the Hercules stream have a mixture of ages as seen in the thin and thick disk, but the Arcturus moving group have few stars younger than 10 Gyr and ages are generally similar to the thick disk \citep{Bensby2014}. In our limited sample   and given our specific kinematic definitions, which as discussed above may inherently include more thick disk stars,  we find both kinematic groups to be similar in age to the thick disk, with age distribution parameters displayed in table \ref{tab:agepop}.

\begin{figure}
      	\centering
	\includegraphics[width=0.75\linewidth]{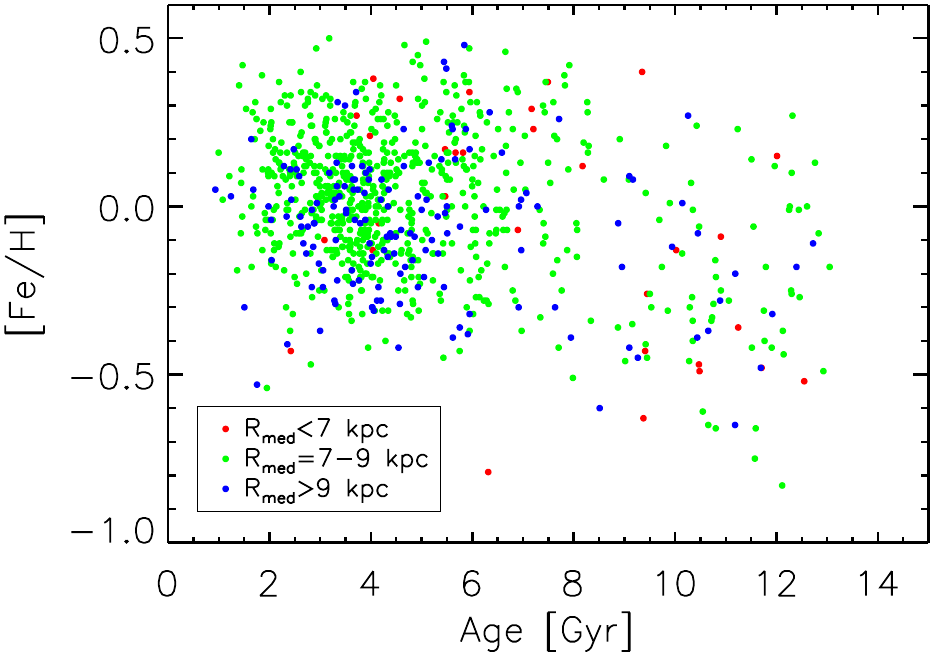}
	\includegraphics[width=0.75\linewidth]{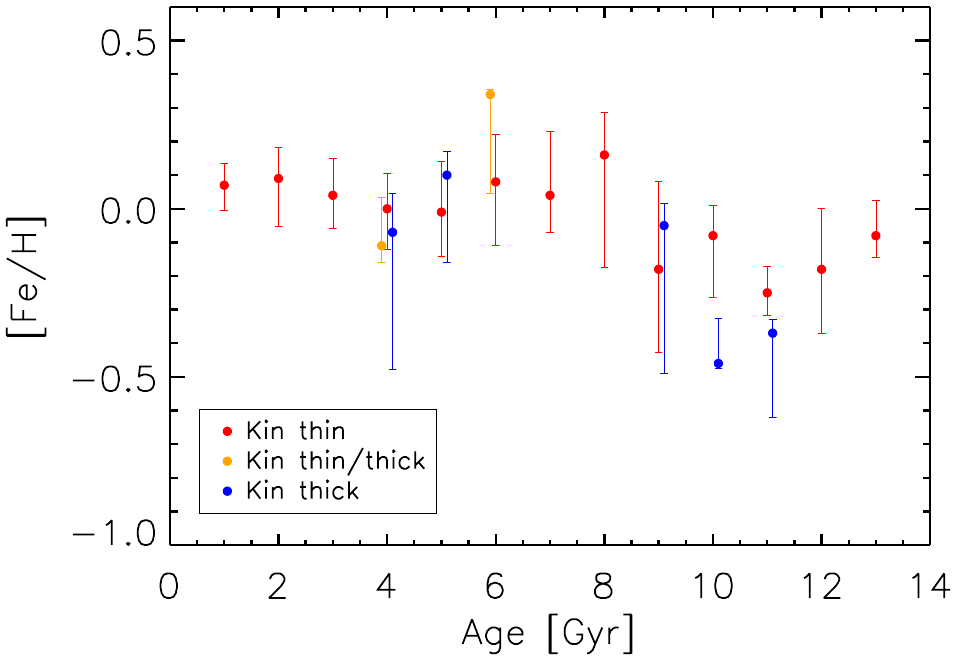}
\caption{\textit{Top}: Metallicity and stellar ages for the  1125 stars with confident ages used for our analysis . Symbols are colored based on $R_{med}$. \textit{Bottom}: Median metallicity for stellar age distributions of the kinematically-defined thin, thin/thick, and thick disk binned at 1 Gyr. Error bars display the first and third quartiles of the metallicity for each age bin.  A bin must contain 4 stars to be displayed. }
\label{fig:age_feh}
\end{figure}

The top of figure \ref{fig:age_feh} displays the relationship between metallicity and age for the overall sample and is color-coded by $R_{med}$. We find that outer disk stars ($R_{med}$ $>$ 9 kpc) generally have lower metallicity across all ages, as expected from inside-out chemical evolution models \citep[e.g.,][]{Chiappini2001,Chiappini2009} combined with radial migration \citep[e.g., Fig 4 of][]{Minchev2013}. The bottom of figure \ref{fig:age_feh} shows the kinematically-defined populations binned at 1 Gyr, which  exhibits an enrichment in metallicity from $\sim$11 to 8 Gyr.  As previously discussed, the large metallicity dispersion in the thin disk at fixed age can be explained by a combination of inward and outward radial migration and blurring. In section \ref{sec:blurclean} we show a blurring cleaned age-metallicity relationship for the thin disk.

\begin{figure}
      	\centering
	\includegraphics[width=0.85\linewidth]{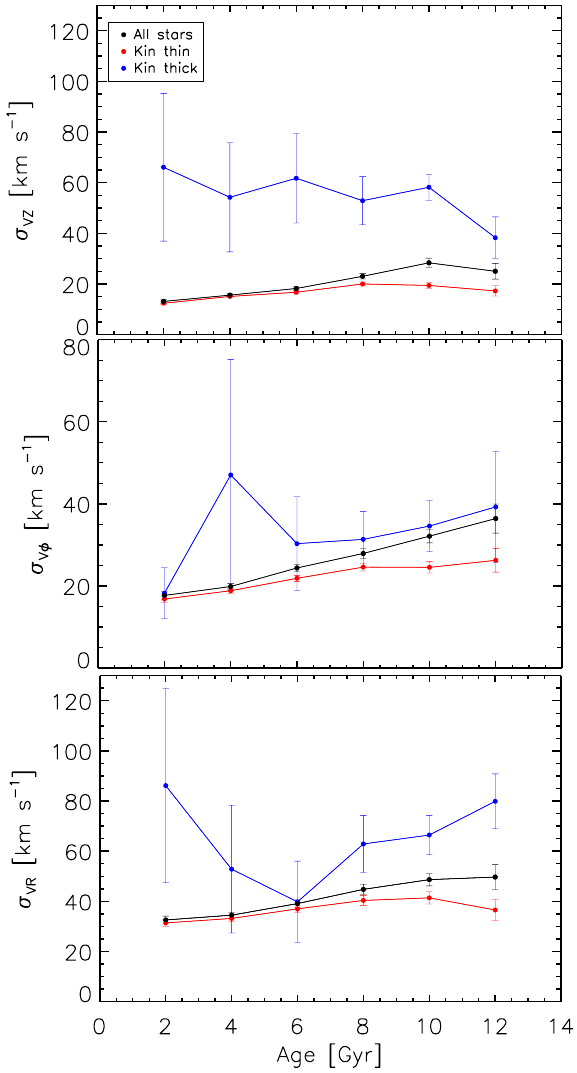}
\caption{Age vs velocity dispersion binned at 2 Gyr for velocity dispersions of $V_{Z}$, $V_{\phi}$, and $V_{R}$ (from top to bottom). Black symbols represent all stars, while red and blue symbols represent stars in the kinematically-defined thin and thick disk, respectively. Errors were estimated using 1000 bootstrap realizations.}
\label{fig:vs_disp_age}
\end{figure}

Figure \ref{fig:vs_disp_age} displays velocity dispersions for $V_{Z}$, $V_{\phi}$, and $V_{R}$ binned at 2 Gyr. Using the Geneva-Copenhagen survey \citet{Casagrande2011} found a prominent rise in velocity dispersion with age. Using 494 main sequence turnoff and subgiant stars from the AMBRE:HARPS survey with accurate astrometric information from \textit{Gaia} DR1, \citet{Hayden2017} analyzed vertical velocity dispersion as a function of age. \citet{Hayden2017} found that both low- and high-[Mg/Fe] star sequences have vertical velocity dispersion that generally increases with age. In figure 6 of \citet{Minchev2013} they display expected age-velocity relations for their chemo-dynamical model for the radial and vertical velocity dispersions. Examining the total population \citet{Minchev2013} find that for ages older than $\sim$3 Gyr $\sigma_{VR}$ flattens significantly and rises again after 8 Gyr, and they find similar behavior for $\sigma_{VZ}$, which is consistent with a violent origin for the hottest stellar population in the solar neighborhood.

Figure \ref{fig:vs_disp_age} shows that in general the kinematically defined thin disk stars have low velocity dispersion, as expected from how we defined the thin disk. The black curve (all stars) in figure \ref{fig:vs_disp_age} gives a better comparison to \citet{Minchev2013}'s results, which do not have thick disk stars. We find that all stars have velocity dispersions that increase with age, which is generally expected from \citet{Minchev2013}'s model, but we do not find a steep increase in $\sigma_{VR}$ and $\sigma_{VZ}$ at 8 Gyr  for the sample of all stars (black curve).  However, these plots are affected by selection effects, i.e., the proportions of thick and thin disk stars in the sample.

\begin{figure}
      	\centering
	\includegraphics[width=0.85\linewidth]{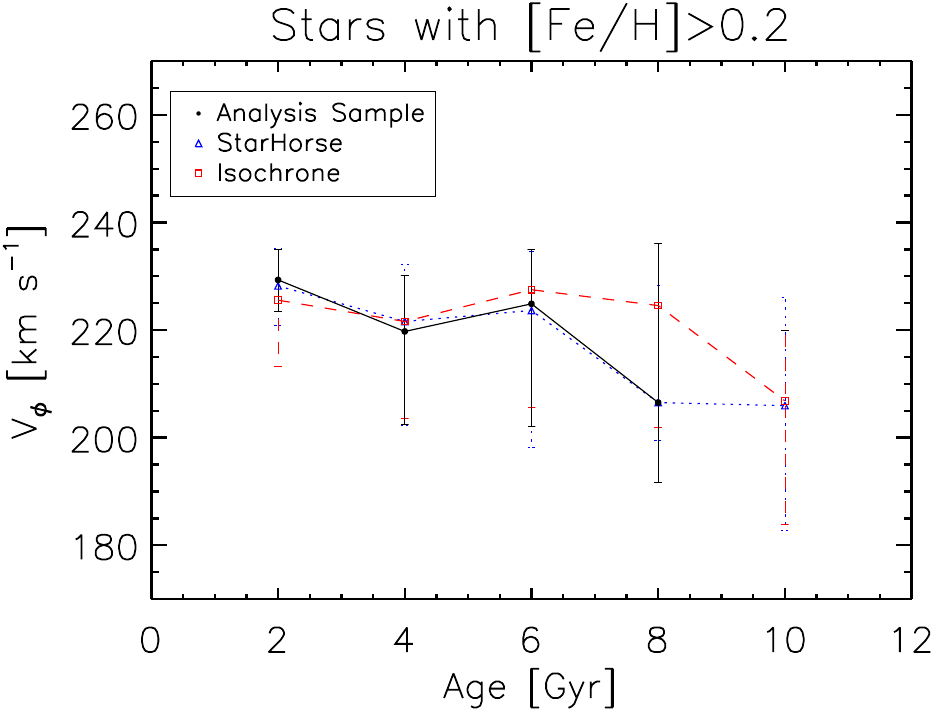}
\caption{Median $V_{\phi}$ as function of age for metal rich ([Fe/H] $>$ 0.2 dex) stars.  Filled circles and solid lines display our sample of ages used for analysis throughout the paper, blue triangles display only ages estimated from the \texttt{StarHorse} code and red squares represent only ages estimated from the Isochrone maximum-likelihood method.  $V_{\phi}$ are binned at 2 Gyr and we require at least 4 stars for each age bin. Error bars display the first and third quartile values of the $V_{\phi}$ distribution for each age bin.}
\label{fig:vphi_age_fehrich}
\end{figure}

Figure \ref{fig:vphi_age_fehrich} displays median $V_{\phi}$ as a function of age for metal rich ([Fe/H] $>$ 0.2 dex) stars. \citet{Hayden2018} found clear differences with $V_{\phi}$ at different metallicities and [Mg/Fe] abundances, which has been previously observed \citep[e.g.,][]{Lee2011b,Recio-Blanco2014,Guiglion2015}. Figure 6 of \citet{Minchev2013} also displays their model expectations for $V_{\phi}$ and age. For our sample we find a very similar relation to \citet{Minchev2013}'s results for $V_{\phi}$ and age, even for only metal rich stars. Notably this shows that we have a mix of populations in the metal rich end, similar to the mixing seen for thin disk stars.

\subsection{Galactic Orbit Analysis} \label{sec:galorbanal}

\subsubsection{Age and Metallicity Distributions}

\begin{figure*}
      	\centering
	\includegraphics[width=0.95\linewidth]{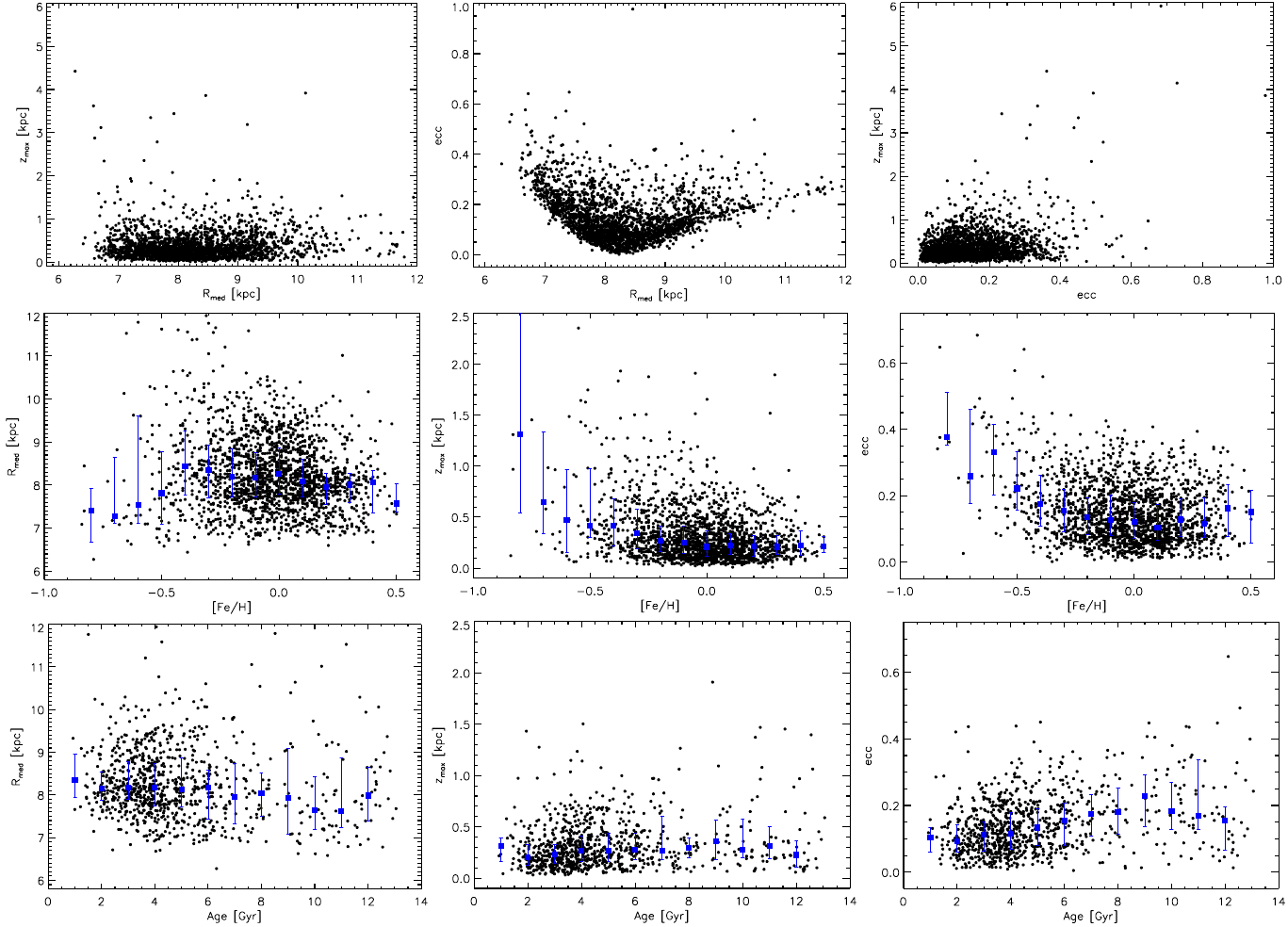}
\caption{Galactic orbital parameter distributions for $R_{med}$, $e$, and $z_{max}$ and their distributions with [Fe/H] and age. Blue symbols show median values with the first and third quartiles as errors for the data binned at 0.1 dex for [Fe/H] plots and 1 Gyr for age plots.}
\label{fig:galorb_fehage}
\end{figure*}

\begin{figure*}
      	\centering
	\includegraphics[width=0.98\linewidth]{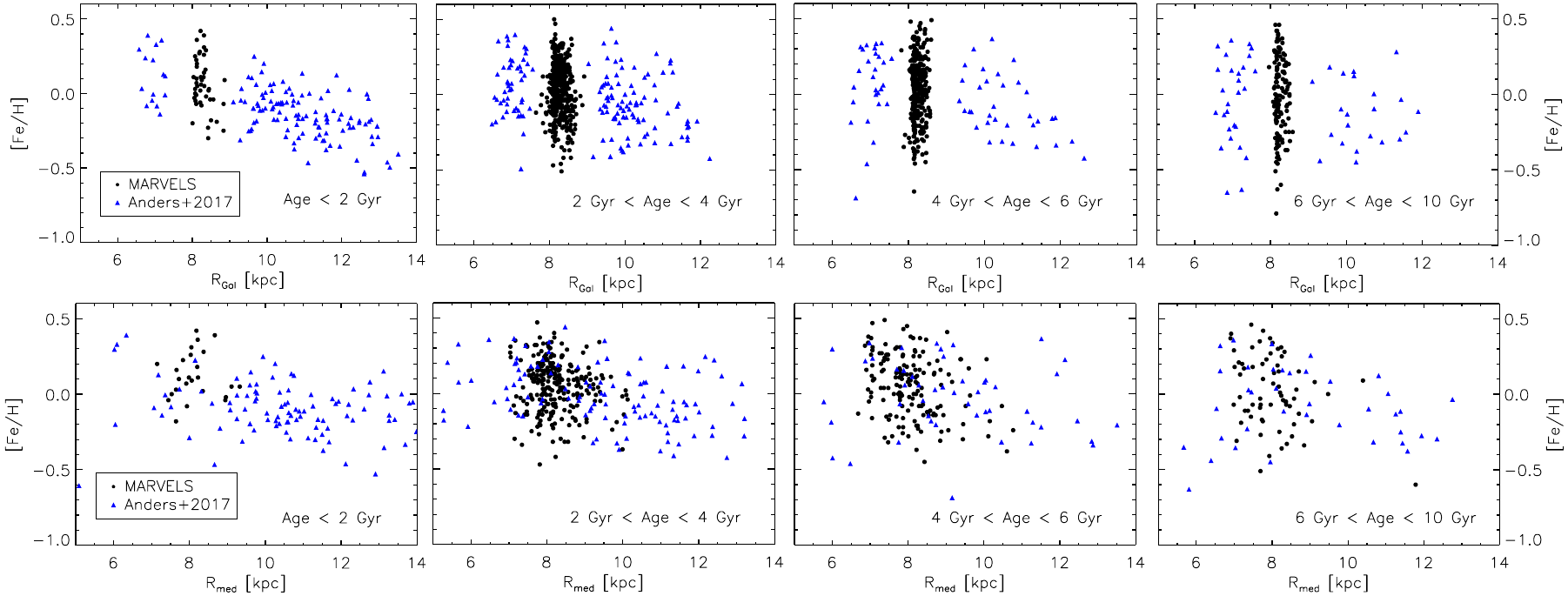}
\caption{\textit{Top}: Galactic orbital radius ($R_{Gal}$) and [Fe/H] for MARVELS stars with various ages in the thin disk geometrically defined as $z$ $<$ 0.3 kpc. APOGEE giant stars from \citet{Anders2017b} are shown with blue triangles. \textit{Bottom}: Same as above now using the orbital parameters $R_{med}$ and $z_{max}$ instead of $R_{Gal}$ and $z$ for MARVELS stars. For APOGEE stars we now display $R_{guide}$ values.}
\label{fig:rgal_feh_ages}
\end{figure*}

Here we analyze Galactic orbital parameters of our stars and how they compare with age and [Fe/H]. Figure \ref{fig:galorb_fehage} shows Galactic orbital parameters $R_{med}$, $e$, and $z_{max}$ for the MARVELS stars with derived Galactic orbits and compares each individually as a function of [Fe/H] and age. The distributions display that [Fe/H] decreases with larger $z_{max}$ and higher $e$. The [Fe/H] and $R_{med}$ distribution displays a slightly lower metallicity distribution for smaller $R_{med}$; we analyze $R_{med}$ and [Fe/H] in further detail by including scale height below. The age of MARVELS stars displays little dependence with $R_{med}$, but increases with both larger $z_{max}$ and higher $e$ values. These increases of age with $z_{max}$ and $e$ are expected as the age-$e$ and age-$z_{max}$ distributions are similar to the age-velocity dispersion relation and are directly related to the heating rate of stellar orbits in the disc.

In figure \ref{fig:rgal_feh_ages} we display Galactocentric orbital radii using both $R_{Gal}$ and [Fe/H] for stars with various ages in a geomtrically thin disk defined as $z$ $<$ 0.3 kpc as in \citet{Anders2017b}. The MARVELS data show a clear increase in metallicity range to more negative [Fe/H] values with larger age. We overlay the 418 stars from the \citet{Anders2017b} study and find the MARVELS stars to agree with the general trends of the giant stars from APOGEE, which display a negative slope in [Fe/H]/$R_{Gal}$ for ages $<$ 4 Gyr and a flat distribution for older ages. \citet{Anders2017b} find their results with [Fe/H]-$R_{Gal}$ are compatible with the $N$-body chemo-dynamical Milky-Way model by \citet{Minchev2013,Minchev2014}.

We also show these distributions using $R_{med}$ and $z_{max}$ in figure \ref{fig:rgal_feh_ages}, which allows us to look at a larger radial range with this local sample. Using $R_{med}$ and $z_{max}$ also allows the abundance gradient to be cleaned from the effect of ``blurring" (stars passing by the solar vicinity with high eccentricity) as opposed to ``churning" or genuine radial migration. If the initial [Fe/H] gradient of the interstellar medium is smooth, then scatter across the gradient through age is due to observational uncertainties, blurring (radial heating), and churning (radial migration). As expected we see slightly less scatter in the [Fe/H] gradient per age bin when observing $R_{med}$ as we remove the effect of blurring.

\begin{figure}
      	\centering
	\includegraphics[width=0.8\linewidth]{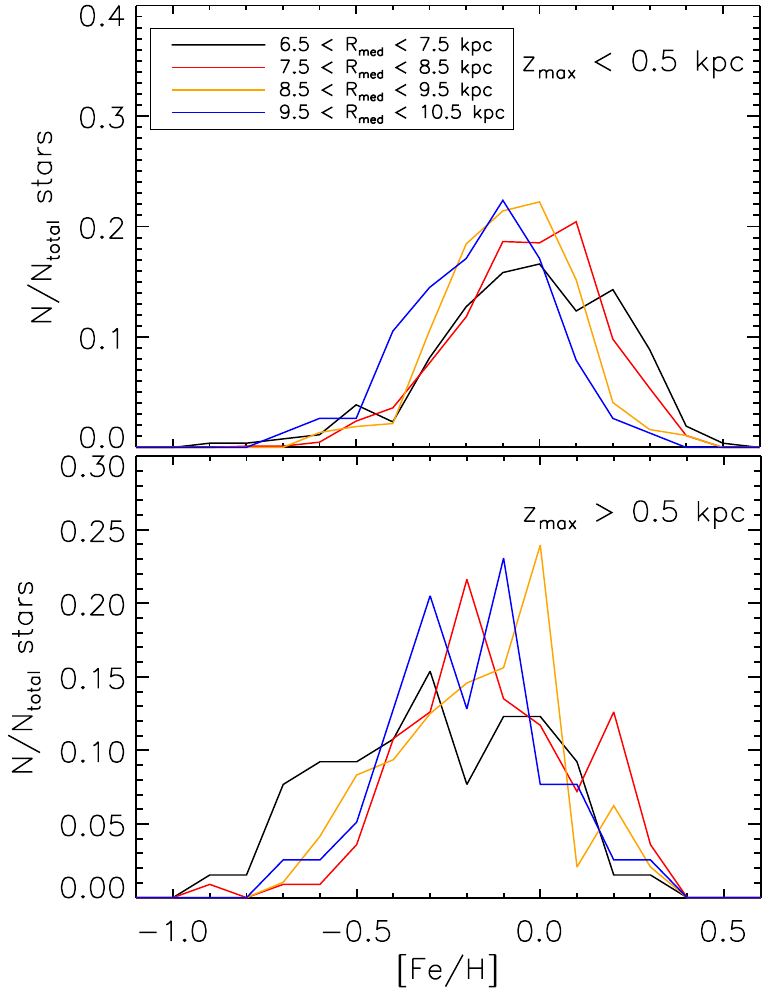}
\caption{\textit{Top}: Metallicity ([Fe/H]) distribution functions for various galactic orbital radii ($R_{med}$) for stars with $z_{max}$ $<$ 0.5 kpc. \textit{Bottom}: Same distributions for stars with $z_{max}$ $>$ 0.5 kpc.}
\label{fig:rgal_mdf}
\end{figure}

When analyzing the metallicity distribution function (MDF) across the disk of the Galaxy \citet{Hayden2015} found more metal-rich populations in the inner Galaxy and that the shape and skewness of the MDF in the midplane of the Galaxy are dependent on location, where the inner disk has a large negative skewness, the solar neighborhood MDF is roughly Gaussian, and the outer disk has a positive skewness. Figure \ref{fig:rgal_mdf} shows the MARVELS MDFs for stars with maximum Galactic scale heights above and below 0.5 kpc. Here we again use $R_{med}$ and $z_{max}$, which cleans the MDFs from the blurring of stellar orbits compared to the MDFs of \citet{Hayden2015}. Similarly to \citet{Anders2014}, who also looked at the MDFs over $R_{med}$ ranges, we find that the MDF is broader in the inner regions when compared to the outer ones, which is in agreement with pure chemical-evolution models for the thin disk \citep[e.g.,][]{Chiappini2001} and predictions of the chemo-dynamical model of \citet{Minchev2013,Minchev2014}.  We note that analyzing stars based on $R_{med}$ may only represent stars that are blurred to the solar neighborhood, which is likely biased towards older stellar populations relative to the in situ populations at those $R_{med}$. 

\begin{figure}
      	\centering
	\includegraphics[width=0.85\linewidth]{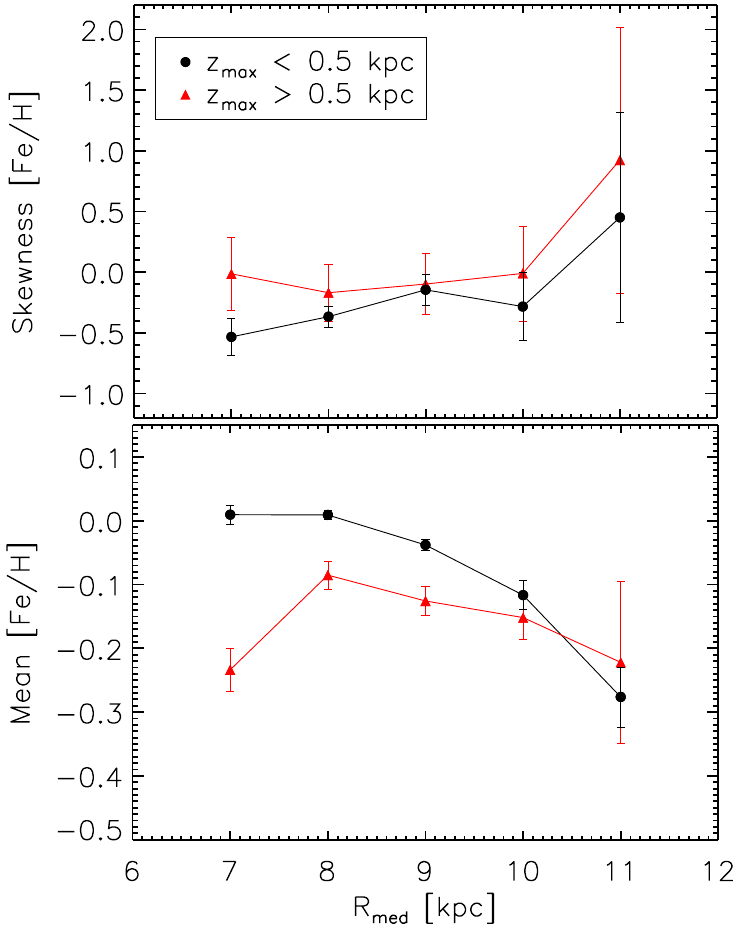}
\caption{\textit{Top}:  Galactic orbital radius ($R_{med}$) with skewness of [Fe/H] for MARVELS stars in bins of 1 kpc from 6.5 kpc $<$ $R_{med}$ $<$ 11.5 kpc. \textit{Bottom}: Galactic orbital radius ($R_{med}$) with mean of the [Fe/H] distribution for MARVELS stars in bins of 1 kpc from 6.5 kpc $<$ $R_{med}$ $<$ 11.5 kpc.}
\label{fig:rgal_skewmean}
\end{figure}

\begin{table}
\caption{Metallicity distribution functions for MARVELS stars across Galactocentric radii $R_{med}$ of the Milky Way}
\label{tab:mdf}
\begin{center}
\begin{tabular}{lcccc}
\hline
\hline
$R_{med}$ range (kpc) &  N stars & $\langle$[Fe/H]$\rangle$ & $\sigma_{\text{[Fe/H]}}$ & Skewness \\		
\hline	
& & $z_{max}$ $<$ 0.5 kpc & & \\
\hline	
6.5 - 7.5 &  259 &  0.01 &  0.24 &  $-$0.53 $\pm$ 0.15 \\
7.5 - 8.5 &  837 &  0.01 &  0.20 &  $-$0.37 $\pm$ 0.08 \\
8.5 - 9.5 &  369 &  $-$0.04 &  0.17 &  $-$0.15 $\pm$ 0.13 \\
9.5 - 10.5 &  76 &  $-$0.12 &  0.20 &  $-$0.28 $\pm$ 0.28 \\
10.5 - 11.5 &  8 &  $-$0.28 &  0.13 &  0.45 $\pm$ 0.87 \\
\hline
& & $z_{max}$ $>$ 0.5 kpc & & \\
\hline
6.5 - 7.5 &  65 &  $-$0.23 &  0.28 &  $-$0.01 $\pm$ 0.30 \\
7.5 - 8.5 &  111 &  $-$0.09 &  0.23 &  $-$0.17 $\pm$ 0.23 \\
8.5 - 9.5 &  96 &  $-$0.13 &  0.22 &  $-$0.10 $\pm$ 0.25 \\
9.5 - 10.5 &  39 &  $-$0.15 &  0.21 &  $-$0.01 $\pm$ 0.39 \\
10.5 - 11.5 &  5 &  $-$0.22 &  0.28 &  0.39 $\pm$ 1.10 \\
\hline
\hline
\end{tabular} \\
\end{center}
\end{table}
 
In figure \ref{fig:rgal_skewmean} we analyze the mean and skewness of the [Fe/H] distribution through Galactic orbital radii in bins of 1 kpc from 6.5 kpc $<$ $R_{med}$ $<$ 11.5 kpc for stars with $z_{max}$ above 0.5 kpc and those with $z_{max}$ below 0.5 kpc. These values are presented in table \ref{tab:mdf}. For stars closer to the Galactic disk ($z_{max}$ $<$ 0.5 kpc) we also find a strong correlation between mean [Fe/H] and Galactic radii where the inner disk stars are more metal-rich; however, stars farther from the Galactic disk ($z_{max}$ $>$ 0.5 kpc) do not display as clear of trend with the most metal poor population residing around $R_{med}$ $\sim$7 kpc. We also see a variance in skewness across Galactic radii where the outerdisk stars show a more positive [Fe/H] skewness than inner disk stars. As suggested by \citet{Hayden2015}, the positively skewed MDFs of the outer disk could be due to radial migration with the high-metallicity tail caused by stars that were born in the inner Galaxy.   We note again that analyzing stars based on $R_{med}$ may only represent stars that are blurred to the solar neighborhood, and the thick disk is likely the dominant population that is blurred to the solar location from the inner Galaxy. This may not be representative of the in situ populations in the inner disk. 

\subsubsection{Blurring Clean Thin Disk} \label{sec:blurclean}

\begin{figure}
      	\centering
	\includegraphics[width=0.8\linewidth]{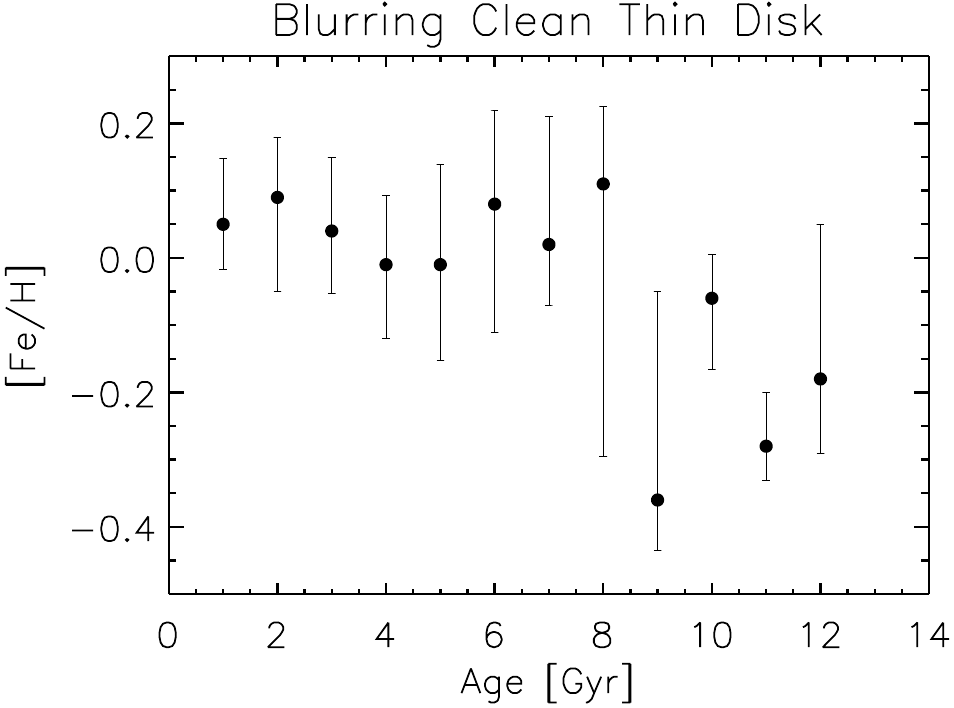}
	\includegraphics[width=0.75\linewidth]{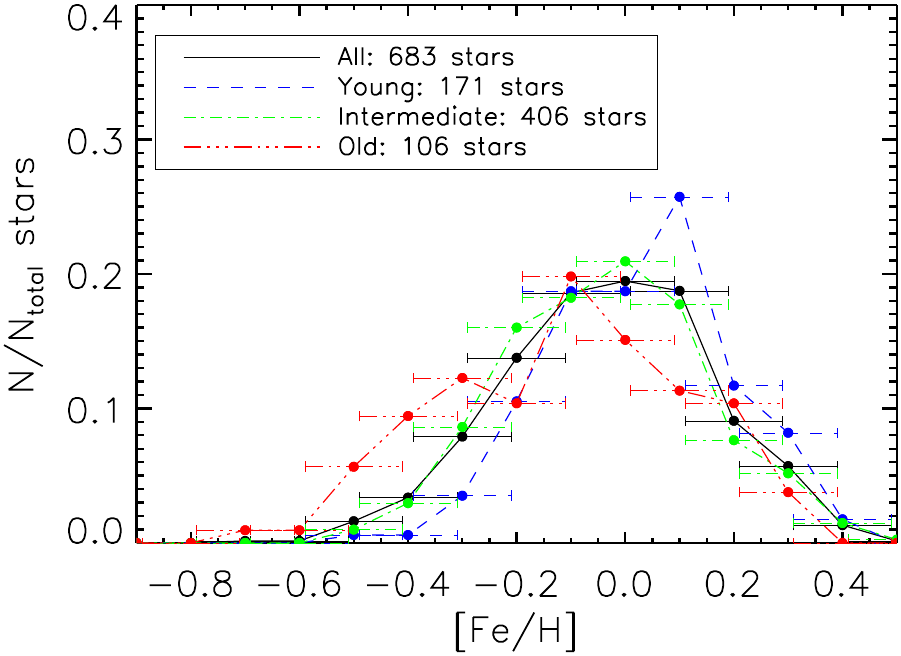}
\caption{\textit{Top}: Age-metallicity distribution for a blurring clean kinematically-defined thin disk where we only consider stars with $R_{guide}$ = 8.2 $\pm$ 1 kpc.  \textit{Bottom}: Metallicity distribution functions for the blurring cleaned kinematically-defined thin disk for three different age bins: young stars (age $<$ 2 Gyr), intermediate stars (2 Gyr $\le$ age $\le$ 6 Gyr), and old stars (age $>$ 6 Gyr). Error bars display typical median errors in [Fe/H] of $\sim$0.09 dex.}
\label{fig:blurclean}
\end{figure}

Metallicity dispersion with age can be caused by blurring, or high eccentricity stars passing in the local volume. Galactic radial position ($R_{Gal}$) gives a snapshot of stars at the present time, which covers a limited range in distance; however, guiding radii ($R_{guide}$) probe larger distances and are more suitable for studying the metallicity gradient of the thin disk. We present a ``blurring clean sample" for our thin disk stars where we only consider stars with $R_{guide}$ = 8.2 $\pm$ 1 kpc, to observe ``local" stars. This cut reduces the kinematically-defined thin disk population with ages and metallicities from  1690 to 1381  stars. 

Figure \ref{fig:blurclean} shows the age-metallicity relationship for the blurring cleaned kinematically-defined thin disk population. As expected, we still observe scatter across the age-metallicity relationship, but this shows the scatter is mainly due to radial migration as the blurring has been taken out. Figure \ref{fig:blurclean} also displays metallicity distributions of the thin disk in different age bins. Slicing the MDF into different age bins \citet{Casagrande2011} found that young stars have a considerably narrower distribution than old stars, though the peak always remains around the solar value. Previous studies have found that the most metal rich thin disk stars are not the youngest ones and therefore are not a product of the chemical enrichment of the local snapshot, but include migrated stars \cite[e.g,][]{Casagrande2011,Trevisan2011}. Our MDF shows again, in agreement with the velocity age relation for metal rich stars, that the metal rich end has stars of all ages. Figure \ref{fig:blurclean} displays that most of the youngest stars are more metal rich, but we find that there is an important contribution of old and intermediate age stars around [Fe/H] = 0.2.

\subsection{Previous MARVELS Companions}

Of the 19 previously published MARVELS low-mass stellar and substellar companions \citep{Lee2011a,Fleming2012,Wisniewski2012,Ma2013,Jiang2013,DeLee2013,Ma2016,Grieves2017} we obtained Galactic space velocities for  seven  stars: HD 87646 \citep{Ma2016}, TYC 4955-00369-1, TYC 3469-00492-1,  GSC 03467-00030,  TYC 3547-01007-1,  TYC 3556-03602-1,  and TYC 3148-02071-1 \citep{Grieves2017}. All  seven  stars likely reside in the thin disk and have total space velocities $v_{tot}$ of  24 to 71  km s$^{-1}$. Although this sample size is extremely limited, the lack of brown dwarf hosting thick disk stars is interesting as \citet{Adibekyan2012a,Adibekyan2012b} reported the frequency of planet hosting stars to be higher in the chemically assigned thick disk than in the thin disk, with eight of 65 stars hosting planets in the thick disk but only three of 136 stars hosting planets in the thin disk.  Our age analysis sample includes 10 of the 19 MARVELS companions. These stars range in age from 2.4 - 7.5 Gyr with a mean age of 3.9 Gyr . We do not find any trends with brown dwarf companion mass and host star age.

\section{Conclusions} \label{sec:conc}

We analyzed 3075 stars in the MARVELS radial velocity survey and obtained absolute radial velocities for 2610 stars and atmospheric parameters, radii, and masses for 2343 dwarf stars, of which 1971 have both absolute RV and atmospheric parameters. Our absolute RV values agree with previous high-resolution results to within 0.210 km s$^{-1}$. Our atmospheric parameter results agree to previous parameters found from high-resolution data and analysis to within 84 K for T$_{\text{eff}}$, 0.16 dex for log \textit{g}, and 0.06 dex for [Fe/H]. Our sample of atmospheric parameters have median values of 5780 K for T$_{\text{eff}}$, 4.38 dex for log \textit{g}, and $-$0.03 dex for [Fe/H]. Using a surface gravity and effective temperature relationship we analyzed all 3075 stars and designated 2358 as dwarfs and 717 as giants.

With our sample of absolute RVs we determined Galactic space velocities for  2504  stars using external sources for parallax and proper motion values. By assigning Gaussian distributions to populations within the Galaxy, we identified likely kinematic population assignments for each of these stars. We designated  2244   thin disk stars,  117   thin/thick disk stars, and  143  thick disk stars. Of these stars  three  likely reside in the halo,  17  may be part of the Hercules stream, and 19 may be associated with the Arcturus moving group. Of the  2504  stars in our sample,  1701  dwarfs and  543  giants are thin disk stars,  92  dwarfs and  25  giants are thin/thick disk stars, and  98  dwarfs and  45  giants are thick disk stars. We also assigned age-defined thin and thick disk populations using an 8 Gyr cut and age probability distributions to assign  851  stars to the thin disk and  57  stars to the thick disk, which displayed a clear separation in metallicity distributions.

By analyzing the  1878  stars with both space velocities and atmospheric parameters we determined median metallicities ([Fe/H]) of $-$0.01 dex for the kinematically-defined thin disk and  $-$0.31  dex for the thick disk. We determined stellar ages using both the spectro-photometric distance code \texttt{StarHorse} and the isochronal age-dating method and obtained median ages of  4.0  Gyr and  9.1  Gyr for the kinematically-defined thin and thick disk, respectively. These results agree with previous findings that the thick disk is likely populated with older and more metal poor stars. Our kinematically identified Arcturus moving group consists of thick disk stars with a median metallicity of  $-$0.49  dex, while the Hercules stream had a mixture of thin and thick disk stars with a median metallicity of  $-$0.10  dex. We found large dispersions in metallicity for both of these populations, which suggests they originated from dynamical perturbations rather than coming from the remnants of open clusters. With our limited sample we find age distributions similar to the thick disk for both of these populations with median ages of  10.5  Gyr and  9.2  Gyr for the Hercules stream and Arcturus moving group, respectively.

We find a likely negative trend in metallicity with total space velocity ($v_{tot}$) for stars with $v_{tot}$ $>$ 50 km s$^{-1}$. Analyzing the total velocity dispersion and metallicity, we found total velocity dispersion to decrease with higher metallicity for thick disk stars, while the total velocity dispersion for thin disk stars is relatively flat. Our kinematically-defined thin disk and all stars exhibit a negative trend in $\sigma_{VZ}$, $\sigma_{V\phi}$, and $\sigma_{VR}$ with increasing metallicity, while the kinematically-defined thick disk displays a negative trend with metallicity for $\sigma_{V\phi}$ as well. Unlike previous findings, our thick disk stars do not exhibit a trend of decreasing median eccentricity with [Fe/H]; however, this is likely due to biases of our kinematic population selection. We find that our total sample is compatible with the predictions of \citet{Minchev2013}. When analyzing velocity dispersion and age, we find that all stars have velocity dispersion that generally increases with age and that the kinematically-defined thick disk is larger in velocity dispersion across all ages. We observe $V_{\phi}$ to decrease with age for metal rich stars for both our age estimation methods, showing that we have a mix of populations in the metal rich end of our sample.

Analyzing the Galactic orbital parameters of the MARVELS stars we find distributions between age, [Fe/H], and Galactocentric radius for stars with Galactic scale heights of less than 0.3 kpc to agree with the trends of \citet{Anders2017b}, which are compatible with the chemo-dynamical Milky-Way model by \citet{Minchev2013,Minchev2014}. We observe stars closer to the Galactic disk ($z_{max}$ $<$ 0.5 kpc) to be more metal rich for smaller $R_{med}$ and have a more positive skewness in [Fe/H] for larger $R_{med}$. Stars farther from the Galactic disk ($z_{max}$ $>$ 0.5 kpc) do not display as clear of trend in mean metallicity across $R_{med}$, but still exhibit a high [Fe/H] skewness for larger $R_{med}$. The positive [Fe/H] skewness of the outer disk may be caused by stars migrating from the inner to outer disk over time. We find that outer disk stars ($R_{med}$ $>$ 9 kpc) generally have lower metallicity across all ages, as expected from inside-out chemical evolution models \citep[e.g.,][]{Chiappini2001,Chiappini2009} combined with radial migration \citep[e.g., Fig 4 of][]{Minchev2013}. Our blurring clean thin disk displays that stars of all ages are in the metal rich end, which agrees with our velocity-age relation for metal rich stars, suggesting radial migration dominates metallicity scatter in the thin disk.

This work corresponds to a large and homogeneous sample of stars with precise radial velocities. The parameters from this work are critical for future analysis of substellar companions orbiting these stars and ideal for a homogeneous study of substellar companions. These results may be used for future analysis of brown dwarf and planet completeness around solar-type stars.

\section*{Acknowledgements}

We acknowledge the generous support from the W.M. Keck Foundation for developing the MARVELS survey instruments. The MARVELS project has also been supported by NSF, NASA and the University of Florida. Funding for SDSS-III has been provided by the Alfred P. Sloan Foundation, the Participating Institutions, the National Science Foundation, and the U.S. Department of Energy Office of Science. SDSS-III is managed by the Astrophysical Research Consortium for the Participating Institutions of the SDSS-III Collaboration.  We thank the anonymous referee for useful comments that helped improve this study and manuscript.  NG acknowledges support from the University of Florida CLAS Dissertation Fellowship. DLO acknowledges the support from FAPESP (2016/20667-8). CC acknowledges support from DFG Grant CH1188/2-1 and from the ChETEC COST Action (CA16117), supported by COST (European Cooperation in Science and Technology). GFPM acknowledges the Brazilian National Council for Scientific and Technological Development (CPNq) research grant 474972/2009-7.



\bibliographystyle{mnras}
\bibliography{marvels} 

\begin{thebibliography}{}
\makeatletter
\relax
\def\mn@urlcharsother{\let\do\@makeother \do\$\do\&\do\#\do\^\do\_\do\%\do\~}
\def\mn@doi{\begingroup\mn@urlcharsother \@ifnextchar [ {\mn@doi@}
  {\mn@doi@[]}}
\def\mn@doi@[#1]#2{\def\@tempa{#1}\ifx\@tempa\@empty \href
  {http://dx.doi.org/#2} {doi:#2}\else \href {http://dx.doi.org/#2} {#1}\fi
  \endgroup}
\def\mn@eprint#1#2{\mn@eprint@#1:#2::\@nil}
\def\mn@eprint@arXiv#1{\href {http://arxiv.org/abs/#1} {{\tt arXiv:#1}}}
\def\mn@eprint@dblp#1{\href {http://dblp.uni-trier.de/rec/bibtex/#1.xml}
  {dblp:#1}}
\def\mn@eprint@#1:#2:#3:#4\@nil{\def\@tempa {#1}\def\@tempb {#2}\def\@tempc
  {#3}\ifx \@tempc \@empty \let \@tempc \@tempb \let \@tempb \@tempa \fi \ifx
  \@tempb \@empty \def\@tempb {arXiv}\fi \@ifundefined
  {mn@eprint@\@tempb}{\@tempb:\@tempc}{\expandafter \expandafter \csname
  mn@eprint@\@tempb\endcsname \expandafter{\@tempc}}}

\bibitem[\protect\citeauthoryear{{Adibekyan}, {Sousa}, {Santos}, {Delgado
  Mena}, {Gonz{\'a}lez Hern{\'a}ndez}, {Israelian}, {Mayor}  \&
  {Khachatryan}}{{Adibekyan} et~al.}{2012a}]{Adibekyan2012a}
{Adibekyan} V.~Z.,  {Sousa} S.~G.,  {Santos} N.~C.,  {Delgado Mena} E.,
  {Gonz{\'a}lez Hern{\'a}ndez} J.~I.,  {Israelian} G.,  {Mayor} M.,
  {Khachatryan} G.,  2012a, \mn@doi [\aap] {10.1051/0004-6361/201219401}, \href
  {http://adsabs.harvard.edu/abs/2012A%26A...545A..32A} {545, A32}

\bibitem[\protect\citeauthoryear{{Adibekyan}, {Delgado Mena}, {Sousa},
  {Santos}, {Israelian}, {Gonz{\'a}lez Hern{\'a}ndez}, {Mayor}  \&
  {Hakobyan}}{{Adibekyan} et~al.}{2012b}]{Adibekyan2012b}
{Adibekyan} V.~Z.,  {Delgado Mena} E.,  {Sousa} S.~G.,  {Santos} N.~C.,
  {Israelian} G.,  {Gonz{\'a}lez Hern{\'a}ndez} J.~I.,  {Mayor} M.,
  {Hakobyan} A.~A.,  2012b, \mn@doi [\aap] {10.1051/0004-6361/201220167}, \href
  {http://adsabs.harvard.edu/abs/2012A%26A...547A..36A} {547, A36}

\bibitem[\protect\citeauthoryear{{Adibekyan} et~al.,}{{Adibekyan}
  et~al.}{2013}]{Adibekyan2013}
{Adibekyan} V.~Z.,  et~al., 2013, \mn@doi [\aap] {10.1051/0004-6361/201321520},
  \href {http://adsabs.harvard.edu/abs/2013A%26A...554A..44A} {554, A44}

\bibitem[\protect\citeauthoryear{{Alam} et~al.,}{{Alam}
  et~al.}{2015}]{Alam2015}
{Alam} S.,  et~al., 2015, \mn@doi [\apjs] {10.1088/0067-0049/219/1/12}, \href
  {http://adsabs.harvard.edu/abs/2015ApJS..219...12A} {219, 12}

\bibitem[\protect\citeauthoryear{{Anders} et~al.,}{{Anders}
  et~al.}{2014}]{Anders2014}
{Anders} F.,  et~al., 2014, \mn@doi [\aap] {10.1051/0004-6361/201323038}, \href
  {http://adsabs.harvard.edu/abs/2014A%26A...564A.115A} {564, A115}

\bibitem[\protect\citeauthoryear{{Anders} et~al.,}{{Anders}
  et~al.}{2017a}]{Anders2017a}
{Anders} F.,  et~al., 2017a, \mn@doi [\aap] {10.1051/0004-6361/201527204},
  \href {http://adsabs.harvard.edu/abs/2017A%26A...597A..30A} {597, A30}

\bibitem[\protect\citeauthoryear{{Anders} et~al.,}{{Anders}
  et~al.}{2017b}]{Anders2017b}
{Anders} F.,  et~al., 2017b, \mn@doi [\aap] {10.1051/0004-6361/201629363},
  \href {http://adsabs.harvard.edu/abs/2017A%26A...600A..70A} {600, A70}

\bibitem[\protect\citeauthoryear{{Andrae} et~al.,}{{Andrae}
  et~al.}{2018}]{Andrae2018}
{Andrae} R.,  et~al., 2018, preprint, \href
  {http://adsabs.harvard.edu/abs/2018arXiv180409374A} {} (\mn@eprint {arXiv}
  {1804.09374})

\bibitem[\protect\citeauthoryear{{Antoja}, {Figueras}, {Fern{\'a}ndez}  \&
  {Torra}}{{Antoja} et~al.}{2008}]{Antoja2008}
{Antoja} T.,  {Figueras} F.,  {Fern{\'a}ndez} D.,   {Torra} J.,  2008, \mn@doi
  [\aap] {10.1051/0004-6361:200809519}, \href
  {http://adsabs.harvard.edu/abs/2008A%26A...490..135A} {490, 135}

\bibitem[\protect\citeauthoryear{{Arifyanto} \& {Fuchs}}{{Arifyanto} \&
  {Fuchs}}{2006}]{Arifyanto2006}
{Arifyanto} M.~I.,  {Fuchs} B.,  2006, \mn@doi [\aap]
  {10.1051/0004-6361:20054355}, \href
  {http://adsabs.harvard.edu/abs/2006A%26A...449..533A} {449, 533}

\bibitem[\protect\citeauthoryear{{Bensby}, {Feltzing}  \&
  {Lundstr{\"o}m}}{{Bensby} et~al.}{2003}]{Bensby2003}
{Bensby} T.,  {Feltzing} S.,   {Lundstr{\"o}m} I.,  2003, \mn@doi [\aap]
  {10.1051/0004-6361:20031213}, \href
  {http://adsabs.harvard.edu/abs/2003A%26A...410..527B} {410, 527}

\bibitem[\protect\citeauthoryear{{Bensby}, {Feltzing}, {Lundstr{\"o}m}  \&
  {Ilyin}}{{Bensby} et~al.}{2005}]{Bensby2005}
{Bensby} T.,  {Feltzing} S.,  {Lundstr{\"o}m} I.,   {Ilyin} I.,  2005, \mn@doi
  [\aap] {10.1051/0004-6361:20040332}, \href
  {http://adsabs.harvard.edu/abs/2005A%26A...433..185B} {433, 185}

\bibitem[\protect\citeauthoryear{{Bensby}, {Feltzing}  \& {Oey}}{{Bensby}
  et~al.}{2014}]{Bensby2014}
{Bensby} T.,  {Feltzing} S.,   {Oey} M.~S.,  2014, \mn@doi [\aap]
  {10.1051/0004-6361/201322631}, \href
  {http://adsabs.harvard.edu/abs/2014A%26A...562A..71B} {562, A71}

\bibitem[\protect\citeauthoryear{{Bland-Hawthorn} \&
  {Gerhard}}{{Bland-Hawthorn} \& {Gerhard}}{2016}]{Bland-Hawthorn2016}
{Bland-Hawthorn} J.,  {Gerhard} O.,  2016, \mn@doi [\araa]
  {10.1146/annurev-astro-081915-023441}, \href
  {http://adsabs.harvard.edu/abs/2016ARA%26A..54..529B} {54, 529}

\bibitem[\protect\citeauthoryear{{Bovy}}{{Bovy}}{2015}]{Bovy2015}
{Bovy} J.,  2015, \mn@doi [\apjs] {10.1088/0067-0049/216/2/29}, \href
  {http://adsabs.harvard.edu/abs/2015ApJS..216...29B} {216, 29}

\bibitem[\protect\citeauthoryear{{Bressan}, {Marigo}, {Girardi}, {Salasnich},
  {Dal Cero}, {Rubele}  \& {Nanni}}{{Bressan} et~al.}{2012}]{Bressan2012}
{Bressan} A.,  {Marigo} P.,  {Girardi} L.,  {Salasnich} B.,  {Dal Cero} C.,
  {Rubele} S.,   {Nanni} A.,  2012, \mn@doi [\mnras]
  {10.1111/j.1365-2966.2012.21948.x}, \href
  {http://adsabs.harvard.edu/abs/2012MNRAS.427..127B} {427, 127}

\bibitem[\protect\citeauthoryear{{Brunetti}, {Chiappini}  \&
  {Pfenniger}}{{Brunetti} et~al.}{2011}]{Brunetti2011}
{Brunetti} M.,  {Chiappini} C.,   {Pfenniger} D.,  2011, \mn@doi [\aap]
  {10.1051/0004-6361/201117566}, \href
  {http://adsabs.harvard.edu/abs/2011A%26A...534A..75B} {534, A75}

\bibitem[\protect\citeauthoryear{{Casagrande}, {Ram{\'{\i}}rez},
  {Mel{\'e}ndez}, {Bessell}  \& {Asplund}}{{Casagrande}
  et~al.}{2010}]{Casagrande2010}
{Casagrande} L.,  {Ram{\'{\i}}rez} I.,  {Mel{\'e}ndez} J.,  {Bessell} M.,
  {Asplund} M.,  2010, \mn@doi [\aap] {10.1051/0004-6361/200913204}, \href
  {http://adsabs.harvard.edu/abs/2010A%26A...512A..54C} {512, A54}

\bibitem[\protect\citeauthoryear{{Casagrande}, {Sch{\"o}nrich}, {Asplund},
  {Cassisi}, {Ram{\'{\i}}rez}, {Mel{\'e}ndez}, {Bensby}  \&
  {Feltzing}}{{Casagrande} et~al.}{2011}]{Casagrande2011}
{Casagrande} L.,  {Sch{\"o}nrich} R.,  {Asplund} M.,  {Cassisi} S.,
  {Ram{\'{\i}}rez} I.,  {Mel{\'e}ndez} J.,  {Bensby} T.,   {Feltzing} S.,
  2011, \mn@doi [\aap] {10.1051/0004-6361/201016276}, \href
  {http://adsabs.harvard.edu/abs/2011A%26A...530A.138C} {530, A138}

\bibitem[\protect\citeauthoryear{{Chabrier}}{{Chabrier}}{2003}]{Chabrier2003}
{Chabrier} G.,  2003, \mn@doi [\pasp] {10.1086/376392}, \href
  {http://adsabs.harvard.edu/abs/2003PASP..115..763C} {115, 763}

\bibitem[\protect\citeauthoryear{{Chen}, {Bressan}, {Girardi}, {Marigo}, {Kong}
   \& {Lanza}}{{Chen} et~al.}{2015}]{Chen2015}
{Chen} Y.,  {Bressan} A.,  {Girardi} L.,  {Marigo} P.,  {Kong} X.,   {Lanza}
  A.,  2015, \mn@doi [\mnras] {10.1093/mnras/stv1281}, \href
  {http://adsabs.harvard.edu/abs/2015MNRAS.452.1068C} {452, 1068}

\bibitem[\protect\citeauthoryear{{Chiappini}, {Matteucci}  \&
  {Romano}}{{Chiappini} et~al.}{2001}]{Chiappini2001}
{Chiappini} C.,  {Matteucci} F.,   {Romano} D.,  2001, \mn@doi [\apj]
  {10.1086/321427}, \href {http://adsabs.harvard.edu/abs/2001ApJ...554.1044C}
  {554, 1044}

\bibitem[\protect\citeauthoryear{{Chiappini}, {G{\'o}rny}, {Stasi{\'n}ska}  \&
  {Barbuy}}{{Chiappini} et~al.}{2009}]{Chiappini2009}
{Chiappini} C.,  {G{\'o}rny} S.~K.,  {Stasi{\'n}ska} G.,   {Barbuy} B.,  2009,
  \mn@doi [\aap] {10.1051/0004-6361:200810849}, \href
  {http://adsabs.harvard.edu/abs/2009A%26A...494..591C} {494, 591}

\bibitem[\protect\citeauthoryear{{Chubak}, {Marcy}, {Fischer}, {Howard},
  {Isaacson}, {Johnson}  \& {Wright}}{{Chubak} et~al.}{2012}]{Chubak2012}
{Chubak} C.,  {Marcy} G.,  {Fischer} D.~A.,  {Howard} A.~W.,  {Isaacson} H.,
  {Johnson} J.~A.,   {Wright} J.~T.,  2012, preprint, \href
  {http://adsabs.harvard.edu/abs/2012arXiv1207.6212C} {} (\mn@eprint {arXiv}
  {1207.6212})

\bibitem[\protect\citeauthoryear{{Ciardi} et~al.,}{{Ciardi}
  et~al.}{2011}]{Ciardi2011}
{Ciardi} D.~R.,  et~al., 2011, \mn@doi [\aj] {10.1088/0004-6256/141/4/108},
  \href {http://adsabs.harvard.edu/abs/2011AJ....141..108C} {141, 108}

\bibitem[\protect\citeauthoryear{{Collier Cameron} et~al.,}{{Collier Cameron}
  et~al.}{2007}]{Collier2007}
{Collier Cameron} A.,  et~al., 2007, \mn@doi [\mnras]
  {10.1111/j.1365-2966.2007.12195.x}, \href
  {http://adsabs.harvard.edu/abs/2007MNRAS.380.1230C} {380, 1230}

\bibitem[\protect\citeauthoryear{{De Lee} et~al.,}{{De Lee}
  et~al.}{2013}]{DeLee2013}
{De Lee} N.,  et~al., 2013, \mn@doi [\aj] {10.1088/0004-6256/145/6/155}, \href
  {http://adsabs.harvard.edu/abs/2013AJ....145..155D} {145, 155}

\bibitem[\protect\citeauthoryear{{De Silva} et~al.,}{{De Silva}
  et~al.}{2015}]{DeSilva2015}
{De Silva} G.~M.,  et~al., 2015, \mn@doi [\mnras] {10.1093/mnras/stv327}, \href
  {http://adsabs.harvard.edu/abs/2015MNRAS.449.2604D} {449, 2604}

\bibitem[\protect\citeauthoryear{{Dehnen}}{{Dehnen}}{1998}]{Dehnen1998}
{Dehnen} W.,  1998, \mn@doi [\aj] {10.1086/300364}, \href
  {http://adsabs.harvard.edu/abs/1998AJ....115.2384D} {115, 2384}

\bibitem[\protect\citeauthoryear{{Dehnen}}{{Dehnen}}{2000}]{Dehnen2000}
{Dehnen} W.,  2000, \mn@doi [\aj] {10.1086/301226}, \href
  {http://adsabs.harvard.edu/abs/2000AJ....119..800D} {119, 800}

\bibitem[\protect\citeauthoryear{{Eggen}}{{Eggen}}{1996}]{Eggen1996}
{Eggen} O.~J.,  1996, \mn@doi [\aj] {10.1086/118126}, \href
  {http://adsabs.harvard.edu/abs/1996AJ....112.1595E} {112, 1595}

\bibitem[\protect\citeauthoryear{{Eisenstein} et~al.,}{{Eisenstein}
  et~al.}{2011}]{Eisenstein2011}
{Eisenstein} D.~J.,  et~al., 2011, \mn@doi [\aj] {10.1088/0004-6256/142/3/72},
  \href {http://adsabs.harvard.edu/abs/2011AJ....142...72E} {142, 72}

\bibitem[\protect\citeauthoryear{{Fleming} et~al.,}{{Fleming}
  et~al.}{2012}]{Fleming2012}
{Fleming} S.~W.,  et~al., 2012, \mn@doi [\aj] {10.1088/0004-6256/144/3/72},
  \href {http://adsabs.harvard.edu/abs/2012AJ....144...72F} {144, 72}

\bibitem[\protect\citeauthoryear{{Freeman} \& {Bland-Hawthorn}}{{Freeman} \&
  {Bland-Hawthorn}}{2002}]{Freeman2002}
{Freeman} K.,  {Bland-Hawthorn} J.,  2002, \mn@doi [\araa]
  {10.1146/annurev.astro.40.060401.093840}, \href
  {http://adsabs.harvard.edu/abs/2002ARA%26A..40..487F} {40, 487}

\bibitem[\protect\citeauthoryear{{Fuhrmann}}{{Fuhrmann}}{1998}]{Fuhrmann1998}
{Fuhrmann} K.,  1998, \aap, \href
  {http://adsabs.harvard.edu/abs/1998A%26A...338..161F} {338, 161}

\bibitem[\protect\citeauthoryear{{Fuhrmann}}{{Fuhrmann}}{2004}]{Fuhrmann2004}
{Fuhrmann} K.,  2004, \mn@doi [Astronomische Nachrichten]
  {10.1002/asna.200310173}, \href
  {http://adsabs.harvard.edu/abs/2004AN....325....3F} {325, 3}

\bibitem[\protect\citeauthoryear{{Fuhrmann}}{{Fuhrmann}}{2011}]{Fuhrmann2011}
{Fuhrmann} K.,  2011, \mn@doi [\mnras] {10.1111/j.1365-2966.2011.18476.x},
  \href {http://adsabs.harvard.edu/abs/2011MNRAS.414.2893F} {414, 2893}

\bibitem[\protect\citeauthoryear{{Gaia Collaboration}, {Brown}, {Vallenari},
  {Prusti}, {de Bruijne}, {Babusiaux}  \& {Bailer-Jones}}{{Gaia Collaboration}
  et~al.}{2018}]{Gaia2018}
{Gaia Collaboration} {Brown} A.~G.~A.,  {Vallenari} A.,  {Prusti} T.,  {de
  Bruijne} J.~H.~J.,  {Babusiaux} C.,   {Bailer-Jones} C.~A.~L.,  2018,
  preprint, \href {http://adsabs.harvard.edu/abs/2018arXiv180409365G} {}
  (\mn@eprint {arXiv} {1804.09365})

\bibitem[\protect\citeauthoryear{{Garc{\'{\i}}a P{\'e}rez}
  et~al.,}{{Garc{\'{\i}}a P{\'e}rez} et~al.}{2016}]{GarciaPerez2016}
{Garc{\'{\i}}a P{\'e}rez} A.~E.,  et~al., 2016, \mn@doi [\aj]
  {10.3847/0004-6256/151/6/144}, \href
  {http://adsabs.harvard.edu/abs/2016AJ....151..144G} {151, 144}

\bibitem[\protect\citeauthoryear{{Ge} et~al.,}{{Ge} et~al.}{2008}]{Ge2008}
{Ge} J.,  et~al., 2008, in {Fischer} D.,  {Rasio} F.~A.,  {Thorsett} S.~E.,
  {Wolszczan} A.,  eds,  Astronomical Society of the Pacific Conference Series
  Vol. 398, Extreme Solar Systems. p.~449

\bibitem[\protect\citeauthoryear{{Ge} et~al.,}{{Ge} et~al.}{2009}]{Ge2009}
{Ge} J.,  et~al., 2009, in Techniques and Instrumentation for Detection of
  Exoplanets IV. p. 74400L, \mn@doi{10.1117/12.826651}

\bibitem[\protect\citeauthoryear{{Ghezzi} et~al.,}{{Ghezzi}
  et~al.}{2014}]{Ghezzi2014}
{Ghezzi} L.,  et~al., 2014, \mn@doi [\aj] {10.1088/0004-6256/148/6/105}, \href
  {http://adsabs.harvard.edu/abs/2014AJ....148..105G} {148, 105}

\bibitem[\protect\citeauthoryear{{Grieves} et~al.,}{{Grieves}
  et~al.}{2017}]{Grieves2017}
{Grieves} N.,  et~al., 2017, \mn@doi [\mnras] {10.1093/mnras/stx334}, \href
  {http://adsabs.harvard.edu/abs/2017MNRAS.467.4264G} {467, 4264}

\bibitem[\protect\citeauthoryear{{Guiglion} et~al.,}{{Guiglion}
  et~al.}{2015}]{Guiglion2015}
{Guiglion} G.,  et~al., 2015, \mn@doi [\aap] {10.1051/0004-6361/201525883},
  \href {http://adsabs.harvard.edu/abs/2015A%26A...583A..91G} {583, A91}

\bibitem[\protect\citeauthoryear{{Gunn} et~al.,}{{Gunn}
  et~al.}{2006}]{Gunn2006}
{Gunn} J.~E.,  et~al., 2006, \mn@doi [\aj] {10.1086/500975}, \href
  {http://adsabs.harvard.edu/abs/2006AJ....131.2332G} {131, 2332}

\bibitem[\protect\citeauthoryear{{Hayden} et~al.,}{{Hayden}
  et~al.}{2015}]{Hayden2015}
{Hayden} M.~R.,  et~al., 2015, \mn@doi [\apj] {10.1088/0004-637X/808/2/132},
  \href {http://adsabs.harvard.edu/abs/2015ApJ...808..132H} {808, 132}

\bibitem[\protect\citeauthoryear{{Hayden}, {Recio-Blanco}, {de Laverny},
  {Mikolaitis}  \& {Worley}}{{Hayden} et~al.}{2017}]{Hayden2017}
{Hayden} M.~R.,  {Recio-Blanco} A.,  {de Laverny} P.,  {Mikolaitis} S.,
  {Worley} C.~C.,  2017, \mn@doi [\aap] {10.1051/0004-6361/201731494}, \href
  {http://adsabs.harvard.edu/abs/2017A%26A...608L...1H} {608, L1}

\bibitem[\protect\citeauthoryear{{Hayden} et~al.,}{{Hayden}
  et~al.}{2018}]{Hayden2018}
{Hayden} M.~R.,  et~al., 2018, \mn@doi [\aap] {10.1051/0004-6361/201730412},
  \href {http://adsabs.harvard.edu/abs/2018A%26A...609A..79H} {609, A79}

\bibitem[\protect\citeauthoryear{{Haywood}, {Di Matteo}, {Lehnert}, {Katz}  \&
  {G{\'o}mez}}{{Haywood} et~al.}{2013}]{Haywood2013}
{Haywood} M.,  {Di Matteo} P.,  {Lehnert} M.~D.,  {Katz} D.,   {G{\'o}mez} A.,
  2013, \mn@doi [\aap] {10.1051/0004-6361/201321397}, \href
  {http://adsabs.harvard.edu/abs/2013A%26A...560A.109H} {560, A109}

\bibitem[\protect\citeauthoryear{{Heiter}, {Jofr{\'e}}, {Gustafsson}, {Korn},
  {Soubiran}  \& {Th{\'e}venin}}{{Heiter} et~al.}{2015}]{Heiter2015}
{Heiter} U.,  {Jofr{\'e}} P.,  {Gustafsson} B.,  {Korn} A.~J.,  {Soubiran} C.,
   {Th{\'e}venin} F.,  2015, \mn@doi [\aap] {10.1051/0004-6361/201526319},
  \href {http://adsabs.harvard.edu/abs/2015A%26A...582A..49H} {582, A49}

\bibitem[\protect\citeauthoryear{{Hernquist}}{{Hernquist}}{1990}]{Hernquist1990}
{Hernquist} L.,  1990, \mn@doi [\apj] {10.1086/168845}, \href
  {http://adsabs.harvard.edu/abs/1990ApJ...356..359H} {356, 359}

\bibitem[\protect\citeauthoryear{{Holmberg}, {Nordstr{\"o}m}  \&
  {Andersen}}{{Holmberg} et~al.}{2007}]{Holmberg2007}
{Holmberg} J.,  {Nordstr{\"o}m} B.,   {Andersen} J.,  2007, \mn@doi [\aap]
  {10.1051/0004-6361:20077221}, \href
  {http://adsabs.harvard.edu/abs/2007A%26A...475..519H} {475, 519}

\bibitem[\protect\citeauthoryear{{Holtzman} et~al.,}{{Holtzman}
  et~al.}{2015}]{Holtzman2015}
{Holtzman} J.~A.,  et~al., 2015, \mn@doi [\aj] {10.1088/0004-6256/150/5/148},
  \href {http://adsabs.harvard.edu/abs/2015AJ....150..148H} {150, 148}

\bibitem[\protect\citeauthoryear{{Huber} et~al.,}{{Huber}
  et~al.}{2013}]{Huber2013}
{Huber} D.,  et~al., 2013, \mn@doi [\apj] {10.1088/0004-637X/767/2/127}, \href
  {http://adsabs.harvard.edu/abs/2013ApJ...767..127H} {767, 127}

\bibitem[\protect\citeauthoryear{{Jiang} et~al.,}{{Jiang}
  et~al.}{2013}]{Jiang2013}
{Jiang} P.,  et~al., 2013, \mn@doi [\aj] {10.1088/0004-6256/146/3/65}, \href
  {http://adsabs.harvard.edu/abs/2013AJ....146...65J} {146, 65}

\bibitem[\protect\citeauthoryear{{Johnson} \& {Soderblom}}{{Johnson} \&
  {Soderblom}}{1987}]{Johnson1987}
{Johnson} D.~R.~H.,  {Soderblom} D.~R.,  1987, \mn@doi [\aj] {10.1086/114370},
  \href {http://adsabs.harvard.edu/abs/1987AJ.....93..864J} {93, 864}

\bibitem[\protect\citeauthoryear{{Kilic}, {Munn}, {Harris}, {von Hippel},
  {Liebert}, {Williams}, {Jeffery}  \& {DeGennaro}}{{Kilic}
  et~al.}{2017}]{Kilic2017}
{Kilic} M.,  {Munn} J.~A.,  {Harris} H.~C.,  {von Hippel} T.,  {Liebert} J.~W.,
   {Williams} K.~A.,  {Jeffery} E.,   {DeGennaro} S.,  2017, \mn@doi [\apj]
  {10.3847/1538-4357/aa62a5}, \href
  {http://adsabs.harvard.edu/abs/2017ApJ...837..162K} {837, 162}

\bibitem[\protect\citeauthoryear{{Kubryk}, {Prantzos}  \&
  {Athanassoula}}{{Kubryk} et~al.}{2015}]{Kubryk2015}
{Kubryk} M.,  {Prantzos} N.,   {Athanassoula} E.,  2015, \mn@doi [\aap]
  {10.1051/0004-6361/201424171}, \href
  {http://adsabs.harvard.edu/abs/2015A%26A...580A.126K} {580, A126}

\bibitem[\protect\citeauthoryear{{Kunder} et~al.,}{{Kunder}
  et~al.}{2017}]{Kunder2017}
{Kunder} A.,  et~al., 2017, \mn@doi [\aj] {10.3847/1538-3881/153/2/75}, \href
  {http://adsabs.harvard.edu/abs/2017AJ....153...75K} {153, 75}

\bibitem[\protect\citeauthoryear{{Kushniruk}, {Schirmer}  \&
  {Bensby}}{{Kushniruk} et~al.}{2017}]{Kushniruk2017}
{Kushniruk} I.,  {Schirmer} T.,   {Bensby} T.,  2017, \mn@doi [\aap]
  {10.1051/0004-6361/201731147}, \href
  {http://adsabs.harvard.edu/abs/2017A%26A...608A..73K} {608, A73}

\bibitem[\protect\citeauthoryear{{Lachaume}, {Dominik}, {Lanz}  \&
  {Habing}}{{Lachaume} et~al.}{1999}]{Lachaume1999}
{Lachaume} R.,  {Dominik} C.,  {Lanz} T.,   {Habing} H.~J.,  1999, \aap, \href
  {http://adsabs.harvard.edu/abs/1999A%26A...348..897L} {348, 897}

\bibitem[\protect\citeauthoryear{{Lasker} et~al.,}{{Lasker}
  et~al.}{2008}]{Lasker2008}
{Lasker} B.~M.,  et~al., 2008, \mn@doi [\aj] {10.1088/0004-6256/136/2/735},
  \href {http://adsabs.harvard.edu/abs/2008AJ....136..735L} {136, 735}

\bibitem[\protect\citeauthoryear{{Lee} et~al.,}{{Lee} et~al.}{2008}]{Lee2008}
{Lee} Y.~S.,  et~al., 2008, \mn@doi [\aj] {10.1088/0004-6256/136/5/2022}, \href
  {http://adsabs.harvard.edu/abs/2008AJ....136.2022L} {136, 2022}

\bibitem[\protect\citeauthoryear{{Lee} et~al.,}{{Lee} et~al.}{2011a}]{Lee2011a}
{Lee} B.~L.,  et~al., 2011a, \mn@doi [\apj] {10.1088/0004-637X/728/1/32}, \href
  {http://adsabs.harvard.edu/abs/2011ApJ...728...32L} {728, 32}

\bibitem[\protect\citeauthoryear{{Lee} et~al.,}{{Lee} et~al.}{2011b}]{Lee2011b}
{Lee} Y.~S.,  et~al., 2011b, \mn@doi [\apj] {10.1088/0004-637X/738/2/187},
  \href {http://adsabs.harvard.edu/abs/2011ApJ...738..187L} {738, 187}

\bibitem[\protect\citeauthoryear{{Ma} et~al.,}{{Ma} et~al.}{2013}]{Ma2013}
{Ma} B.,  et~al., 2013, \mn@doi [\aj] {10.1088/0004-6256/145/1/20}, \href
  {http://adsabs.harvard.edu/abs/2013AJ....145...20M} {145, 20}

\bibitem[\protect\citeauthoryear{{Ma} et~al.,}{{Ma} et~al.}{2016}]{Ma2016}
{Ma} B.,  et~al., 2016, \mn@doi [\aj] {10.3847/0004-6256/152/5/112}, \href
  {http://adsabs.harvard.edu/abs/2016AJ....152..112M} {152, 112}

\bibitem[\protect\citeauthoryear{{Majewski}}{{Majewski}}{1993}]{Majewski1993}
{Majewski} S.~R.,  1993, \mn@doi [\araa] {10.1146/annurev.aa.31.090193.003043},
  \href {http://adsabs.harvard.edu/abs/1993ARA%26A..31..575M} {31, 575}

\bibitem[\protect\citeauthoryear{{Majewski} et~al.,}{{Majewski}
  et~al.}{2017}]{Majewski2017}
{Majewski} S.~R.,  et~al., 2017, \mn@doi [\aj] {10.3847/1538-3881/aa784d},
  \href {http://adsabs.harvard.edu/abs/2017AJ....154...94M} {154, 94}

\bibitem[\protect\citeauthoryear{{Mathur} et~al.,}{{Mathur}
  et~al.}{2017}]{Mathur2017}
{Mathur} S.,  et~al., 2017, \mn@doi [\apjs] {10.3847/1538-4365/229/2/30}, \href
  {http://adsabs.harvard.edu/abs/2017ApJS..229...30M} {229, 30}

\bibitem[\protect\citeauthoryear{{M{\'e}sz{\'a}ros} et~al.,}{{M{\'e}sz{\'a}ros}
  et~al.}{2013}]{Meszaros2013}
{M{\'e}sz{\'a}ros} S.,  et~al., 2013, \mn@doi [\aj]
  {10.1088/0004-6256/146/5/133}, \href
  {http://adsabs.harvard.edu/abs/2013AJ....146..133M} {146, 133}

\bibitem[\protect\citeauthoryear{{Minchev} \& {Famaey}}{{Minchev} \&
  {Famaey}}{2010}]{Minchev2010}
{Minchev} I.,  {Famaey} B.,  2010, \mn@doi [\apj]
  {10.1088/0004-637X/722/1/112}, \href
  {http://adsabs.harvard.edu/abs/2010ApJ...722..112M} {722, 112}

\bibitem[\protect\citeauthoryear{{Minchev}, {Chiappini}  \& {Martig}}{{Minchev}
  et~al.}{2013}]{Minchev2013}
{Minchev} I.,  {Chiappini} C.,   {Martig} M.,  2013, \mn@doi [\aap]
  {10.1051/0004-6361/201220189}, \href
  {http://adsabs.harvard.edu/abs/2013A%26A...558A...9M} {558, A9}

\bibitem[\protect\citeauthoryear{{Minchev}, {Chiappini}  \& {Martig}}{{Minchev}
  et~al.}{2014}]{Minchev2014}
{Minchev} I.,  {Chiappini} C.,   {Martig} M.,  2014, \mn@doi [\aap]
  {10.1051/0004-6361/201423487}, \href
  {http://adsabs.harvard.edu/abs/2014A%26A...572A..92M} {572, A92}

\bibitem[\protect\citeauthoryear{{Miyamoto} \& {Nagai}}{{Miyamoto} \&
  {Nagai}}{1975}]{Miyamoto1975}
{Miyamoto} M.,  {Nagai} R.,  1975, \pasj, \href
  {http://adsabs.harvard.edu/abs/1975PASJ...27..533M} {27, 533}

\bibitem[\protect\citeauthoryear{{Mortier}, {Santos}, {Sousa}, {Fernandes},
  {Adibekyan}, {Delgado Mena}, {Montalto}  \& {Israelian}}{{Mortier}
  et~al.}{2013}]{Mortier2013}
{Mortier} A.,  {Santos} N.~C.,  {Sousa} S.~G.,  {Fernandes} J.~M.,  {Adibekyan}
  V.~Z.,  {Delgado Mena} E.,  {Montalto} M.,   {Israelian} G.,  2013, \mn@doi
  [\aap] {10.1051/0004-6361/201322240}, \href
  {http://adsabs.harvard.edu/abs/2013A%26A...558A.106M} {558, A106}

\bibitem[\protect\citeauthoryear{{Mortier}, {Sousa}, {Adibekyan}, {Brand{\~a}o}
   \& {Santos}}{{Mortier} et~al.}{2014}]{Mortier2014}
{Mortier} A.,  {Sousa} S.~G.,  {Adibekyan} V.~Z.,  {Brand{\~a}o} I.~M.,
  {Santos} N.~C.,  2014, \mn@doi [\aap] {10.1051/0004-6361/201424537}, \href
  {http://adsabs.harvard.edu/abs/2014A%26A...572A..95M} {572, A95}

\bibitem[\protect\citeauthoryear{{Navarro}, {Frenk}  \& {White}}{{Navarro}
  et~al.}{1997}]{Navarro1997}
{Navarro} J.~F.,  {Frenk} C.~S.,   {White} S.~D.~M.,  1997, \mn@doi [\apj]
  {10.1086/304888}, \href {http://adsabs.harvard.edu/abs/1997ApJ...490..493N}
  {490, 493}

\bibitem[\protect\citeauthoryear{{Navarro}, {Helmi}  \& {Freeman}}{{Navarro}
  et~al.}{2004}]{Navarro2004}
{Navarro} J.~F.,  {Helmi} A.,   {Freeman} K.~C.,  2004, \mn@doi [\apjl]
  {10.1086/381751}, \href {http://adsabs.harvard.edu/abs/2004ApJ...601L..43N}
  {601, L43}

\bibitem[\protect\citeauthoryear{{Navarro}, {Abadi}, {Venn}, {Freeman}  \&
  {Anguiano}}{{Navarro} et~al.}{2011}]{Navarro2011}
{Navarro} J.~F.,  {Abadi} M.~G.,  {Venn} K.~A.,  {Freeman} K.~C.,   {Anguiano}
  B.,  2011, \mn@doi [\mnras] {10.1111/j.1365-2966.2010.17975.x}, \href
  {http://adsabs.harvard.edu/abs/2011MNRAS.412.1203N} {412, 1203}

\bibitem[\protect\citeauthoryear{{Nidever}, {Marcy}, {Butler}, {Fischer}  \&
  {Vogt}}{{Nidever} et~al.}{2002}]{Nidever2002}
{Nidever} D.~L.,  {Marcy} G.~W.,  {Butler} R.~P.,  {Fischer} D.~A.,   {Vogt}
  S.~S.,  2002, \mn@doi [\apjs] {10.1086/340570}, \href
  {http://adsabs.harvard.edu/abs/2002ApJS..141..503N} {141, 503}

\bibitem[\protect\citeauthoryear{{Nissen}}{{Nissen}}{2004}]{Nissen2004}
{Nissen} P.~E.,  2004, Origin and Evolution of the Elements, \href
  {http://adsabs.harvard.edu/abs/2004oee..symp..154N} {p.~154}

\bibitem[\protect\citeauthoryear{{Nissen}}{{Nissen}}{2015}]{Nissen2015}
{Nissen} P.~E.,  2015, \mn@doi [\aap] {10.1051/0004-6361/201526269}, \href
  {http://adsabs.harvard.edu/abs/2015A%26A...579A..52N} {579, A52}

\bibitem[\protect\citeauthoryear{{Nordstr{\"o}m} et~al.,}{{Nordstr{\"o}m}
  et~al.}{2004}]{Nordstrom2004}
{Nordstr{\"o}m} B.,  et~al., 2004, \mn@doi [\aap] {10.1051/0004-6361:20035959},
  \href {http://adsabs.harvard.edu/abs/2004A%26A...418..989N} {418, 989}

\bibitem[\protect\citeauthoryear{{Paegert} et~al.,}{{Paegert}
  et~al.}{2015}]{Paegert2015}
{Paegert} M.,  et~al., 2015, \mn@doi [\aj] {10.1088/0004-6256/149/6/186}, \href
  {http://adsabs.harvard.edu/abs/2015AJ....149..186P} {149, 186}

\bibitem[\protect\citeauthoryear{{Pecaut} \& {Mamajek}}{{Pecaut} \&
  {Mamajek}}{2013}]{Pecaut2013}
{Pecaut} M.~J.,  {Mamajek} E.~E.,  2013, \mn@doi [\apjs]
  {10.1088/0067-0049/208/1/9}, \href
  {http://adsabs.harvard.edu/abs/2013ApJS..208....9P} {208, 9}

\bibitem[\protect\citeauthoryear{{Queiroz} et~al.,}{{Queiroz}
  et~al.}{2018}]{Queiroz2018}
{Queiroz} A.~B.~A.,  et~al., 2018, \mn@doi [\mnras] {10.1093/mnras/sty330},
  \href {http://adsabs.harvard.edu/abs/2018MNRAS.tmp..326Q} {}

\bibitem[\protect\citeauthoryear{{Quillen} et~al.,}{{Quillen}
  et~al.}{2018}]{Quillen2018}
{Quillen} A.~C.,  et~al., 2018, \mn@doi [\mnras] {10.1093/mnras/sty865}, \href
  {http://adsabs.harvard.edu/abs/2018MNRAS.478..228Q} {478, 228}

\bibitem[\protect\citeauthoryear{{Raboud}, {Grenon}, {Martinet}, {Fux}  \&
  {Udry}}{{Raboud} et~al.}{1998}]{Raboud1998}
{Raboud} D.,  {Grenon} M.,  {Martinet} L.,  {Fux} R.,   {Udry} S.,  1998, \aap,
  \href {http://adsabs.harvard.edu/abs/1998A%26A...335L..61R} {335, L61}

\bibitem[\protect\citeauthoryear{{Ram{\'{\i}}rez} et~al.,}{{Ram{\'{\i}}rez}
  et~al.}{2014}]{Ramirez2014}
{Ram{\'{\i}}rez} I.,  et~al., 2014, \mn@doi [\aap]
  {10.1051/0004-6361/201424244}, \href
  {http://adsabs.harvard.edu/abs/2014A%26A...572A..48R} {572, A48}

\bibitem[\protect\citeauthoryear{{Ramya}, {Reddy}  \& {Lambert}}{{Ramya}
  et~al.}{2012}]{Ramya2012}
{Ramya} P.,  {Reddy} B.~E.,   {Lambert} D.~L.,  2012, \mn@doi [\mnras]
  {10.1111/j.1365-2966.2012.21677.x}, \href
  {http://adsabs.harvard.edu/abs/2012MNRAS.425.3188R} {425, 3188}

\bibitem[\protect\citeauthoryear{{Ramya}, {Reddy}, {Lambert}  \&
  {Musthafa}}{{Ramya} et~al.}{2016}]{Ramya2016}
{Ramya} P.,  {Reddy} B.~E.,  {Lambert} D.~L.,   {Musthafa} M.~M.,  2016,
  \mn@doi [\mnras] {10.1093/mnras/stw852}, \href
  {http://adsabs.harvard.edu/abs/2016MNRAS.460.1356R} {460, 1356}

\bibitem[\protect\citeauthoryear{{Recio-Blanco} et~al.,}{{Recio-Blanco}
  et~al.}{2014}]{Recio-Blanco2014}
{Recio-Blanco} A.,  et~al., 2014, \mn@doi [\aap] {10.1051/0004-6361/201322944},
  \href {http://adsabs.harvard.edu/abs/2014A%26A...567A...5R} {567, A5}

\bibitem[\protect\citeauthoryear{{Ribas} et~al.,}{{Ribas}
  et~al.}{2010}]{Ribas2010}
{Ribas} I.,  et~al., 2010, \mn@doi [\apj] {10.1088/0004-637X/714/1/384}, \href
  {http://adsabs.harvard.edu/abs/2010ApJ...714..384R} {714, 384}

\bibitem[\protect\citeauthoryear{{Rix} \& {Bovy}}{{Rix} \&
  {Bovy}}{2013}]{Rix2013}
{Rix} H.-W.,  {Bovy} J.,  2013, \mn@doi [\aapr] {10.1007/s00159-013-0061-8},
  \href {http://adsabs.harvard.edu/abs/2013A%26ARv..21...61R} {21, 61}

\bibitem[\protect\citeauthoryear{{Ro{\v s}kar}, {Debattista}, {Stinson},
  {Quinn}, {Kaufmann}  \& {Wadsley}}{{Ro{\v s}kar} et~al.}{2008}]{Roskar2008}
{Ro{\v s}kar} R.,  {Debattista} V.~P.,  {Stinson} G.~S.,  {Quinn} T.~R.,
  {Kaufmann} T.,   {Wadsley} J.,  2008, \mn@doi [\apjl] {10.1086/586734}, \href
  {http://adsabs.harvard.edu/abs/2008ApJ...675L..65R} {675, L65}

\bibitem[\protect\citeauthoryear{{Santiago} et~al.,}{{Santiago}
  et~al.}{2016}]{Santiago2016}
{Santiago} B.~X.,  et~al., 2016, \mn@doi [\aap] {10.1051/0004-6361/201323177},
  \href {http://adsabs.harvard.edu/abs/2016A%26A...585A..42S} {585, A42}

\bibitem[\protect\citeauthoryear{{Sch{\"o}nrich} \& {Binney}}{{Sch{\"o}nrich}
  \& {Binney}}{2009}]{Schonrich2009}
{Sch{\"o}nrich} R.,  {Binney} J.,  2009, \mn@doi [\mnras]
  {10.1111/j.1365-2966.2009.14750.x}, \href
  {http://adsabs.harvard.edu/abs/2009MNRAS.396..203S} {396, 203}

\bibitem[\protect\citeauthoryear{{Sch{\"o}nrich}, {Binney}  \&
  {Dehnen}}{{Sch{\"o}nrich} et~al.}{2010}]{Schonrich2010}
{Sch{\"o}nrich} R.,  {Binney} J.,   {Dehnen} W.,  2010, \mn@doi [\mnras]
  {10.1111/j.1365-2966.2010.16253.x}, \href
  {http://adsabs.harvard.edu/abs/2010MNRAS.403.1829S} {403, 1829}

\bibitem[\protect\citeauthoryear{{Sellwood} \& {Binney}}{{Sellwood} \&
  {Binney}}{2002}]{Sellwood2002}
{Sellwood} J.~A.,  {Binney} J.~J.,  2002, \mn@doi [\mnras]
  {10.1046/j.1365-8711.2002.05806.x}, \href
  {http://adsabs.harvard.edu/abs/2002MNRAS.336..785S} {336, 785}

\bibitem[\protect\citeauthoryear{{Skrutskie} et~al.,}{{Skrutskie}
  et~al.}{2006}]{Skrutskie2006}
{Skrutskie} M.~F.,  et~al., 2006, \mn@doi [\aj] {10.1086/498708}, \href
  {http://adsabs.harvard.edu/abs/2006AJ....131.1163S} {131, 1163}

\bibitem[\protect\citeauthoryear{{Smolinski} et~al.,}{{Smolinski}
  et~al.}{2011}]{Smolinski2011}
{Smolinski} J.~P.,  et~al., 2011, \mn@doi [\aj] {10.1088/0004-6256/141/3/89},
  \href {http://adsabs.harvard.edu/abs/2011AJ....141...89S} {141, 89}

\bibitem[\protect\citeauthoryear{{Soderblom}}{{Soderblom}}{2010}]{Soderblom2010}
{Soderblom} D.~R.,  2010, \mn@doi [\araa]
  {10.1146/annurev-astro-081309-130806}, \href
  {http://adsabs.harvard.edu/abs/2010ARA%26A..48..581S} {48, 581}

\bibitem[\protect\citeauthoryear{{Sousa}}{{Sousa}}{2014}]{Sousa2014}
{Sousa} S.~G.,  2014, {ARES + MOOG: A Practical Overview of an Equivalent Width
  (EW) Method to Derive Stellar Parameters}.
pp 297--310, \mn@doi{10.1007/978-3-319-06956-2_26}

\bibitem[\protect\citeauthoryear{{Steinmetz} et~al.,}{{Steinmetz}
  et~al.}{2006}]{Steinmetz2006}
{Steinmetz} M.,  et~al., 2006, \mn@doi [\aj] {10.1086/506564}, \href
  {http://adsabs.harvard.edu/abs/2006AJ....132.1645S} {132, 1645}

\bibitem[\protect\citeauthoryear{{Takeda}, {Ford}, {Sills}, {Rasio}, {Fischer}
  \& {Valenti}}{{Takeda} et~al.}{2007}]{Takeda2007}
{Takeda} G.,  {Ford} E.~B.,  {Sills} A.,  {Rasio} F.~A.,  {Fischer} D.~A.,
  {Valenti} J.~A.,  2007, \mn@doi [\apjs] {10.1086/509763}, \href
  {http://adsabs.harvard.edu/abs/2007ApJS..168..297T} {168, 297}

\bibitem[\protect\citeauthoryear{{Tang}, {Bressan}, {Rosenfield}, {Slemer},
  {Marigo}, {Girardi}  \& {Bianchi}}{{Tang} et~al.}{2014}]{Tang2014}
{Tang} J.,  {Bressan} A.,  {Rosenfield} P.,  {Slemer} A.,  {Marigo} P.,
  {Girardi} L.,   {Bianchi} L.,  2014, \mn@doi [\mnras]
  {10.1093/mnras/stu2029}, \href
  {http://adsabs.harvard.edu/abs/2014MNRAS.445.4287T} {445, 4287}

\bibitem[\protect\citeauthoryear{{The Astropy Collaboration} et~al.,}{{The
  Astropy Collaboration} et~al.}{2018}]{Astropy2018}
{The Astropy Collaboration} et~al., 2018, preprint, \href
  {http://adsabs.harvard.edu/abs/2018arXiv180102634T} {} (\mn@eprint {arXiv}
  {1801.02634})

\bibitem[\protect\citeauthoryear{Thomas}{Thomas}{2015}]{Thomas2015}
Thomas N.,  2015, PhD thesis, University of Florida, Gainesville, FL

\bibitem[\protect\citeauthoryear{{Thomas}, {Ge}, {Grieves}, {Li}  \&
  {Sithajan}}{{Thomas} et~al.}{2016}]{Thomas2016}
{Thomas} N.,  {Ge} J.,  {Grieves} N.,  {Li} R.,   {Sithajan} S.,  2016, \mn@doi
  [\pasp] {10.1088/1538-3873/128/962/045003}, \href
  {http://adsabs.harvard.edu/abs/2016PASP..128d5003T} {128, 045003}

\bibitem[\protect\citeauthoryear{{Tody}}{{Tody}}{1993}]{Tody1993}
{Tody} D.,  1993, in {Hanisch} R.~J.,  {Brissenden} R.~J.~V.,   {Barnes} J.,
  eds,  Astronomical Society of the Pacific Conference Series Vol. 52,
  Astronomical Data Analysis Software and Systems II. p.~173

\bibitem[\protect\citeauthoryear{{Torres}, {Andersen}  \&
  {Gim{\'e}nez}}{{Torres} et~al.}{2010}]{Torres2010a}
{Torres} G.,  {Andersen} J.,   {Gim{\'e}nez} A.,  2010, \mn@doi [\aapr]
  {10.1007/s00159-009-0025-1}, \href
  {http://adsabs.harvard.edu/abs/2010A%26ARv..18...67T} {18, 67}

\bibitem[\protect\citeauthoryear{{Torres}, {Fischer}, {Sozzetti}, {Buchhave},
  {Winn}, {Holman}  \& {Carter}}{{Torres} et~al.}{2012}]{Torres2012}
{Torres} G.,  {Fischer} D.~A.,  {Sozzetti} A.,  {Buchhave} L.~A.,  {Winn}
  J.~N.,  {Holman} M.~J.,   {Carter} J.~A.,  2012, \mn@doi [\apj]
  {10.1088/0004-637X/757/2/161}, \href
  {http://adsabs.harvard.edu/abs/2012ApJ...757..161T} {757, 161}

\bibitem[\protect\citeauthoryear{{Trevisan}, {Barbuy}, {Eriksson},
  {Gustafsson}, {Grenon}  \& {Pomp{\'e}ia}}{{Trevisan}
  et~al.}{2011}]{Trevisan2011}
{Trevisan} M.,  {Barbuy} B.,  {Eriksson} K.,  {Gustafsson} B.,  {Grenon} M.,
  {Pomp{\'e}ia} L.,  2011, \mn@doi [\aap] {10.1051/0004-6361/201016056}, \href
  {http://adsabs.harvard.edu/abs/2011A%26A...535A..42T} {535, A42}

\bibitem[\protect\citeauthoryear{{Valentini} et~al.,}{{Valentini}
  et~al.}{2017}]{Valentini2017}
{Valentini} M.,  et~al., 2017, \mn@doi [\aap] {10.1051/0004-6361/201629701},
  \href {http://adsabs.harvard.edu/abs/2017A%26A...600A..66V} {600, A66}

\bibitem[\protect\citeauthoryear{{Williams}, {Freeman}, {Helmi}  \& {RAVE
  Collaboration}}{{Williams} et~al.}{2009}]{Williams2009}
{Williams} M.~E.~K.,  {Freeman} K.~C.,  {Helmi} A.,   {RAVE Collaboration}
  2009, in {Andersen} J.,  {Nordstr{\"o}ara} {m} B.,   {Bland-Hawthorn} J.,
  eds,  IAU Symposium Vol. 254, The Galaxy Disk in Cosmological Context. pp
  139--144 (\mn@eprint {arXiv} {0810.2669}), \mn@doi{10.1017/S1743921308027518}

\bibitem[\protect\citeauthoryear{{Wisniewski} et~al.,}{{Wisniewski}
  et~al.}{2012}]{Wisniewski2012}
{Wisniewski} J.~P.,  et~al., 2012, \mn@doi [\aj] {10.1088/0004-6256/143/5/107},
  \href {http://adsabs.harvard.edu/abs/2012AJ....143..107W} {143, 107}

\bibitem[\protect\citeauthoryear{{Wu} et~al.,}{{Wu} et~al.}{2011}]{Wu2011}
{Wu} Y.,  et~al., 2011, \mn@doi [Research in Astronomy and Astrophysics]
  {10.1088/1674-4527/11/8/006}, \href
  {http://adsabs.harvard.edu/abs/2011RAA....11..924W} {11, 924}

\bibitem[\protect\citeauthoryear{{Wu}, {Du}, {Luo}, {Zhao}  \& {Yuan}}{{Wu}
  et~al.}{2014}]{Wu2014}
{Wu} Y.,  {Du} B.,  {Luo} A.,  {Zhao} Y.,   {Yuan} H.,  2014, in {Heavens} A.,
  {Starck} J.-L.,   {Krone-Martins} A.,  eds,  IAU Symposium Vol. 306,
  Statistical Challenges in 21st Century Cosmology. pp 340--342 (\mn@eprint
  {arXiv} {1407.1980}), \mn@doi{10.1017/S1743921314010825}

\bibitem[\protect\citeauthoryear{{Yanny} et~al.,}{{Yanny}
  et~al.}{2009}]{Yanny2009}
{Yanny} B.,  et~al., 2009, \mn@doi [\aj] {10.1088/0004-6256/137/5/4377}, \href
  {http://adsabs.harvard.edu/abs/2009AJ....137.4377Y} {137, 4377}

\bibitem[\protect\citeauthoryear{{Yi}, {Kim}  \& {Demarque}}{{Yi}
  et~al.}{2003}]{Yi2003}
{Yi} S.~K.,  {Kim} Y.-C.,   {Demarque} P.,  2003, \mn@doi [\apjs]
  {10.1086/345101}, \href {http://adsabs.harvard.edu/abs/2003ApJS..144..259Y}
  {144, 259}

\bibitem[\protect\citeauthoryear{{York} et~al.,}{{York}
  et~al.}{2000}]{York2000}
{York} D.~G.,  et~al., 2000, \mn@doi [\aj] {10.1086/301513}, \href
  {http://adsabs.harvard.edu/abs/2000AJ....120.1579Y} {120, 1579}

\bibitem[\protect\citeauthoryear{{Zhao}, {Zhao}, {Chu}, {Jing}  \&
  {Deng}}{{Zhao} et~al.}{2012}]{Zhao2012}
{Zhao} G.,  {Zhao} Y.-H.,  {Chu} Y.-Q.,  {Jing} Y.-P.,   {Deng} L.-C.,  2012,
  \mn@doi [Research in Astronomy and Astrophysics]
  {10.1088/1674-4527/12/7/002}, \href
  {http://adsabs.harvard.edu/abs/2012RAA....12..723Z} {12, 723}

\bibitem[\protect\citeauthoryear{{van Leeuwen}}{{van
  Leeuwen}}{2007}]{vanLeeuwen2007}
{van Leeuwen} F.,  2007, \mn@doi [\aap] {10.1051/0004-6361:20078357}, \href
  {http://adsabs.harvard.edu/abs/2007A%26A...474..653V} {474, 653}

\makeatother
\end{thebibliography}







\bsp	
\label{lastpage}
\end{document}